\documentclass[a4paper,11pt]{article}
\pdfoutput=1 

\usepackage{jheppub}
\makeatletter
\renewcommand{\@fpheader}{}
\makeatother

\usepackage[T1]{fontenc} 

\usepackage[svgnames]{xcolor}
\usepackage{parskip}
\usepackage{breqn}
\allowdisplaybreaks
\usepackage{slashed}
\usepackage{bm}
\usepackage{appendix}
\usepackage{comment}

\title{\boldmath Quality time in extra dimensions: axion quality and radion cosmology}

\author[a]{Falguni Biswas,}
\author[a]{Yeshan Juberi,}
\author[b]{Ahmed Rakin Kamal}

\affiliation[a]{Department of Mathematics and Physical Sciences,
BRAC University, Kha 224 Pragati Sarani, Merul Badda , Dhaka 1212, Bangladesh}
\affiliation[b]{Department of Theoretical Physics and Astrophysics, Faculty of Science,
Masaryk University, Kotl\'a\v{r}sk\'a 2, CS-61137 Brno, Czechia}

\emailAdd{ahmedrakinkamaltunok@gmail.com}

\abstract{Higher-form symmetries can protect extra-dimensional axions, but additional symmetry breaking can still spoil the solution to the strong CP problem. We focus on electrically charged matter and magnetic gauging in a five-dimensional circle compactification where the axion arises from an Abelian gauge field and the radius is dynamical. We discuss two mechanisms for stabilising the radius/radion using Casimir energy from fields neutral under the Abelian gauge field, with and without an additional periodic scalar winding around the compact circle. Both admit Minkowski and de Sitter minima. At the Minkowski minima (with radion masses of $100$--$170\,\mathrm{TeV}$), electrically charged matter induces a residual strong CP angle of order $10^{-17}$ and shifts the radius negligibly. Magnetic gauging remains compatible after including its additional degree of freedom in the Casimir spectrum, provided its coupling is sufficiently small. We observe that increasing the radius suppresses charged-matter breaking relative to QCD but increases magnetic breaking contribution. We also determine the radion decay channels, including the graviphoton, and study reheating from a possible radion-dominated era without scalar winding or magnetic gauging. For a non-minimal Higgs–curvature coupling $\xi=7$, the visible-sector reheating temperature is about $85\,\mathrm{MeV}$ with a dark-radiation contribution of $\Delta N_{\rm eff}\approx0.02$. The entropy released during radion decay dilutes the axion abundance, allowing axions with $f_a=10^{14}\,\mathrm{GeV}$ and an initial misalignment angle $\vartheta_i\approx0.43$ to account for the entire observed dark matter abundance. For the same decay constant, reproducing this abundance in radiation domination requires $\vartheta_i\approx0.065$.}

\begin{document} 
\maketitle
\flushbottom
\section{Introduction}

The discovery of an ever-growing number of hadrons in the 1950s made the strong interactions appear increasingly complicated. A remarkable simplification came in 1961, when Gell-Mann and Ne'eman organised these particles using an approximate flavour symmetry, $SU(3)_f$, in the Eightfold Way \cite{GellMann1961Eightfold,Neeman1961}. Its predictive power became clear with the discovery of the $\Omega^-$ in 1964 \cite{Barnes:1964pd}. In the same year, Gell-Mann and Zweig independently proposed that these patterns arise from three constituents, the $u$, $d$ and $s$ quarks \cite{GellMann1964Quarks,Zweig1964}. This brought a new puzzle as the spin-$3/2$ ground state $\Omega^-\sim sss$ contains three identical quarks with a symmetric spatial, spin and flavour wavefunction, conflicting Fermi statistics. An additional three-valued degree of freedom, colour, allows an antisymmetric factor in the wavefunction and resolves this problem. Greenberg's paraquark proposal was an early step towards this understanding \cite{Greenberg:1964pe}. Han and Nambu then introduced a second $SU(3)$ symmetry and an octet of gauge fields in 1965, providing an important precursor to the gauge theory of colour \cite{Han:1965pf}. Deep-inelastic electron--proton scattering at SLAC in the late 1960s supplied a crucial experimental breakthrough that the proton might have pointlike constituents \cite{Breidenbach:1969kd}. The emerging picture led to quantum chromodynamics (QCD), a gauge theory of $SU(3)_c$ in which quarks interact through eight gluons \cite{Fritzsch:1973pi}. Here $SU(3)_c$ acts on colour states within each flavour, whereas the earlier $SU(3)_f$ relates different flavours. The discovery of asymptotic freedom by Gross, Wilczek and Politzer in 1973 explained how quarks could behave as nearly free constituents at short distances despite participating in the strong interaction \cite{Gross:1973id,Politzer:1973fx}. At low energies, the coupling grows and the theory becomes strongly coupled, with quarks and gluons confined into colour-singlet hadrons.

QCD however leaves an important question unanswered. Its Lagrangian admits a CP-violating term proportional to $G^A_{\mu\nu}\widetilde G^{A\mu\nu}$. Although this is a total derivative, the non-trivial vacuum structure makes its coefficient physically relevant \cite{Jackiw:1976pf}. Including the quark-mass phases, the observable angle is $\bar\theta=\theta+\arg\det M_q$. The neutron electric dipole moment constrains $|\bar\theta|\lesssim10^{-10}$ \cite{Abel:2020gbr,GrillidiCortona:2015jxo}. Why this parameter is so small, despite CP violation in the weak interactions, is the strong CP problem.

In 1977, Peccei and Quinn proposed a dynamical solution based on a spontaneously broken global symmetry with a QCD anomaly, now known as the Peccei--Quinn (PQ) symmetry \cite{Peccei:1977hh,Peccei:1977ur}. Weinberg and Wilczek independently identified the associated light pseudoscalar in 1978, now known as the axion \cite{Weinberg:1977ma,Wilczek:1977pj}. Its QCD potential selects an expectation value that cancels the original strong CP angle. Parity-based solutions were also proposed during this period \cite{MohapatraSenjanovic1978}. Subsequent axion models, including the KSVZ construction, separated the PQ-breaking scale from the electroweak scale \cite{Kim:1979if,Shifman:1979if}. We focus here on the PQ mechanism and the protection needed for this cancellation to remain effective.

This protection is non-trivial because additional PQ-breaking terms generally shift the axion away from the QCD minimum. Quantum gravity is expected to forbid exact global symmetries, and Planck suppression alone need not make the resulting shift sufficiently small \cite{Banks:2010zn,Kamionkowski:1992mf,Holman:1992us}. Maintaining the $\theta-$parameter below the experimental bound is the axion quality problem. 

Extra dimensions provide a way to restrict these contributions. An axion can arise as the Wilson line of a higher-dimensional gauge field around a compact direction. Local gauge-invariant operators constructed from the field strength are insensitive to a constant holonomy and therefore cannot generate a potential for this mode \cite{Choi2004GaugeAxion}. Higher-form symmetries make the connection to axion quality explicit by identifying the possible sources of shift-symmetry breaking \cite{Reece:2023czb,Reece2024Expectations,CraigKongsore2024}.\footnote{Generalised symmetries also helps understand the global form of the Standard Model gauge group \cite{Tong2017LineOperators} and give selection rules in massless QED and QCD despite the axial anomaly \cite{ChoiLamShao2022}. See \cite{Gaiotto2015Generalized} for the general higher-form framework.} Heavy charged particles can generate a potential by propagating around the compact direction, with exponential suppression set by their winding action. Magnetic gauging can generate an additional contribution without this winding exponential. The axion--gluon coupling required for the PQ mechanism arises upon reducing a five-dimensional Chern--Simons term.

Usually in the compactifications considered in existing literatures, the compactification radius is usually considered to be fixed and time-independent. However, the compactification radius itself can itself be a a dynamical field. Its evolution changes the axion kinetic term, the winding action and the relative strengths of the QCD and additional potentials. Conversely, these potentials contribute to the equation determining the radius. Axion quality at a prescribed radius therefore needs to be supplemented by a stabilisation analysis. Motivated also by time-dependent compactifications \cite{KamalMahanta2026,ArkoEtAl2026}, we study this question in a five-dimensional circle model with radion $\beta$.

We construct stabilised Minkowski minima and nearby local dS minima using neutral Casimir contributions and a five-dimensional cosmological constant, with and without scalar winding. Here, scalar winding means that an additional periodic scalar changes by an integer multiple of its period around the compact circle, returning to the same physical value. Lighter neutral fields can support these minima while heavier charged fields generate an exponentially suppressed axion potential. Within the assumed effective-theory regime and subject to conditions on the independent vacuum terms, we exhibit parameters satisfying the strong CP bound with a stable coupled axion--radion minimum and negligible displacement of the radius. Magnetic gauging remains compatible with stabilisation after including its additional degree of freedom in the Casimir spectrum, although its coupling must be small. We also derive the graviphoton couplings and follow radion decay through reheating in the construction without scalar winding or magnetic gauging. The entropy produced during this decay dilutes the coherent axion abundance, allowing axions with $f_a=10^{14}\,\mathrm{GeV}$ to account for all of the observed dark matter with a larger initial misalignment than required for the same abundance in radiation domination. Our benchmark also gives a small dark-radiation contribution, $\Delta N_{\rm eff}\simeq0.0196$.

Section~\ref{sec:strong-cp-quality} reviews strong CP and axion quality. Section~\ref{sec:extra-dimensional-axions} describes higher-dimensional protection and its breaking. Section~\ref{sec:time-dependent-axions} introduces the dynamical radius and constructs the stabilised potentials. Section~\ref{sec:Cosmological and Phenomenological Implications} examines the coupled vacuum, the graviphoton and reheating, while section~\ref{sec:outlookl} discusses ultraviolet embeddings and open questions. Appendices~\ref{app:higher-form-symmetries} and~\ref{app:radion-one-loop} collect the higher-form symmetry background and the one-loop Casimir derivation.

\section{Strong CP Problem and High Quality Axions}
\label{sec:strong-cp-quality}

\subsection{The strong CP problem}

The QCD Lagrangian admits a CP-violating term which is consistent with all the symmetries of the Standard Model. Including the quark mass matrix $M_q$, the relevant terms are
\begin{equation}
    \mathcal{L}_{\rm QCD}\supset
    -\left(\bar q_L M_q q_R+\mathrm{h.c.}\right)
    +\theta\frac{g_s^2}{32\pi^2}
    G^A_{\mu\nu}\widetilde G^{A\mu\nu},
    \qquad
    \widetilde G^{A\mu\nu}=\frac12\epsilon^{\mu\nu\rho\sigma}G^A_{\rho\sigma},
    \label{eq:qcd-theta}
\end{equation}
where $g_s$ is the strong coupling and $A$ is an adjoint colour index. Although $G\widetilde G$ is a total derivative, its integral can be non-zero for topologically non-trivial gauge configurations. The $\theta$ term therefore changes the relative phases with which different topological sectors contribute to the path integral \cite{Jackiw:1976pf}. A common axial rotation of the quark fields shifts $\theta$ through the anomalous Jacobian of the fermion path-integral measure. The same rotation shifts $\arg\det M_q$ by the opposite amount, so removing $\theta$ from the gauge term transfers the CP-violating phase into the quark mass matrix \cite{Fujikawa:1979ay}. The invariant parameter is therefore
\begin{equation}
    \bar\theta=\theta+\arg\det M_q
    \label{eq:physical-theta}
\end{equation}
where $M_q$ are quark mass matrix. Thus, making the quark masses real does not eliminate strong CP violation. It just transfers their phase into the coefficient of $G\widetilde G$. A non-zero $\bar\theta$ induces a neutron electric dipole moment. Its experimental upper bound requires $|\bar\theta|\lesssim10^{-10}$ \cite{Abel:2020gbr,GrillidiCortona:2015jxo}. The weak interactions already violate CP through the complex phase of the CKM matrix, so CP is not an exact symmetry of the Standard Model. However, the CKM phase and $\bar\theta$ are independent sources of CP violation. The Standard Model does not explain why the combination in \eqref{eq:physical-theta} is so small, while allowing comparatively larger CP violation in weak interactions. This unexplained smallness of $\bar\theta$ is the strong CP problem.

\subsection{The Peccei--Quinn mechanism}

The Peccei--Quinn (PQ) mechanism makes the physical QCD angle dynamical, allowing the vacuum to relax to a CP-preserving minimum \cite{Peccei:1977hh,Peccei:1977ur}. Introducing a complex scalar $\Phi$ charged under a global $U(1)_{\rm PQ}$, with a potential whose minimum spontaneously breaks this symmetry. Expanding around the vacuum, we write
\begin{equation}
    \Phi(x)=\frac{f_a+\rho(x)}{\sqrt2}\,e^{ia(x)/f_a},
    \qquad a\sim a+2\pi f_a,
    \label{eq:pq-scalar}
\end{equation}
where $\rho$ is the massive radial excitation and $a$ is the angular mode, identified with the axion \cite{Weinberg:1977ma,Wilczek:1977pj}. Under a PQ transformation, $\Phi\to e^{i\alpha}\Phi$, the axion shifts as $a\to a+\alpha f_a$. At energies below the radial-mode mass, setting $\rho$ to its vacuum value in $|\partial\Phi|^2$ gives the canonically normalised kinetic term $\frac12(\partial a)^2$. In the absence of explicit PQ breaking, the angular direction is flat and the axion is massless. The connection to the strong CP problem comes from the QCD anomaly of the PQ symmetry. For example, the KSVZ construction introduces a heavy vector-like colour-triplet quark $Q$ with the Yukawa interaction $-y\Phi\bar Q_LQ_R+\mathrm{h.c.}$ \cite{Kim:1979if,Shifman:1979if}. When $\Phi$ acquires a vacuum expectation value, this interaction gives $Q$ a mass whose phase depends on $a/f_a$. An axial rotation of $Q$ removes the axion-dependent phase from the mass term, while the anomalous Jacobian of the fermion path-integral measure transfers it to the coefficient of $G\widetilde G$. This generates the axion--gluon coupling. After integrating out $Q$ and choosing a basis in which the ordinary quark masses are real, the low-energy axion Lagrangian contains
\begin{equation}
    \mathcal{L}_a\supset
    \frac12\partial_\mu a\,\partial^\mu a
    +\left(\bar\theta+\frac{a}{f_a}\right)\frac{g_s^2}{32\pi^2}G^A_{\mu\nu}\widetilde G^{A\mu\nu}.
    \label{eq:axion-qcd}
\end{equation}
$f_a$ denotes the decay constant. The physical angle entering QCD is therefore $\vartheta=\bar\theta+a/f_a$, whose value can change through the axion field. Non-perturbative QCD generates a vacuum energy $E_{\rm QCD}(\vartheta)$ which is periodic under $\vartheta\to\vartheta+2\pi$ and minimised at $\vartheta=0$ modulo $2\pi$. This lifts the flat axion direction and gives the axion a mass. Choosing the CP-preserving minimum at $\vartheta=0$, the potential near it is
\begin{equation}
    V_{\rm QCD}(a)=E_{\rm QCD}(\vartheta)-E_{\rm QCD}(0)
    =\frac12\chi_{\rm QCD}\vartheta^2+\mathcal{O}(\vartheta^4),
    \qquad
    \chi_{\rm QCD}=\left.\frac{\partial^2E_{\rm QCD}}{\partial\vartheta^2}\right|_{\vartheta=0}.
    \label{eq:qcd-susceptibility}
\end{equation}
Here $\chi_{\rm QCD}$ is the topological susceptibility, which determines the quadratic dependence of the QCD vacuum energy on $\vartheta$ near its CP-preserving minimum. Since $\partial_aV_{\rm QCD}=f_a^{-1}\partial_\vartheta E_{\rm QCD}$, minimising the axion potential gives
\begin{equation}
    \theta_{\rm eff}\equiv\bar\theta+\frac{\langle a\rangle}{f_a}=0,
    \qquad
    m_a^2=\left.\frac{\partial^2V_{\rm QCD}}{\partial a^2}\right|_{a=\langle a\rangle}
    =\frac{\chi_{\rm QCD}}{f_a^2}.
    \label{eq:pq-minimum}
\end{equation}
Thus the axion expectation value cancels the original CP-violating angle without requiring $\bar\theta$ itself to be small. At zero temperature, $\chi_{\rm QCD}\simeq(75.5\,\mathrm{MeV})^4$, corresponding to $m_a\simeq5.7\,\mu\mathrm{eV}\,(10^{12}\,\mathrm{GeV}/f_a)$ \cite{GrillidiCortona:2015jxo}. This solution assumes that any additional explicit PQ breaking is sufficiently suppressed that it does not appreciably displace the QCD minimum. Maintaining this suppression is the axion quality problem.

\subsection{Planck-suppressed operators and axion quality}

The cancellation in \eqref{eq:pq-minimum} requires any additional PQ-breaking contribution to leave the QCD minimum almost unchanged. This is a non-trivial requirement if PQ is a global symmetry. Quantum-gravity effects are generally expected to violate global symmetries \cite{Banks:2010zn}\footnote{See \cite{vanBeest:2021lhn} for a review of quantum-gravity constraints on effective field theories.}, and Planck suppression alone does not guarantee that their effects are sufficiently small. To see this, suppose the leading PQ-violating operator allowed by the other symmetries has dimension $d>4$ and contributes
\begin{equation}
    \Delta V=-\left(\frac{\kappa_d}{M_{\rm Pl}^{d-4}}\Phi^d+\mathrm{h.c.}\right),
    \label{eq:planck-operator}
\end{equation}
where $\kappa_d$ is a dimensionless complex coefficient and $M_{\rm Pl}\simeq2.4\times10^{18}\,\mathrm{GeV}$ is the reduced Planck mass. Keeping the heavy radial field at $\rho=0$, \eqref{eq:pq-scalar} gives $\Phi=(f_a/\sqrt2)e^{ia/f_a}$. Substituting this into \eqref{eq:planck-operator} and adding the hermitian conjugate gives
\begin{equation}
    \Delta V(a)=-\frac{2|\kappa_d|}{M_{\rm Pl}^{d-4}}\left(\frac{f_a}{\sqrt2}\right)^d\cos\left(d\frac{a}{f_a}+\arg\kappa_d\right),
    \label{eq:quality-potential}
\end{equation}
where $\arg\kappa_d$ denotes the phase of $\kappa_d$. QCD favours $a=-f_a\bar\theta$, where the effective strong CP angle vanishes. The additional term is minimised when the argument of its cosine is an integer multiple of $2\pi$. For generic relative phases, it shifts the axion away from the QCD minimum, leaving a non-zero strong CP angle.

We can calculate the remaining CP angle by finding how far the minimum moves. Near its minimum, the QCD potential is $V_{\rm QCD}\simeq m_a^2(a+f_a\bar\theta)^2/2$, using $m_a^2=\chi_{\rm QCD}/f_a^2$. Setting the derivative of the total potential to zero therefore gives
\begin{equation}
    0\simeq m_a^2\left(\langle a\rangle+f_a\bar\theta\right)
    +\left.\frac{\partial\Delta V}{\partial a}\right|_{a=\langle a\rangle}.
    \label{eq:quality-stationarity}
\end{equation}
The first term is proportional to the displacement from the QCD minimum. At the new minimum, it must balance the derivative of the additional potential. We assume that the displacement is small and that $|\partial_a^2\Delta V|\ll m_a^2$ nearby. To obtain the leading shift, we can then evaluate the derivative of $\Delta V$ at the original minimum, $a=-f_a\bar\theta$. Dividing the resulting displacement by $f_a$ gives the residual strong CP angle:
\begin{align}
    \theta_{\rm eff}
    &=\frac{\langle a\rangle+f_a\bar\theta}{f_a}
    \simeq-\frac{1}{m_a^2f_a}\left.\frac{\partial\Delta V}{\partial a}\right|_{a=-f_a\bar\theta}\nonumber\\
    &=-\frac{2d|\kappa_d|}{\chi_{\rm QCD}M_{\rm Pl}^{d-4}}\left(\frac{f_a}{\sqrt2}\right)^d\sin\left(\arg\kappa_d-d\bar\theta\right),
    \qquad |\theta_{\rm eff}|\lesssim10^{-10}.
    \label{eq:axion-quality-bound}
\end{align}
The sine measures the relative phase between the two contributions. Unless their phases are specially aligned (which requires a specific explanation itself), it is of order one. The experimental bound then requires the coefficient multiplying it to be smaller than about $10^{-10}$. This explains why a small correction to the axion potential can still be problematic. For example, take $f_a=10^{12}\,\mathrm{GeV}$, $|\kappa_d|=1$ and $\chi_{\rm QCD}\simeq(75.5\,\mathrm{MeV})^4$. For an order-one sine factor, \eqref{eq:axion-quality-bound} gives $|\theta_{\rm eff}|\sim3.3\times10^{-6}$ for $d=13$, but $|\theta_{\rm eff}|\sim1.1\times10^{-12}$ for $d=14$. Thus the dimension-thirteen operator moves the minimum too far, whereas the dimension-fourteen operator satisfies the bound. For these parameter choices and generic phases, lower-dimensional operators of the form \eqref{eq:planck-operator} must therefore be absent or sufficiently suppressed. Smaller coefficients or a special phase alignment can relax this requirement, but their origin must be explained. Ensuring that additional PQ breaking does not spoil the cancellation of the strong CP angle is the axion quality problem. This motivates the higher-dimensional gauge origin of axions discussed next.

\section{High Quality Axions From Extra-dimensions}
\label{sec:extra-dimensional-axions}

\subsection{Axions from higher-dimensional gauge fields}

One way to address the axion quality problem is to obtain the axion from a higher-dimensional gauge field. Its gauge symmetry then restricts the operators that can generate an axion potential. Such constructions arise naturally in string theory, where axions descend from antisymmetric tensor fields upon compactification \cite{Witten1984Superstrings,ChoiKim1985Axions,SvrcekWitten2006}.

To see how this works, consider a compact $p$-form gauge potential $C_p$ on an internal manifold $X$. Its components along non-trivial internal $p$-cycles give scalar fields in four dimensions. Expanding in harmonic forms $\omega_I$, we write
\begin{equation}
    C_p=\sum_I\xi^I(x)\omega_I(y)+\cdots,
    \qquad \int_{\Sigma_p^J}\omega_I=\delta_I{}^J,
    \qquad \xi^I\sim\xi^I+2\pi.
    \label{eq:form-axions}
\end{equation}
Here $\xi^I$ is the integral of $C_p$ over the corresponding cycle. We normalise $C_p$ so that a unit charge contributes $\exp(i\int C_p)$ and large gauge transformations then give the displayed periodicity. Reducing the higher-dimensional kinetic term determines the axion kinetic matrix through the integrals $\int_X\omega_I\wedge *_X\omega_J$. The decay constants therefore depend on the internal geometry \cite{SvrcekWitten2006}. String-theory examples include axions obtained from $B_2$, $C_3$ in type IIA, and $C_2$ and $C_4$ in type IIB which have interesting phenomenology for example quintessence as studied in \cite{Chakraborty:2025wqn}.

The strong CP "protection" then follows from the higher-dimensional origin of these scalars. Since $\mathrm{d}\omega_I=0$, the field strength contains $\mathrm{d}\xi^I$ but no undifferentiated $\xi^I$. Local operators constructed only from $\mathrm{d}C_p$ therefore cannot generate a potential for these modes. The free theory consequently has continuous axion shift symmetries, while large gauge transformations identify shifts by integer multiples of $2\pi$.

This protection does not by itself guarantee a high-quality QCD axion. Flux couplings and wrapped instantons can generate additional potentials, and moduli stabilisation can also lift axion directions \cite{SvrcekWitten2006,Conlon2006Axion,Chakraborty:2025wqn}. A viable construction must retain an axion that couples to QCD, with other contributions sufficiently suppressed or aligned to preserve the strong-CP bound. These questions connect the higher-dimensional origin of axions to their quality \cite{BurgessChoiQuevedo2024,BurgessQuevedo2025}.

To illustrate the axion quality "protection" mechanism, we follow \cite{CraigKongsore2024,Reece:2023czb} and consider a compact $U(1)$ gauge field $C$ on $M_4\times S^1$. We consider $y$ to be a proper-length coordinate with period $2\pi R$, and keep $R$ fixed in this section. Writing $F=\mathrm{d}C$, we have
\begin{equation}
    S_C=-\frac{1}{2g_5^2}\int F\wedge *_5F,
    \qquad \frac{a}{f_a}=\oint_{S^1}C,
    \qquad C_y=\frac{a}{2\pi Rf_a},
    \label{eq:circle-axion}
\end{equation}
where $[g_5^2]=\mathrm{mass}^{-1}$. The dimensionless angle is $a/f_a$, while $a$ retains the canonical normalisation of section~\ref{sec:strong-cp-quality}. In the scalar kinetic-sign convention of \eqref{eq:axion-qcd}, integrating over the circle gives
\begin{equation}
    S_{\rm kin}^{(4)}=
    \frac{2\pi R}{2g_5^2}\int\mathrm{d}^4x\sqrt{-\hat g}\,
    \partial_\mu C_y\partial^\mu C_y
    =\frac12\int\mathrm{d}^4x\sqrt{-\hat g}\,(\partial a)^2,
    \qquad f_a^2=\frac{1}{2\pi Rg_5^2}.
    \label{eq:circle-normalisation}
\end{equation}
Thus the factor $2\pi R$ comes from the volume integral, while the Wilson line itself is an integral of the connection.

\subsection{The QCD coupling from extra dimensions}

For the Wilson-line axion to solve the strong CP problem, it must acquire the anomalous coupling to gluons discussed in section~\ref{sec:strong-cp-quality}. This coupling arises from a mixed Chern--Simons term in five dimensions \cite{CraigKongsore2024}.
\begin{equation}
    \begin{aligned}
        S_{\rm CS}
        &=\frac{1}{8\pi^2}\int_{M_4\times S^1}C\wedge\mathrm{Tr}(\mathcal G\wedge\mathcal G)\longrightarrow\int_{M_4}\mathrm{d}^4x\sqrt{-\hat g}\,
        \frac{a}{f_a}\frac{g_s^2}{32\pi^2}G^A_{\mu\nu}\widetilde G^{A\mu\nu}.
    \end{aligned}
    \label{eq:cs-reduction}
\end{equation}
Here $\mathcal G$ is the five-dimensional colour field strength, with the gauge coupling absorbed into the connection.\footnote{With $\mathrm{Tr}(T^AT^B)=\delta^{AB}/2$, the instanton-density four-form is
\begin{equation}
    \mathcal I_4=\frac{1}{8\pi^2}\mathrm{Tr}(\mathcal G\wedge\mathcal G),
    \qquad
    \mathcal I_4\big|_{M_4}=\frac{g_s^2}{32\pi^2}G^A_{\mu\nu}\widetilde G^{A\mu\nu}\sqrt{-\hat g}\,\mathrm{d}^4x.
    \label{eq:instanton-density-form}
\end{equation}
The five-dimensional interaction can therefore also be written as $S_{\rm CS}=\int C\wedge\mathcal I_4$.} We have chosen the quantised Chern--Simons coefficient to be one, matching our earlier anomaly convention. The reduction follows directly from $\oint_{S^1}C=a/f_a$: the component of $C$ along the circle supplies the axion, while the colour field strengths supply the four-dimensional gluon coupling. Including the microscopic QCD angle and the quark mass phase then gives $\vartheta=\bar\theta+a/f_a$. In the absence of other contributions, the QCD potential is minimised at $\vartheta=0$, reproducing \eqref{eq:pq-minimum}.

To identify the symmetry protecting the axion from additional potentials, we first get rid of the Chern--Simons interaction and consider the free gauge field $C$. This theory has an electric one-form symmetry and a magnetic two-form symmetry \cite{Gaiotto2015Generalized,CraigKongsore2024}. In our convention where conserved currents are closed forms, their currents satisfy
\begin{equation}
    J_e=\frac{1}{g_5^2}*_5F,\qquad
    J_m=\frac{F}{2\pi},\qquad
    \mathrm{d}J_e=0,\qquad
    \mathrm{d}J_m=0.
    \label{eq:five-dimensional-currents}
\end{equation}
The first conservation equation follows from the equation of motion for $C$, while the second follows from the Bianchi identity. The electric symmetry acts as $C\to C+\lambda$, with $\mathrm{d}\lambda=0$. This leaves $F$ unchanged but shifts the Wilson line by $\oint_{S^1}\lambda$. Since this integral can vary continuously, the electric symmetry becomes the four-dimensional PQ shift symmetry and forbids an axion potential. Its charged operators are Wilson lines. The magnetic symmetry instead gives the axion winding current $\mathrm{d}(a/f_a)/(2\pi)$. This current can remain conserved even when an axion potential is present, so it does not provide the same protection.

Restoring the Chern--Simons interaction modifies the equation of motion to
\begin{equation}
    \mathrm{d}J_e=\frac{1}{8\pi^2}\mathrm{Tr}(\mathcal G\wedge\mathcal G).
    \label{eq:five-dimensional-abj}
\end{equation}
The electric current is therefore no longer conserved. This is the five-dimensional counterpart of the Adler--Bell--Jackiw anomaly \cite{CraigKongsore2024}, which becomes the QCD anomaly of the PQ symmetry after compactification. Although subtracting a colour Chern--Simons three-form would make the current locally conserved, the resulting current would not be gauge invariant and would not restore the continuous symmetry. The QCD anomaly supplies the breaking needed to select the CP-preserving axion vacuum. Other sources of electric-symmetry breaking can generate additional potentials whose minima need not coincide with it. Axion quality therefore requires these contributions to leave the QCD minimum sufficiently undisturbed. We next examine two such sources which are due to electrically charged matter and gauging of the magnetic two-form symmetry. The breaking of all global (including higher form) symmetries is something that is motivated by the absence of no global symmetries in quantum gravity. Hence, the study of the effects of breaking the global symmetries by possible ways is a must. 

\subsection{Breaking the electric one-form symmetry by charged matter}

We now introduce a bulk particle of mass $m_5$ and non-zero integer charge $q$ under $C$. Charged matter allows a charge-$q$ Wilson line to end on a particle insertion, so the corresponding electric charge can be screened. Considering this source of breaking alone, the current equation becomes $\mathrm{d}J_e=j_{\rm matter}$. The continuous electric one-form symmetry is therefore explicitly broken, with only the subgroup $\mathbb Z_{|q|}^{(1)}$ surviving \cite{CraigKongsore2024}. After compactification, this breaking allows a potential for the Wilson-line axion.

The origin of this potential can be understood using the Euclidean worldline description \cite{Reece:2023czb,CraigKongsore2024}. For a charged particle following a closed trajectory $\gamma$, the action is
\begin{equation}
    S_E[\gamma]=m_5\int_\gamma\mathrm{d}\ell-iq\int_\gamma C,
    \label{eq:charged-particle-worldline-action}
\end{equation}
where $\mathrm{d}\ell$ is the proper Euclidean length. The first term suppresses propagation over long distances, while the second supplies the Wilson factor $\exp(iq\int_\gamma C)$ in the path-integral weight $e^{-S_E}$. Closed trajectories are classified by their integer winding number $w$ around the circle. At constant $a$, the shortest trajectory in a non-zero winding sector remains at a fixed position in the four non-compact directions and satisfies
\begin{equation}
    \int_{\gamma_w}\mathrm{d}\ell=2\pi R|w|,
    \qquad
    \int_{\gamma_w}C=w\oint_{S^1}C=w\frac{a}{f_a}.
    \label{eq:wrapped-worldline-integrals}
\end{equation}
Its action is consequently
\begin{equation}
    S_E(w)=2\pi Rm_5|w|-iqw\frac{a}{f_a}.
    \label{eq:wrapped-worldline-action}
\end{equation}
The absolute value appears in the length term because reversing the orientation leaves the length unchanged. The gauge integral instead changes sign. Summing the two orientations therefore gives
\begin{equation}
    e^{-S_E(w)}+e^{-S_E(-w)}
    =2e^{-2\pi Rm_5w}
    \cos\left(wq\frac{a}{f_a}\right),
    \qquad w>0.
    \label{eq:wrapped-worldline-pair}
\end{equation}
These wrapped trajectories are often called worldline instantons. They explain the exponential suppression and the harmonics of the axion potential, while the one-loop determinant determines their prefactors and overall sign. Since the phase depends only on the charge under $C$, the particle need not carry QCD colour to generate this potential. To calculate the coefficient, we take the particle to be a periodic complex scalar. In a constant Wilson-line background, its Kaluza--Klein masses are
\begin{equation}
    m_k^2(a)=m_5^2+\frac{1}{R^2}
    \left(k+\frac{qa}{2\pi f_a}\right)^2,
    \qquad k\in\mathbb Z.
    \label{eq:charged-kk-masses}
\end{equation}
Evaluating the one-loop determinant by the proper-time and Poisson-resummation method of appendix~\ref{app:radion-one-loop} gives the finite winding contribution \cite{ArkaniHamed2003Extranatural,Reece:2023czb,CraigKongsore2024},
\begin{equation}
    \Delta V_{\rm el}(a)=
    -\frac{3}{32\pi^6R^4}\sum_{w=1}^{\infty}
    \frac{e^{-z_w}}{w^5}
    \left(1+z_w+\frac{z_w^2}{3}\right)
    \cos\left(wq\frac{a}{f_a}\right),
    \qquad z_w=2\pi Rm_5w.
    \label{eq:charged-scalar-potential}
\end{equation}
The coefficient includes both real degrees of freedom of the complex scalar, and the sum includes the paired contributions from positive and negative winding.\footnote{The zero-winding contribution is independent of $a$ and renormalises the local bulk vacuum energy. It does not affect the axion potential at fixed radius, but must be included through the renormalised bulk cosmological constant when studying radion stabilisation.} Each harmonic is invariant under $a\to a+2\pi f_a/|q|$, reflecting the discrete shift symmetry inherited from $\mathbb Z_{|q|}^{(1)}$ when this source of breaking is considered alone. For $m_5R\gg1$, the first winding dominates. Its amplitude and leading large-mass behaviour are
\begin{align}
    \Delta V_{\rm el}(a)
    &\simeq-\Lambda_{\rm el}^4
    \cos\left(q\frac{a}{f_a}\right),
    \nonumber\\
    \Lambda_{\rm el}^4
    &=\frac{3e^{-z_1}}{32\pi^6R^4}
    \left(1+z_1+\frac{z_1^2}{3}\right)
    \simeq\frac{m_5^2}{8\pi^4R^2}e^{-2\pi Rm_5}.
    \label{eq:heavy-electric-potential}
\end{align}
The final approximation keeps the leading $z_1^2/3$ term in the polynomial. The suppression is controlled by the dimensionless winding action $z_1=2\pi Rm_5$.

This potential favours $qa/f_a=2\pi\ell$, with $\ell\in\mathbb Z$, independently of $\bar\theta$. Its preferred values therefore need not coincide with the QCD minimum at $a=-f_a\bar\theta$. In the regime $q^2\Lambda_{\rm el}^4\ll\chi_{\rm QCD}$, applying \eqref{eq:axion-quality-bound} gives
\begin{equation}
    \theta_{\rm eff}\simeq
    \frac{q\Lambda_{\rm el}^4}{\chi_{\rm QCD}}
    \sin(q\bar\theta).
    \label{eq:fixed-radius-electric-quality}
\end{equation}
For a generic relative phase, the strong CP bound thus requires $|q|\Lambda_{\rm el}^4\lesssim10^{-10}\chi_{\rm QCD}$. Using the leading heavy-mass expression in \eqref{eq:heavy-electric-potential}, this condition becomes
\begin{equation}
    z_1-2\log z_1
    \gtrsim
    \log\left(
        \frac{10^{10}|q|}
        {32\pi^6R^4\chi_{\rm QCD}}
    \right),
    \qquad z_1\gg1.
    \label{eq:electric-log-quality}
\end{equation}
The required hierarchy therefore enters logarithmically in the lower bound on the winding action, with the term $2\log z_1$ accounting for the polynomial prefactor. This is the sense in which extra-dimensional axions ``take the log'' of the axion quality problem \cite{Reece:2023czb,CraigKongsore2024}. Sufficiently heavy charged states can satisfy the bound through their winding suppression. States with $m_5R\lesssim1$ have no comparable exponential suppression. 

\subsection{Gauging the magnetic two-form symmetry}

A second way to generate an axion potential is to gauge the magnetic two-form symmetry. Since its current is $J_m=F/(2\pi)$, this amounts to introducing a compact three-form gauge field $K_3$ coupled to $J_m$ and making it dynamical. Writing $H_4=\mathrm{d}K_3$, the additional five-dimensional action is
\begin{equation}
    S_K^{(5)}=\int\left[-\frac{1}{2e_{K,5}^2}H_4\wedge *_5H_4
    +\frac{M}{2\pi}K_3\wedge F\right],
    \qquad M\in\mathbb Z,
    \label{eq:magnetic-gauging-five}
\end{equation}
where $[e_{K,5}]=\mathrm{mass}^{3/2}$. This coupling preserves the gauge invariance of $C$, but modifies its equation of motion. Temporarily omitting the QCD coupling, we obtain $\mathrm{d}J_e=MH_4/(2\pi)$. Although $J_e-MK_3/(2\pi)$ is locally closed, it depends on the gauge choice for $K_3$. Requiring the corresponding symmetry operators to be invariant under large $K_3$ gauge transformations leaves only $\mathbb Z_{|M|}^{(1)}$ for $M\ne0$. Thus magnetic gauging breaks the continuous electric symmetry that protects the axion \cite{CraigKongsore2024}. Keeping the four-dimensional three-form zero mode, again denoted by $K_3$, and using $\oint_{S^1}C=a/f_a$, the action reduces to
\begin{equation}
    S_K^{(4)}=\int\left[-\frac{1}{2e_{K,4}^2}H_4\wedge *_4H_4
    +\frac{M}{2\pi f_a}K_3\wedge\mathrm{d}a\right],
    \qquad e_{K,4}^2=\frac{e_{K,5}^2}{2\pi R}.
    \label{eq:magnetic-gauging-four}
\end{equation}
A massless three-form in four dimensions has no propagating degrees of freedom, but its four-form flux can carry energy. The coupling to $a$ makes this energy depend on the axion. Indeed, the three-form equation integrates to $*_4H_4/e_{K,4}^2-Ma/(2\pi f_a)=\mathrm{constant}$. Eliminating $H_4$ at fixed flux, gives \cite{KaloperSorbo2009,BurgessQuevedo2025}
\begin{equation}
    V_{K,n}(a)=\frac{e_{K,4}^2}{2}
    \left(n+\frac{Ma}{2\pi f_a}\right)^2,
    \qquad
    m_K^2=\frac{M^2e_{K,4}^2}{4\pi^2f_a^2}.
    \label{eq:magnetic-branch-potential}
\end{equation}
Here $n\in\mathbb Z$ labels the quantised flux branches. Each branch is quadratic, but the shift $a\to a+2\pi f_a$ must be accompanied by $n\to n-M$, to preserve the periodic identification of the full theory. The mass also has a simple five-dimensional interpretation which is that dualising $K_3$ gives a compact scalar which is eaten by $C$ through a St\"uckelberg coupling. After compactification, the component of this massive vector along the circle survives as the massive axion. For $M\ne0$, the selected branch has its minimum at $a_n=-2\pi f_a n/M$, so $V_{K,n}=m_K^2(a-a_n)^2/2$. QCD instead favours $a=-f_a\bar\theta$. Near the QCD minimum, its potential is $m_a^2(a+f_a\bar\theta)^2/2$, with $m_a^2=\chi_{\rm QCD}/f_a^2$. Minimising the sum gives
\begin{equation}
    \theta_{\rm eff}\simeq
    \frac{m_K^2}{m_a^2+m_K^2}
    \left(\bar\theta+\frac{a_n}{f_a}\right),
    \qquad m_a^2=\frac{\chi_{\rm QCD}}{f_a^2},
    \label{eq:magnetic-quality-shift}
\end{equation}
valid near $\theta_{\rm eff}=0$. The factor in parentheses measures the mismatch between the two preferred minima. If it is of order one, the strong-CP bound requires $m_K^2\lesssim10^{-10}m_a^2$. Unlike the charged-particle contribution in \eqref{eq:heavy-electric-potential}, this mass arises at tree level and has no exponential suppression. The coupling must therefore be sufficiently weak or absent, as for $M=0$, unless the selected branch is suitably aligned with QCD \cite{CraigKongsore2024}.

For comparison, gauging the electric one-form symmetry changes which fields remain physical \cite{CraigKongsore2024}. Introducing a dynamical two-form $B_2$ replaces $F$ by the gauge-invariant combination $F-kB_2$, where $k\in\mathbb Z$. Upon circle reduction, $B_{\mu y}$ supplies a four-dimensional vector $\mathcal A_\mu$, and the axion kinetic term becomes
\begin{equation}
    \mathcal{L}_{\rm kin}
    =\frac12\left(\partial_\mu a-kf_a\mathcal A_\mu\right)
    \left(\partial^\mu a-kf_a\mathcal A^\mu\right).
\end{equation}
For non-zero $k$, the vector eats the axion through the St\"uckelberg mechanism, leaving no independent axion scalar to relax the strong CP angle. Moreover, the axion--gluon coupling in \eqref{eq:cs-reduction} is not invariant under this gauged shift and requires additional anomaly-cancelling ingredients. Electric gauging alone therefore does not provide a protected physical QCD axion. The no-global-symmetries conjecture rules out exact global symmetries, but does not by itself determine the mechanism or strength of symmetry breaking. Hence, the exact UV theory could potentially select the actual posssible mechanism out of the above possibilities. 

\subsection{Higher-form symmetry protection improves axion quality}

The higher-dimensional origin of the axion restricts how its potential can be generated. Local operators built from $F$ and its derivatives are insensitive to a constant Wilson line. Terms such as $F^4$, and kinetic mixing with another Abelian field, can modify the conserved electric current without breaking the continuous symmetry. They therefore do not generate an axion potential by themselves. Charged particles do break this symmetry, but their contribution can depend on the Wilson line only through propagation around the circle. For sufficiently heavy states, this produces an exponential suppression, providing stronger protection than the polynomial Planck suppression discussed in section~\ref{sec:strong-cp-quality}.

Good axion quality is therefore compatible with the expected absence of exact global symmetries in quantum gravity. Heavy charged states can break the higher-form symmetry while leaving an approximate PQ shift symmetry at low energies \cite{Banks:2010zn,BurgessChoiQuevedo2024,CraigKongsore2024}. Nevertheless, magnetic gauging, light charged states and additional instanton sectors can generate contributions that are not sufficiently suppressed. Their effect must be assessed by how far they move the QCD minimum. For an additional potential $\Delta V$ producing a small displacement,
\begin{equation}
    \theta_{\rm eff}\simeq
    -\frac{f_a\Delta V'(-f_a\bar\theta)}{\chi_{\rm QCD}},
    \qquad
    \big|f_a\Delta V'(-f_a\bar\theta)\big|
    \lesssim10^{-10}\chi_{\rm QCD},
    \label{eq:general-extra-dimensional-quality}
\end{equation}
provided $|f_a^2\Delta V''|\ll\chi_{\rm QCD}$ near the QCD minimum. Higher-form symmetry restricts the possible sources of breaking, while axion quality requires their combined contribution to satisfy this bound.

\section{Time Dependent Background and Axions}
\label{sec:time-dependent-axions}

The discussion in section~\ref{sec:extra-dimensional-axions} assumed a fixed compactification radius. This will not hold before the internal geometry is stabilised. A dynamical radius changes the four-dimensional kinetic terms, the Kaluza--Klein spectrum and the action of a particle winding around the compact direction. It is therefore necessary to determine how the mechanisms discussed above depend on the radion and whether the resulting axion potential remains compatible with the strong CP bound.

Time-dependent compactifications provide a motivation for this question. In \cite{KamalMahanta2026}, families of type IIB flux backgrounds were constructed in which the internal Calabi--Yau metric carries an overall time-dependent scale factor, together with an evolving four-dimensional geometry. More recently, \cite{ArkoEtAl2026} constructed doubly warped type II flux backgrounds on smooth compact internal spaces, with four-dimensional Einstein-frame cosmologies evolving from a Big Bang through a maximum size to a Big Crunch. These constructions motivate studying the effect of an time-dependent internal geometry on extra-dimensional axions. Here, as a toy model, we consider a five-dimensional circle compactification. We keep the coordinate period $y\sim y+2\pi R$ used in \eqref{eq:circle-axion} and parametrise the metric directly in the four-dimensional Einstein frame as
\begin{equation}
    \mathrm{d}s_5^2=e^{-b\beta(x)/2}g^E_{\mu\nu}(x)\mathrm{d}x^\mu\mathrm{d}x^\nu-e^{b\beta(x)}\mathrm{d}y^2,
    \qquad y\sim y+2\pi R.
    \label{eq:beta-einstein-metric}
\end{equation}
Here $\beta$ is dimensionless and the non-zero constant $b$ fixes its normalisation. We use mostly-minus signature, consistently with \eqref{eq:axion-qcd}. The proper circumference is $2\pi R e^{b\beta/2}$, whereas $R$ remains a fixed reference length. Thus $y$ is no longer a proper-length coordinate when $\beta\ne0$. Moving to Einstein frame, we have
\begin{equation}
    \mathcal R_5=e^{b\beta/2}\left[\mathcal R_E+\frac b2\Box_E\beta-\frac{3b^2}{8}(\partial\beta)^2_E\right].
    \label{eq:beta-curvature-reduction}
\end{equation}
The $\Box_E\beta$ term is a total derivative in four dimensions. Before including the axion-dependent potentials, the scalar and gravitational terms therefore give
\begin{equation}
    S_4=\int\mathrm{d}^4x\sqrt{-g_E}\left[-\frac{M_{\rm Pl}^2}{2}\mathcal R_E+\frac{3b^2M_{\rm Pl}^2}{16}(\partial\beta)^2_E+\frac12e^{-b\beta}(\partial a)^2_E-V(\beta)\right].
    \label{eq:beta-four-dimensional-action}
\end{equation}
Here $V(\beta)$ denotes the axion-independent stabilising potential, which will be studied in section~\ref{sec:beta-stabilisation}. For the axion term, $\sqrt{|g_5|}g_5^{\mu\nu}g_5^{yy}=-e^{-b\beta}\sqrt{-g_E}g_E^{\mu\nu}$, which gives the displayed kinetic coefficient upon substituting $C_y=a/(2\pi Rf_a)$. We retain the axion definition and the decay constant of \eqref{eq:circle-normalisation} that is, 
\begin{equation}
    \frac{a}{f_a}=\oint_{S^1}C,
    \qquad a\sim a+2\pi f_a,
    \qquad f_a^2=\frac{1}{2\pi Rg_5^2}.
    \label{eq:beta-fixed-period}
\end{equation}
Thus $a$ is canonically normalised at $\beta=0$. At a fixed value of $\beta$, 
\begin{equation}
    f_{\rm eff}^2(\beta)=f_a^2e^{-b\beta}.
    \label{eq:beta-effective-decay-constant}
\end{equation}
The radius dependence of $f_{\rm eff}$ does not change the periodicity of $a/f_a$. A spacetime-dependent rescaling of $a$ would also introduce derivative couplings to $\beta$, so we keep the field convention \eqref{eq:beta-fixed-period} throughout. The four-dimensional description applies below the Einstein-frame KK scale
\begin{equation}
    m_{\rm KK,E}(\beta)=\frac1R e^{-3b\beta/4}.
    \label{eq:beta-kk-scale}
\end{equation}
We first determine the radion dependence of the axion couplings and potentials, and then study stabilisation of the radius.

\subsection{Axion quality in a time-dependent background}
\label{sec:beta-axion-quality}

When the radius evolves, the QCD potential and the additional axion potentials may not change in the same way. A contribution that is sufficiently suppressed at one value of $\beta$ may therefore become important at another. These potentials also contribute to the radion equation, so their effect on axion quality cannot be separated completely from stabilisation. For a stationary late-time vacuum, we need a stable value of $\beta$ at which the axion minimum satisfies the strong CP bound. We first determine how the couplings and potentials of section~\ref{sec:extra-dimensional-axions} depend on $\beta$.

\subsubsection{The QCD coupling}

The Chern--Simons interaction does not contain the metric. Its reduction therefore depends only on the Wilson line and is unchanged by the evolving radius as
\begin{equation}
    \begin{aligned}
        S_{\rm CS}
        &=\frac{1}{8\pi^2}\int_{M_4\times S^1}C\wedge\mathrm{Tr}(\mathcal G\wedge\mathcal G)\longrightarrow\frac{1}{8\pi^2}\int_{M_4}\frac{a}{f_a}\mathrm{Tr}(\mathcal G\wedge\mathcal G),
        \qquad \vartheta=\bar\theta+\frac{a}{f_a}.
    \end{aligned}
    \label{eq:beta-cs-reduction}
\end{equation}
In particular, no factor depending on $\beta$ multiplies the periodic angle $a/f_a$. The varying metric also does not explicitly break the electric one-form symmetry. The transformation $C\to C+\lambda$, with $\mathrm{d}\lambda=0$, still leaves $F$ unchanged. Although the Hodge star in the electric current now depends on $\beta$, the current remains conserved whenever the corresponding equation of motion has no symmetry-breaking source. For the four-dimensional axion, the kinetic term in \eqref{eq:beta-four-dimensional-action} gives
\begin{equation}
    J_{\rm PQ}=f_a e^{-b\beta}*_E\mathrm{d}a,
    \qquad \mathrm{d}J_{\rm PQ}=0,
    \label{eq:beta-pq-current}
\end{equation}
when the symmetry-breaking interactions are omitted. Restoring the Chern--Simons coupling gives $\mathrm{d}J_{\rm PQ}=(8\pi^2)^{-1}\mathrm{Tr}(\mathcal G\wedge\mathcal G)|_{M_4}$, with the convention of \eqref{eq:five-dimensional-abj}. The winding current remains $\mathrm{d}(a/f_a)/(2\pi)$. Thus the radius dependence changes the kinetic coefficient, but not the identification of the symmetry that protects the axion. The strength of QCD can nevertheless depend on the radius. This is because
\begin{equation}
    \frac{1}{g_s^2(\beta)}=\frac{e^{b\beta/2}}{g_s^2(0)}.
    \label{eq:beta-colour-coupling}
\end{equation}
At zero temperature and sufficiently low spacetime curvature, the QCD vacuum energy near its CP-preserving minimum is
\begin{equation}
    E_{\rm QCD}(\beta,\vartheta)=E_{\rm QCD}(\beta,0)+\frac12\chi_{\rm QCD}(\beta)\vartheta^2+\mathcal O(\vartheta^4).
    \label{eq:beta-qcd-potential}
\end{equation}
We hold $\bar\theta$ fixed throughout. Changing $\beta$ can then change the strength of the potential, while its minimum remains at $\vartheta=0$. Dividing the second derivative with respect to $a$ by the kinetic coefficient gives the axion mass at fixed $\beta$
\begin{equation}
    m_a^2(\beta)=\frac{e^{b\beta}\chi_{\rm QCD}(\beta)}{f_a^2}
    =\frac{\chi_{\rm QCD}(\beta)}{f_{\rm eff}^2(\beta)}.
    \label{eq:beta-qcd-axion-mass}
\end{equation}
The term $E_{\rm QCD}(\beta,0)$ is independent of the axion, but its $\beta$ dependence still contributes to the radion potential. It must therefore be retained when studying $\beta$ stabilisation. 

\subsubsection{Electric one-form breaking by charged matter}

Consider the periodic charged scalar of \eqref{eq:charged-scalar-potential}, keeping its proper five-dimensional mass $m_5$ fixed. In the metric \eqref{eq:beta-einstein-metric}, its four-dimensional Einstein-frame KK masses are
\begin{equation}
    m_{k,E}^2(\beta,a)=m_5^2e^{-b\beta/2}
    +\frac{e^{-3b\beta/2}}{R^2}\left(k+\frac{qa}{2\pi f_a}\right)^2,
    \qquad k\in\mathbb Z.
    \label{eq:beta-charged-kk-masses}
\end{equation}
The first term is the Einstein-frame contribution of the proper mass. The second comes from momentum along the circle, shifted by the Wilson line. The same radius dependence can be understood from a particle worldline winding around the compact direction. At fixed background fields, its Euclidean action is
\begin{equation}
    S_E(w)=2\pi Rm_5e^{b\beta/2}|w|-iqw\frac{a}{f_a}.
    \label{eq:beta-wrapped-worldline}
\end{equation}
The real part is the mass multiplied by the proper length of the winding trajectory. The imaginary part depends only on the Wilson line and is unchanged. Consequently, this charged sector still breaks the electric symmetry to $\mathbb Z_{|q|}^{(1)}$ when considered by itself \cite{CraigKongsore2024}. Applying the proper-time calculation of appendix~\ref{app:radion-one-loop} to \eqref{eq:beta-charged-kk-masses}, the shifted KK momenta introduce the phase $\exp(iwqa/f_a)$ after Poisson resummation. Combining the two winding orientations gives
\begin{align}
    \Delta V_{\rm el}(\beta,a)
    &=-\frac{3e^{-3b\beta}}{32\pi^6R^4}
    \sum_{w=1}^{\infty}\frac{e^{-z_w}}{w^5}
    \left(1+z_w+\frac{z_w^2}{3}\right)
    \cos\left(wq\frac{a}{f_a}\right),\nonumber\\
    z_w&=2\pi wm_5R e^{b\beta/2}.
    \label{eq:beta-charged-potential}
\end{align}
Here $z_w$ is the winding action appearing in \eqref{eq:charged-scalar-potential}, now evaluated using the proper circumference. The coefficient includes the two real degrees of freedom of the complex scalar. The factor $e^{-3b\beta}$ accounts for the radius dependence of the Einstein-frame prefactor, while $e^{-z_w}$ gives the suppression associated with propagation around the circle. At $\beta=0$, the result reduces to \eqref{eq:charged-scalar-potential}. For evolving backgrounds, this expression is the leading potential term in a derivative expansion and applies when background variations are sufficiently slow compared with the masses of the modes integrated out. The zero-winding term is independent of $a$ and renormalises the bulk cosmological constant. It contributes to the stabilising potential through
\begin{equation}
    V(\beta)\supset2\pi R\,\delta\Lambda_5\,e^{-b\beta/2}.
    \label{eq:beta-charged-local-term}
\end{equation}
As explained in appendix~\ref{app:radion-one-loop}, this correction is absorbed into the renormalised bulk cosmological constant which is however irrelevant for QCD in this section. Its effect on the radion is not suppressed by $e^{-z_w}$. The finite Casimir contribution must also be counted only once: a charged field included through \eqref{eq:beta-charged-potential} should not be included again as a separate contribution to the stabilising spectrum. For $z_1\gg1$, the first winding dominates. Using the notation of \eqref{eq:heavy-electric-potential}, but retaining the full polynomial prefactor, we obtain
\begin{align}
    \Delta V_{\rm el}(\beta,a)
    &\simeq-\Lambda_{\rm el}^4(\beta)\cos\left(q\frac{a}{f_a}\right),\nonumber\\
    \Lambda_{\rm el}^4(\beta)
    &=\frac{3e^{-3b\beta-z_1}}{32\pi^6R^4}
    \left(1+z_1+\frac{z_1^2}{3}\right)
    \simeq\frac{m_5^2}{8\pi^4R^2}e^{-2b\beta-z_1}.
    \label{eq:beta-electric-amplitude}
\end{align}
The last expression retains only the leading term for large $z_1$. To see how this contribution changes with the radius, we use $\partial_\beta z_1=bz_1/2$. The full first-winding amplitude then satisfies
\begin{equation}
    \frac1b\partial_\beta\log\Lambda_{\rm el}^4
    =-3-\frac{z_1^2(1+z_1)}{6(1+z_1+z_1^2/3)}.
    \label{eq:beta-electric-log-derivative}
\end{equation}
Thus increasing the proper radius suppresses the charged-particle contribution. Whether axion quality improves also depends on how $\chi_{\rm QCD}(\beta)$ changes, since the displacement of the minimum is determined by the relative strength of the two potentials. For $q^2\Lambda_{\rm el}^4(\beta)\ll\chi_{\rm QCD}(\beta)$, minimising the QCD and electric potentials near $\vartheta=0$ gives
\begin{equation}
    \theta_{\rm eff}\simeq
    \frac{q\Lambda_{\rm el}^4(\beta)}{\chi_{\rm QCD}(\beta)}
    \sin(q\bar\theta).
    \label{eq:beta-electric-quality-shift}
\end{equation}
For a generic relative phase, the strong CP bound therefore requires
\begin{equation}
    |q|\Lambda_{\rm el}^4(\beta)\lesssim10^{-10}\chi_{\rm QCD}(\beta).
    \label{eq:beta-electric-quality}
\end{equation}
The kinetic coefficient determines the mass whereas this condition concerns the position of the minimum in the periodic angle $a/f_a$.

\subsubsection{Gauging the magnetic two-form symmetry}

The magnetic coupling has a different radius dependence. For the entirely external four-form $H_4$, the four inverse metrics in its kinetic term contribute $e^{2b\beta}$, while the five-dimensional volume element contributes $e^{-b\beta/2}$. Reducing \eqref{eq:magnetic-gauging-five} therefore gives
\begin{equation}
    S_K^{(4)}=\int\left[
    -\frac{e^{3b\beta/2}}{2e_{K,4}^2}H_4\wedge *_E H_4
    +\frac{M}{2\pi f_a}K_3\wedge\mathrm{d}a\right],
    \qquad e_{K,4}^2=\frac{e_{K,5}^2}{2\pi R}.
    \label{eq:beta-magnetic-action}
\end{equation}
Here $e_{K,4}$ remains the fixed reference coupling of \eqref{eq:magnetic-gauging-four}; the radius dependence is displayed explicitly. The topological mixing contains no metric and is unchanged. Varying $K_3$ gives
\begin{equation}
    \mathrm{d}\left(
    \frac{e^{3b\beta/2}}{e_{K,4}^2}*_E H_4
    -\frac{Ma}{2\pi f_a}\right)=0.
    \label{eq:beta-four-form-equation}
\end{equation}
This integrates to
\begin{equation}
    \frac{e^{3b\beta/2}}{e_{K,4}^2}*_E H_4
    -\frac{Ma}{2\pi f_a}=n,
    \qquad n\in\mathbb Z.
    \label{eq:beta-four-form-flux}
\end{equation}
It is the integration constant $n$ that remains fixed as the fields evolve. The four-form density changes with $\beta$ and $a$ so that this equation continues to hold. Eliminating $H_4$ at fixed $n$, with the boundary term appropriate to fixed flux, gives
\begin{equation}
    V_{K,n}(\beta,a)=\frac{e_{K,4}^2}{2}e^{-3b\beta/2}
    \left(n+\frac{Ma}{2\pi f_a}\right)^2.
    \label{eq:beta-magnetic-potential}
\end{equation}
The factor $e^{-3b\beta/2}$ is the inverse of the four-form kinetic coefficient. It follows from solving \eqref{eq:beta-four-form-flux}, rather than holding $*_E H_4$ fixed. The branch identification $a\to a+2\pi f_a$, $n\to n-M$ is unchanged as the $\beta$ dependent part is an overall factor here. Magnetic gauging considered by itself also leaves the same discrete electric symmetry as at fixed radius. The second derivative of \eqref{eq:beta-magnetic-potential} must again be divided by the axion kinetic coefficient to obtain the mass at fixed $\beta$. This gives
\begin{equation}
    m_K^2(\beta)=e^{b\beta}\partial_a^2V_{K,n}
    =\frac{M^2e_{K,4}^2}{4\pi^2f_a^2}e^{-b\beta/2}.
    \label{eq:beta-magnetic-mass}
\end{equation}
For $M\ne0$, the branch minimum remains $a_n=-2\pi f_a n/M$. Thus the evolving radius changes the strength of this potential but does not move its preferred axion value. Combining it with the QCD potential gives
\begin{equation}
    \theta_{\rm eff}\simeq
    \frac{m_K^2(\beta)}{m_a^2(\beta)+m_K^2(\beta)}
    \left(\bar\theta+\frac{a_n}{f_a}\right),
    \label{eq:beta-magnetic-quality}
\end{equation}
provided the resulting minimum lies sufficiently close to $\vartheta=0$. As at fixed radius, the bound depends on the mismatch between the branch minimum and the QCD minimum. For $|\bar\theta+a_n/f_a|$ of order one, it requires $m_K^2(\beta)\lesssim10^{-10}m_a^2(\beta)$. A suitably aligned branch can weaken this requirement. The radius dependence is not exponentially suppressed in the proper radius and does not provide the winding exponential present in the charged-particle contribution. The magnetic potential also acts on the radion. At fixed $a$ and $n$,
\begin{equation}
    \partial_\beta V_{K,n}=-\frac{3b}{2}V_{K,n}.
    \label{eq:beta-magnetic-force}
\end{equation}
For $M\ne0$, minimising this contribution alone sets $a=a_n$ and makes its energy vanish. It then supplies no independent stabilising potential for $\beta$. When QCD is also present, the axion need not sit at $a_n$, and the remaining branch energy contributes to the radion equation. For $M=0$, the potential is independent of the axion, but a non-zero flux $n$ still contributes to the radion potential.

For comparison, electric one-form gauging changes the axion kinetic term to
\begin{equation}
    \mathcal L_{\rm kin}
    =\frac12e^{-b\beta}
    \left(\partial_\mu a-kf_a\mathcal A_\mu\right)
    \left(\partial^\mu a-kf_a\mathcal A^\mu\right).
    \label{eq:beta-electric-gauging}
\end{equation}
For $k\ne0$, the vector still eats the axion through the St\"uckelberg mechanism. The radius dependence changes the kinetic coefficient but does not restore an independent axion scalar. The axion--gluon coupling also remains non-invariant under the gauged shift, so the additional anomaly-cancelling ingredients discussed in section~\ref{sec:extra-dimensional-axions} are still required.

\subsubsection{The combined potential and the combined system}

When the electric symmetry is not gauged, adding the contributions above to the stabilising potential in \eqref{eq:beta-four-dimensional-action} gives
\begin{equation}
    V(\beta,a)=V(\beta)+E_{\rm QCD}(\beta,\vartheta)
    +\Delta V_{\rm el}(\beta,a)+V_{K,n}(\beta,a).
    \label{eq:beta-combined-axion-potential}
\end{equation}
Here $V(\beta)$ retains its earlier meaning as the axion-independent stabilising contribution, with local vacuum corrections included in its renormalised parameters. The QCD vacuum energy is retained in $E_{\rm QCD}(\beta,\vartheta)$ and should not also be included in $V(\beta)$. Any absent symmetry-breaking sector is omitted from the sum.

Applying \eqref{eq:general-extra-dimensional-quality} at a fixed value of $\beta$, the additional potential $\Delta V$ must satisfy
\begin{equation}
    \left|f_a\partial_a\Delta V(\beta,-f_a\bar\theta)\right|
    \lesssim10^{-10}\chi_{\rm QCD}(\beta),
    \qquad
    \left|f_a^2\partial_a^2\Delta V\right|
    \ll\chi_{\rm QCD}(\beta),
    \label{eq:beta-general-quality}
\end{equation}
with the second condition imposed near the QCD minimum. In the present model, $\Delta V$ contains the electric and magnetic contributions above. The second inequality specifies when the leading displacement formula applies and it is not a necessary condition for an exactly aligned potential. The experimental bound constrains the late-time strong CP angle. Imposing the same bound at every point along an earlier radion trajectory is an additional requirement.

These conditions locate instantaneous potential minima, but do not determine whether the fields follow them. For homogeneous fields in an Einstein-frame FLRW background, the action gives
\begin{align}
    \ddot a+(3H-b\dot\beta)\dot a
    +e^{b\beta}\partial_aV&=0,\nonumber\\
    \frac{3b^2M_{\rm Pl}^2}{8}
    \left(\ddot\beta+3H\dot\beta\right)
    +\frac b2e^{-b\beta}\dot a^2
    +\partial_\beta V&=0,
    \label{eq:beta-axion-background-equations}
\end{align}
where dots denote derivatives with respect to Einstein-frame cosmic time and $H$ is the corresponding Hubble parameter. The first equation shows that radion motion changes both the axion kinetic term and its effective friction. The second shows that the axion kinetic energy and the $\beta$ dependence of its potentials contribute to radion evolution. A small displacement of the axion minimum therefore does not, by itself, establish that the effect on radion stabilisation is negligible.

A stationary late-time vacuum requires the radion to settle at a stable value where the axion minimum also satisfies the strong CP bound. Stabilising $\beta$ fixes the effective decay constant and the relative strengths of the potentials, but does not automatically ensure the required vacuum alignment. We next construct the stabilising potential $V(\beta)$ and then examine whether its minimum remains stable after including the QCD and additional axion contributions.

\subsection{$\beta$ stabilisation}
\label{sec:beta-stabilisation}

We now construct the axion-independent potential $V(\beta)$ appearing in \eqref{eq:beta-four-dimensional-action}. We consider two possibilities: stabilisation by the bulk cosmological constant and Casimir energy alone, and stabilisation with an additional scalar winding around the circle. In both cases, the fields responsible for stabilisation are taken to be neutral under $C$, so they do not directly generate a Wilson-line potential. The QCD and additional axion contributions remain separate terms in \eqref{eq:beta-combined-axion-potential}. For a stationary background with constant axion, the radion minimum must satisfy
\begin{equation}
    V'(\beta_0)=0,\qquad V''(\beta_0)>0,\qquad
    m_\beta^2=\frac{8V''(\beta_0)}{3b^2M_{\rm Pl}^2},
    \label{eq:beta-ds-conditions}
\end{equation}
where primes denote derivatives with respect to $\beta$. The mass follows from the kinetic coefficient in \eqref{eq:beta-four-dimensional-action}. A Minkowski vacuum additionally requires $V(\beta_0)=0$, while a dS vacuum requires $V(\beta_0)>0$, with $H^2=V(\beta_0)/(3M_{\rm Pl}^2)$.

Both of our stabilisation constructions include a bulk cosmological constant. Since $\sqrt{|g_5|}=e^{-b\beta/2}\sqrt{-g_E}$, its reduction gives
\begin{equation}
    V_\Lambda(\beta)=2\pi R\Lambda_5e^{-b\beta/2}.
    \label{eq:beta-bulk-potential}
\end{equation}
Here $\Lambda_5$ is the renormalised five-dimensional vacuum energy density, of mass dimension five. It includes the local zero-winding corrections discussed around \eqref{eq:beta-charged-local-term}. Although $\Lambda_5$ is constant, its four-dimensional Einstein-frame contribution depends on the radius. The second ingredient is the finite Casimir energy. A field propagating around this circle has discrete momenta in the compact direction. These are its Kaluza–Klein modes. Each mode has quantum fluctuations even when no particles are present, and the corresponding vacuum energy depends on the KK masses. For one neutral real bosonic degree of freedom with proper mass $m$ and periodic boundary conditions, the potential is
\begin{equation}
    V_B(\beta;m)
    =-\frac{3e^{-3b\beta}}{64\pi^6R^4}
    \sum_{w=1}^{\infty}\frac{e^{-z_w}}{w^5}
    \left(1+z_w+\frac{z_w^2}{3}\right),
    \qquad z_w=2\pi wmR e^{b\beta/2}.
    \label{eq:beta-finite-casimir}
\end{equation}
The derivation is given in appendix~\ref{app:radion-one-loop}, with the result in \eqref{eq:beta-casimir-evaluation}. The same calculation gives \eqref{eq:beta-charged-potential} after including the Wilson-line phase and the two real degrees of freedom of a complex scalar. In the massless limit, 
\begin{equation}
    V_B(\beta;0)
    =-\frac{3\zeta(5)}{64\pi^6R^4}e^{-3b\beta}.
    \label{eq:beta-massless-casimir}
\end{equation}
A periodic fermionic degree of freedom contributes with the opposite sign. In particular, a five-dimensional Dirac fermion contributes $-4V_B(\beta;m)$. The different radius dependence of massless and massive contributions allows their competition to produce a minimum \cite{RubinRoth1983,PontonPoppitz2001,GreeneLevin2007}.

\subsubsection{Stabilisation without scalar winding}
\label{sec:stabilisation-without-winding}

We first consider a spectrum containing $N_b>0$ massless bosonic degrees of freedom and neutral massive fermions with proper masses $m_i>0$ and multiplicities $N_i$. All fields are periodic around the circle, and the proper masses are independent of $\beta$. The potential is
\begin{equation}
    V(\beta)=V_\Lambda(\beta)+V_{\rm Cas}(\beta),
    \qquad
    V_{\rm Cas}(\beta)=N_bV_B(\beta;0)-\sum_iN_iV_B(\beta;m_i).
    \label{eq:fluxless-potential}
\end{equation}
The multiplicities count physical degrees of freedom, so one Dirac fermion has $N_i=4$. At sufficiently small radius, the massive fermions behave approximately as massless fields and give a positive Casimir contribution. As the radius increases, their contribution becomes exponentially suppressed, while the negative massless bosonic contribution remains. A positive bulk cosmological constant eventually dominates at large radius because its contribution decreases more slowly than the Casimir energy.

To find a Minkowski vacuum, choose the reference radius so that the candidate minimum lies at $\beta_0=0$. The conditions $V(0)=V'(0)=0$ then give
\begin{equation}
    2\pi R\Lambda_5=\frac2bV_{\rm Cas}'(0),
    \qquad
    V_{\rm Cas}(0)+\frac2bV_{\rm Cas}'(0)=0.
    \label{eq:fluxless-minkowski-conditions}
\end{equation}
The second equation determines the radius for a specified massive spectrum, while the first fixes the required bulk cosmological constant. Substituting \eqref{eq:beta-finite-casimir} and writing $z_{i,w}=2\pi wm_iR$ gives
\begin{equation}
    N_b\zeta(5)
    =\sum_iN_i\sum_{w=1}^{\infty}
    \frac{e^{-z_{i,w}}}{w^5}
    \left(1+z_{i,w}+\frac25z_{i,w}^2+\frac1{15}z_{i,w}^3\right).
    \label{eq:fluxless-radius-equation}
\end{equation}
For fixed positive masses, the right-hand side decreases strictly as $R$ increases.\footnote{Since $\mathrm{d}z_{i,w}/\mathrm{d}\log R=z_{i,w}$,
\begin{equation}
    \begin{aligned}
        &\frac{\mathrm{d}}{\mathrm{d}\log R}
        \left[e^{-z_{i,w}}
        \left(1+z_{i,w}+\frac25z_{i,w}^2+\frac1{15}z_{i,w}^3\right)\right]\\
        &\qquad=-\frac{e^{-z_{i,w}}z_{i,w}^2}{15}
        \left(z_{i,w}^2+3z_{i,w}+3\right)<0.
    \end{aligned}
    \label{eq:fluxless-radius-monotonicity}
\end{equation}
} It approaches $(\sum_iN_i)\zeta(5)$ as $R\to0$ and vanishes as $R\to\infty$. Hence a unique finite Minkowski radius exists precisely when
\begin{equation}
    \sum_iN_i>N_b.
    \label{eq:fluxless-existence-condition}
\end{equation}
The fermionic degrees of freedom must therefore outnumber the massless bosonic ones, allowing their positive contribution to dominate at small radius before becoming suppressed. Equal fermion masses are not required. At this radius, \eqref{eq:fluxless-minkowski-conditions} fixes
\begin{equation}
    2\pi R\Lambda_5
    =\frac{1}{320\pi^6R^4}
    \sum_iN_i\sum_{w=1}^{\infty}
    \frac{e^{-z_{i,w}}}{w^5}
    z_{i,w}^2(1+z_{i,w})>0.
    \label{eq:fluxless-bulk-tuning}
\end{equation}
Thus the Minkowski solution requires a positive renormalised bulk cosmological constant. To check stability, we use \eqref{eq:beta-casimir-derivatives} and eliminate the massless contribution with \eqref{eq:fluxless-radius-equation}, obtaining
\begin{equation}
    V''(0)
    =\frac{b^2}{256\pi^6R^4}
    \sum_iN_i\sum_{w=1}^{\infty}
    \frac{e^{-z_{i,w}}}{w^5}
    z_{i,w}^2\left(z_{i,w}^2+3z_{i,w}+3\right)>0.
    \label{eq:fluxless-positive-curvature}
\end{equation}
Every term is positive, so the solution is a local minimum in the radion direction. Its radion mass follows from \eqref{eq:beta-ds-conditions}. A nearby dS minimum follows by replacing $\Lambda_5$ with $\Lambda_5+\delta\Lambda_5$, where $\delta\Lambda_5>0$ is sufficiently small, while keeping the masses and reference radius fixed. Expanding about the Minkowski solution gives
\begin{equation}
    \delta\beta\simeq\frac{\pi bR\,\delta\Lambda_5}{V''(0)},
    \qquad
    V(\delta\beta)\simeq2\pi R\,\delta\Lambda_5>0.
    \label{eq:fluxless-small-uplift}
\end{equation}
Here $V''(0)$ refers to the original Minkowski solution, while $V(\delta\beta)$ is the shifted vacuum energy after changing $\Lambda_5$. The second derivative remains positive for a sufficiently small change, giving a local dS minimum. Since the potential approaches zero at infinite proper radius, its local stability alone does not determine the vacuum lifetime.

\begin{figure}
    \centering
    \includegraphics[width=0.5\linewidth]{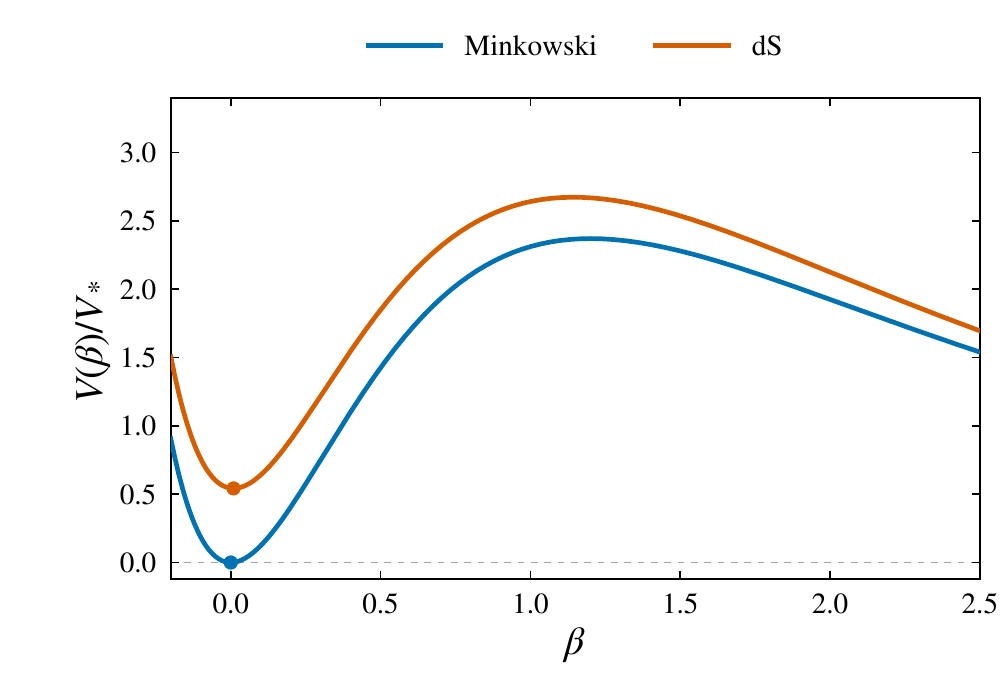}
    \caption{Leading-order radion stabilisation without scalar winding. The massless bosonic count is \(N_b=5+3+24=32\), from the five-dimensional graviton, \(C\), and eight gluons. Twelve equal-mass Dirac spectators, neutral under \(C\) and colour singlets, give \(N_f=48>N_b\). We assume periodic boundary conditions, unbroken colour, trivial colour holonomy, and no magnetic three-form sector. With \(b=R=1\), \(2\pi mR\simeq2.05226\), and \(V_*=3/(64\pi^6R^4)\), tuning \(\Lambda_5\) gives a Minkowski minimum at \(\beta=0\) (blue). Increasing \(\Lambda_5\) by \(10\%\), at fixed mass and reference radius, gives a local de Sitter minimum at \(\beta\simeq0.00923\) (orange-red). Dots mark the minima, both with \(V''>0\).}
    \label{fig:withoutwindingstabilisedbeta}
\end{figure}
\subsubsection{Stabilisation with scalar winding}
\label{sec:stabilisation-with-winding}

We now include an independent compact scalar $\sigma\sim\sigma+2\pi$, with a positive kinetic term
\begin{equation}
    S_\sigma=\frac{f_{\sigma,5}^3}{2}
    \int\mathrm{d}^4x\,\mathrm{d}y\sqrt{|g_5|}\,
    g_5^{AB}\partial_A\sigma\partial_B\sigma,
    \qquad f_{\sigma,5}^3>0.
    \label{eq:beta-winding-action}
\end{equation}
Since it has shift symmetry, it is natural to assume that a potential in the effective theory is forbidden. The scalar is neutral under $C$ and is independent of the scalar dual to the magnetically gauged three-form. In particular, its kinetic term contains $\mathrm{d}\sigma$, rather than a St\"uckelberg coupling to $C$, so this additional sector does not give the Wilson-line axion a classical mass. A non-zero winding configuration is
\begin{equation}
    \sigma=\frac{n_\sigma y}{R}\pmod{2\pi},
    \qquad n_\sigma\in\mathbb Z\setminus\{0\},
    \qquad \oint_{S^1}\mathrm{d}\sigma=2\pi n_\sigma.
    \label{eq:beta-winding-background}
\end{equation}
As $y$ goes once around the spatial circle, $\sigma$ winds $n_\sigma$ times around its periodic field space, returning smoothly to the same physical value. Most importantly, $\partial_y\sigma=n_\sigma/R$ gives a $y$-independent stress tensor \cite{BlancoPilladoShlaer2010}. Substituting it into \eqref{eq:beta-winding-action} gives
\begin{equation}
    V_\sigma(\beta)
    =\frac{\pi f_{\sigma,5}^3n_\sigma^2}{R}e^{-3b\beta/2}>0.
    \label{eq:beta-winding-potential}
\end{equation}
The factor $e^{-3b\beta/2}$ combines the volume element with the inverse internal metric. The winding energy therefore grows as the circle shrinks and opposes contraction. Before adding the Casimir contribution, the potential is $V=V_\Lambda+V_\sigma$. Its stationary-point equation gives
\begin{equation}
    V'(\beta_0)
    =-\frac b2\left(V_\Lambda+3V_\sigma\right)_{\beta_0}=0,
    \qquad
    V_\Lambda(\beta_0)=-3V_\sigma(\beta_0).
    \label{eq:beta-classical-stationarity}
\end{equation}
Since $V_\sigma>0$, this requires $\Lambda_5<0$. The radius, vacuum energy and second derivative are then
\begin{align}
    R^2e^{b\beta_0}
    &=-\frac{3f_{\sigma,5}^3n_\sigma^2}{2\Lambda_5},\nonumber\\
    V(\beta_0)&=-2V_\sigma(\beta_0)<0,
    \qquad
    V''(\beta_0)=\frac{3b^2}{2}V_\sigma(\beta_0)>0.
    \label{eq:beta-classical-ads}
\end{align}
Thus scalar winding and a negative bulk cosmological constant stabilise the radius, but the resulting vacuum is AdS. We next include the finite Casimir energy:
\begin{equation}
    V(\beta)=V_\Lambda(\beta)+V_\sigma(\beta)+V_{\rm Cas}(\beta).
    \label{eq:beta-total-potential}
\end{equation}
The spectrum entering $V_{\rm Cas}$ must now include the fluctuations of $\sigma$ as well as the other bulk fields.\footnote{The winding background can mix the scalar fluctuations with the Kaluza--Klein graviphoton. The free-field Casimir expressions apply when the induced mass splittings are negligible compared with the relevant compactification scales as otherwise the mixed fluctuation spectrum must be used.} A purely massless Casimir spectrum is insufficient to obtain a stable Minkowski or dS vacuum from these contributions. In that case $V_{\rm Cas}\propto e^{-3b\beta}$, and eliminating $V_\Lambda$ with the stationary-point equation gives
\begin{equation}
    m_\beta^2=-\frac4{M_{\rm Pl}^2}
    \left[V(\beta_0)+V_\sigma(\beta_0)\right].
    \label{eq:beta-massless-obstruction}
\end{equation}
For $V(\beta_0)\geq0$, this is negative. Massive fields are therefore needed within this construction as well. Their winding actions depend on $\beta$, so their Casimir contribution is not a single exponential and can change both the vacuum energy and the sign of the radion mass. For a general massive spectrum, stationarity fixes the bulk term as
\begin{equation}
    V_\Lambda(\beta_0)
    =-3V_\sigma(\beta_0)+\frac2bV_{\rm Cas}'(\beta_0).
    \label{eq:beta-general-stationarity}
\end{equation}
Substituting this relation into the potential and its second derivative gives
\begin{align}
    V(\beta_0)
    &=-2V_\sigma(\beta_0)+V_{\rm Cas}(\beta_0)
    +\frac2bV_{\rm Cas}'(\beta_0),\nonumber\\
    V''(\beta_0)
    &=\frac{3b^2}{2}V_\sigma(\beta_0)
    +\frac b2V_{\rm Cas}'(\beta_0)
    +V_{\rm Cas}''(\beta_0).
    \label{eq:beta-general-energy-curvature}
\end{align}
For a specified spectrum and candidate radius, evaluating $V_{\rm Cas}$ using \eqref{eq:beta-finite-casimir} and its derivatives using \eqref{eq:beta-casimir-derivatives}, with the multiplicities and signs of the chosen spectrum. The winding contribution must then give positive vacuum energy and positive $V''$, while \eqref{eq:beta-general-stationarity} determines $\Lambda_5$. A stable dS vacuum with $\Lambda_5<0$ therefore requires
\begin{equation}
    \begin{aligned}
        \max\left\{
        0,\,
        \frac{2V_{\rm Cas}'}{3b},\,
        -\frac{V_{\rm Cas}'}{3b}
        -\frac{2V_{\rm Cas}''}{3b^2}
        \right\}
        &<V_\sigma<\frac12V_{\rm Cas}+\frac1bV_{\rm Cas}'.
    \end{aligned}
    \label{eq:beta-winding-ds-window}
\end{equation}
All quantities are evaluated at $\beta_0$. The lower bounds enforce positive winding energy, negative $\Lambda_5$ and positive radion mass; the upper bound ensures positive vacuum energy. A non-empty interval gives an allowed range of $f_{\sigma,5}$ for each non-zero $n_\sigma$ through \eqref{eq:beta-winding-potential}, while \eqref{eq:beta-general-stationarity} fixes $\Lambda_5$. Saturating only the upper bound gives a Minkowski vacuum.

\begin{figure}
    \centering
    \includegraphics[width=0.5\linewidth]{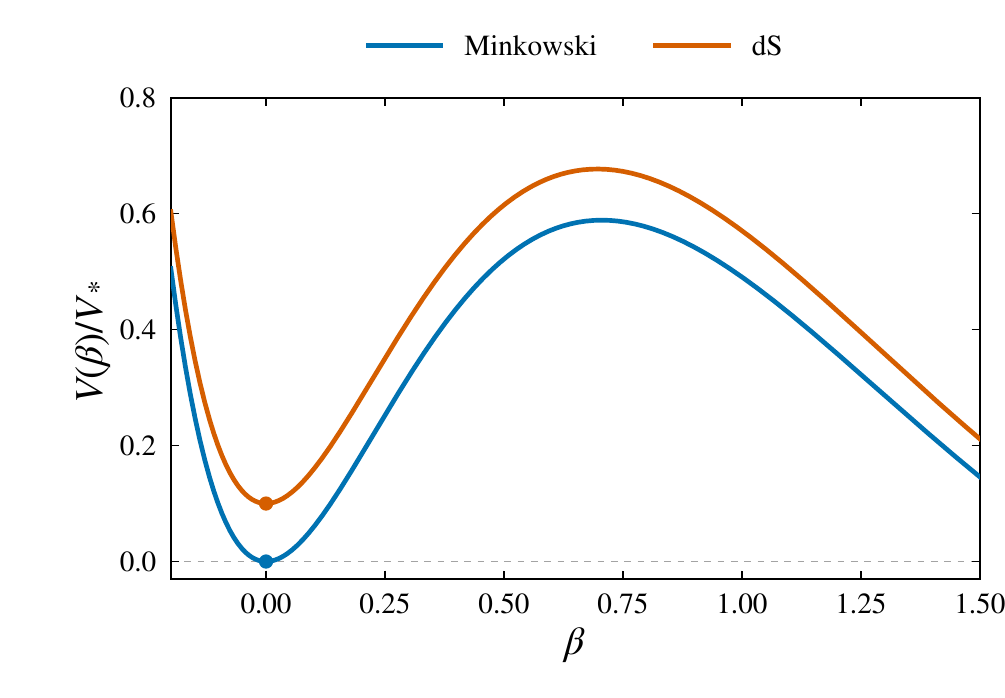}
    \caption{Leading-order radion stabilisation with scalar winding and massive Casimir energy. Including the winding scalar gives \(N_b=5+3+24+1=33\), twelve equal-mass Dirac spectators, neutral under \(C\) and colour singlets, give \(N_f=48\). We assume periodic boundary conditions, unbroken colour, trivial colour holonomy and no magnetic three-form sector. We take \(b=R=1\), \(n_\sigma=1\), \(2\pi mR=12/5\) and \(V_*=3/(64\pi^6R^4)\). The blue curve uses \(V_\sigma(0)=V_{\rm Cas}(0)/2+V'_{\rm Cas}(0)/b\), giving a local Minkowski minimum. Reducing this coefficient by \(V_*/20\) and retuning \(\Lambda_5\) through Eq. (4.50) gives a local de Sitter minimum (orange-red), with \(V(0)=0.1V_*\). The bulk cosmological constant is negative for both curves. Both minima lie at \(\beta=0\) and satisfy \(V''(0)>0\). No vacuum energy is subtracted. A deeper AdS minimum lies outside the displayed range.}
    \label{fig:betawithwindingpotential}
\end{figure}
Lastly, both of our constructions use the leading Casimir potential. They establish stabilisation in the radion direction for the specified spectrum as illustrated in fig. \ref{fig:withoutwindingstabilisedbeta} and fig. \ref{fig:betawithwindingpotential}. Compatibility with the axion vacuum must subsequently be checked using the full potential \eqref{eq:beta-combined-axion-potential} as this could be in tension with maintaining axion quality, including the axion-independent QCD vacuum energy and the additional symmetry-breaking contributions which we do in the following section. 

\subsection{$\beta$ stabilisation and axion quality}
\label{sec:beta-quality-compatibility}

We now derive the conditions under which the radion minima constructed in section~\ref{sec:beta-stabilisation} remain compatible with axion quality after including the QCD and additional axion contributions in \eqref{eq:beta-combined-axion-potential}. The neutral and charged masses are independent parameters, so lighter neutral fields can support the Casimir minimum while heavier fields charged under $C$ generate an exponentially suppressed axion potential at the same proper radius. We must then check that the added contributions induce only a small radion displacement and preserve the stability of the coupled axion--radion minimum. These checks require comparing their radion derivatives with the \(V''\) of the stabilising potential. They also apply separately to the QCD vacuum energy and any magnetic contribution, which do not share the charged-particle winding suppression.

We first discuss how to see this displacement. Before including the additional potentials, the radion sits at $\beta_0$, where
\begin{equation}
    V'(\beta_0)=0,
    \qquad
    V''(\beta_0)=\frac{3b^2M_{\rm Pl}^2m_\beta^2}{8}>0.
    \label{eq:beta-stabilisation-stiffness}
\end{equation}
If the radion moves to $\beta_0+\delta\beta$, the slope of its original potential becomes $V'(\beta_0+\delta\beta)\simeq V''(\beta_0)\delta\beta$. At the new minimum, this slope must cancel the slope supplied by the additional potentials. Thus a larger $V''(\beta_0)$ means a smaller displacement for the same added contribution. The value of $V(\beta_0)$ determines the vacuum energy and does not enter this balance. The required second derivative is given by \eqref{eq:fluxless-positive-curvature} for the Minkowski construction without scalar winding, and by \eqref{eq:beta-general-energy-curvature} when winding is included.

To evaluate the added contribution, we work in the perturbative regime of \eqref{eq:beta-general-quality}, where the axion remains close to its QCD minimum $a=-f_a\bar\theta$. At this value, $\vartheta=0$, so the QCD contribution is $E_{\rm QCD}(\beta,0)$. The electric and magnetic contributions are contained in $\Delta V(\beta,-f_a\bar\theta)$. Treating their effect and the QCD vacuum term as small perturbations of the radion minimum, the stationarity condition becomes
\begin{equation*}
    0\simeq V''(\beta_0)\delta\beta
    +\left.
    \frac{\partial}{\partial\beta}
    \left[
        E_{\rm QCD}(\beta,0)
        +\Delta V(\beta,-f_a\bar\theta)
    \right]\right|_{\beta=\beta_0}.
\end{equation*}
The first term is the change in the slope of the stabilising potential and the second is the slope of the added contributions, evaluated at the original minimum. Solving for the displacement gives
\begin{equation}
    \delta\beta\simeq
    -\frac{1}{V''(\beta_0)}
    \left.
    \frac{\partial}{\partial\beta}
    \left[
        E_{\rm QCD}(\beta,0)
        +\Delta V(\beta,-f_a\bar\theta)
    \right]\right|_{\beta=\beta_0}.
    \label{eq:beta-backreaction-displacement}
\end{equation}
These derivatives hold the axion fixed. Its small displacement from the QCD minimum affects this result only at higher order in the additional breaking. The proper radius changes by the factor $e^{b\delta\beta/2}$, so negligible displacement requires $|b\delta\beta|\ll1$.\footnote{The expansion also assumes that the additional radion derivatives are perturbatively small and that $|V'''(\beta_0)\delta\beta|\ll V''(\beta_0)$, so the second derivative of the original potential changes little across the displacement.} We will check these requirements for the individual contributions below.

A small displacement does not yet establish that the new stationary point is stable. The potential must increase under small changes of either field, including changes in which both fields move together. For the full potential $V(\beta,a)$, this requires
\begin{equation}
    \partial_a^2V>0,
    \qquad
    \partial_\beta^2V
    -\frac{(\partial_\beta\partial_aV)^2}{\partial_a^2V}>0.
    \label{eq:beta-coupled-stability}
\end{equation}
All derivatives here are evaluated at the new stationary point. The first condition ensures stability when the axion varies at fixed radius. The second tests the radion direction while allowing the axion to readjust. This readjustment lowers the quadratic energy cost of a radion displacement by the displayed mixing term. If $\partial_\beta\partial_aV=0$, the conditions reduce to positive second derivatives in the two separate directions. We must therefore check that the added potentials neither displace the radius appreciably nor spoil these stability conditions. 

\subsubsection{The charged contribution}

Consider first the charged scalar, with magnetic gauging absent. For $z_1\gg1$, its potential is dominated by the first winding in \eqref{eq:beta-electric-amplitude}. If $q^2\Lambda_{\rm el}^4\ll\chi_{\rm QCD}$, the axion minimum is given by \eqref{eq:beta-electric-quality-shift}. Substituting that minimum into the sum of the QCD and charged potentials gives, through second order in the charged contribution,
\begin{align}
    \left[
        E_{\rm QCD}(\beta,\vartheta)
        +\Delta V_{\rm el}(\beta,a)
    \right]_{a=\langle a\rangle}
    &\simeq E_{\rm QCD}(\beta,0)
    -\Lambda_{\rm el}^4(\beta)\cos(q\bar\theta)
    \nonumber\\
    &\quad
    -\frac{q^2\bigl(\Lambda_{\rm el}^4(\beta)\bigr)^2}
    {2\chi_{\rm QCD}(\beta)}
    \sin^2(q\bar\theta).
    \label{eq:beta-backreaction-electric-relaxed-potential}
\end{align}
The term linear in $\Lambda_{\rm el}^4$ remains even when the residual strong CP angle is very small. It therefore contributes to the radion equation. The last term accounts for the small displacement of the axion and is suppressed by an additional power of the charged amplitude.

The relevant question is consequently how the derivatives of $\Lambda_{\rm el}^4$ compare with the restoring force from $V(\beta)$. Equation~\eqref{eq:beta-electric-log-derivative} gives its first derivative, while the second follows from \eqref{eq:beta-casimir-derivatives}. At large $z_1$, they behave as
\begin{align}
    \frac{\partial\Lambda_{\rm el}^4}{\partial\beta}
    &=-\frac{bz_1}{2}\Lambda_{\rm el}^4
    \left[1+\mathcal O(z_1^{-1})\right],
    \nonumber\\
    \frac{\partial^2\Lambda_{\rm el}^4}{\partial\beta^2}
    &=\frac{b^2z_1^2}{4}\Lambda_{\rm el}^4
    \left[1+\mathcal O(z_1^{-1})\right].
    \label{eq:beta-backreaction-electric-derivatives}
\end{align}
Differentiation introduces powers of the winding action, but leaves the factor $e^{-z_1}$ intact. A sufficiently heavy charged field therefore gives an exponentially small radion force as well as an exponentially small axion potential. Keeping the leading powers of $z_1$, the charged contribution to the displacement and the direct second-derivative correction satisfy
\begin{align}
    |b\delta\beta_{\rm el}|
    &\lesssim
    \frac{z_1}{2}
    \frac{b^2\Lambda_{\rm el}^4}{V''(\beta_0)},
    \nonumber\\
    \frac{|\partial_\beta^2\Delta V_{\rm el}|}
    {V''(\beta_0)}
    &\lesssim
    \frac{z_1^2}{4}
    \frac{b^2\Lambda_{\rm el}^4}{V''(\beta_0)}.
    \label{eq:beta-backreaction-electric-bounds}
\end{align}
The quantities on the right-hand side are evaluated at $\beta_0$. For a generic relative phase, the quality bound \eqref{eq:beta-electric-quality} then gives
\begin{align}
    |b\delta\beta_{\rm el}|
    &\lesssim
    \frac{10^{-10}z_1}{2|q|}
    \frac{b^2\chi_{\rm QCD}}{V''(\beta_0)},
    \nonumber\\
    \frac{|\partial_\beta^2\Delta V_{\rm el}|}
    {V''(\beta_0)}
    &\lesssim
    \frac{10^{-10}z_1^2}{4|q|}
    \frac{b^2\chi_{\rm QCD}}{V''(\beta_0)}.
    \label{eq:beta-quality-implies-small-backreaction}
\end{align}
These expressions make the required comparison explicit. The charged contribution can satisfy axion quality and have a negligible effect on the radion whenever the displayed ratios are small. This depends on the strength of stabilisation as well as the charged amplitude. In particular, a minimum close to $V''(\beta_0)=0$ can be sensitive even to a small extra potential. An accidentally aligned phase also requires separate care as it can suppress the strong CP angle without suppressing $\Lambda_{\rm el}^4$ or its radion derivatives. The mixing term in \eqref{eq:beta-coupled-stability} can be checked. At the perturbed axion minimum, to leading order,
\begin{equation}
    \frac{(\partial_\beta\partial_aV)^2}
    {\partial_a^2V\,V''(\beta_0)}
    \simeq
    \frac{\chi_{\rm QCD}\theta_{\rm eff}^2}
    {V''(\beta_0)}
    \left[
        \frac{\partial}{\partial\beta}
        \log\left(
            \frac{\Lambda_{\rm el}^4}{\chi_{\rm QCD}}
        \right)
    \right]^2.
    \label{eq:beta-backreaction-mixing-bound}
\end{equation}
The strong CP bound supplies a factor $\theta_{\rm eff}^2\lesssim10^{-20}$ which is a substancial suppression. The mixing correction is therefore small provided the radion minimum and the radius dependence of $\chi_{\rm QCD}$ do not overcome this suppression. The same assumption is needed when neglecting derivatives of the higher-order term in \eqref{eq:beta-backreaction-electric-relaxed-potential}. The susceptibility must retain its $\beta$ dependence throughout this comparison.

Finally, the quality bound must hold at the shifted radion minimum. Evaluating it at $\beta_0$ is reliable if
\begin{equation}
    \left|
        \delta\beta\,
        \frac{\partial}{\partial\beta}
        \log\left(
            \frac{\Lambda_{\rm el}^4}{\chi_{\rm QCD}}
        \right)
    \right|_{\beta=\beta_0}\ll1,
    \label{eq:beta-backreaction-quality-change}
\end{equation}
where $\delta\beta$ includes all contributions to the radion displacement. This condition ensures that the ratio controlling the axion displacement changes little when the radion minimum moves.

\subsubsection{Magnetic gauging and the QCD vacuum energy}

The magnetic contribution requires a separate check because it has no winding exponential. Consider a fixed branch with $M\ne0$ and first assume $m_K^2\ll m_a^2$, as required for a generic branch misalignment. At the unperturbed QCD minimum, \eqref{eq:beta-magnetic-potential} becomes
\begin{equation}
    V_{K,n}(\beta,-f_a\bar\theta)
    =\frac12f_{\rm eff}^2(\beta)m_K^2(\beta)
    \left(\bar\theta+\frac{a_n}{f_a}\right)^2.
    \label{eq:beta-backreaction-magnetic-energy}
\end{equation}
Allowing the axion to relax changes this expression only at higher order in $m_K^2/m_a^2$. Since $f_{\rm eff}^2m_K^2\propto e^{-3b\beta/2}$, the leading radion corrections are
\begin{align}
    |b\delta\beta_K|
    &\simeq
    \frac{3b^2}{2V''(\beta_0)}
    V_{K,n}(\beta_0,-f_a\bar\theta),
    \nonumber\\
    \frac{|\partial_\beta^2V_{K,n}|}
    {V''(\beta_0)}
    &=\frac{9b^2}{4V''(\beta_0)}
    V_{K,n}(\beta_0,-f_a\bar\theta).
    \label{eq:beta-backreaction-magnetic-bounds}
\end{align}
The derivative in the second line is evaluated at the same point. For an order-one mismatch $\bar\theta+a_n/f_a$, the quality bound $m_K^2\lesssim10^{-10}m_a^2$ also suppresses the branch energy to $V_{K,n}\lesssim10^{-10}\chi_{\rm QCD}$. The resulting radion displacement and direct second-derivative correction are therefore negligible if $10^{-10}b^2\chi_{\rm QCD}\ll V''(\beta_0)$. The mixing correction follows \eqref{eq:beta-backreaction-mixing-bound}, with $\Lambda_{\rm el}^4$ replaced by $f_{\rm eff}^2m_K^2$, and is suppressed by $\theta_{\rm eff}^2\lesssim10^{-20}$ from the same quality bound. The numerator satisfies $\partial_\beta\log(f_{\rm eff}^2m_K^2)=-3b/2$, while the $\beta$ dependence of $\chi_{\rm QCD}$ must also be retained.

Unlike the charged contribution, this suppression must come from a sufficiently small magnetic coupling or a suitable branch alignment. For exact alignment, $\bar\theta+a_n/f_a=0$, the QCD and magnetic potentials have a common axion minimum. The magnetic branch energy and its radion derivatives then vanish at that minimum, even when $m_K^2$ is large. For $M=0$, however, a non-zero flux produces an axion-independent contribution to the radion potential, and axion quality places no bound on it.

The QCD vacuum energy $E_{\rm QCD}(\beta,0)$ is also independent of the strong CP bound. Although QCD alone keeps the axion at $\vartheta=0$, the value of its vacuum energy can change with the radius. Negligible displacement and a small correction to the radion second derivative therefore require
\begin{equation}
    \left|
        \frac{\partial E_{\rm QCD}(\beta,0)}{\partial\beta}
    \right|_{\beta=\beta_0}
    \ll\frac{V''(\beta_0)}{|b|},
    \qquad
    \left|
        \frac{\partial^2E_{\rm QCD}(\beta,0)}{\partial\beta^2}
    \right|_{\beta=\beta_0}
    \ll V''(\beta_0).
    \label{eq:beta-backreaction-qcd-vacuum}
\end{equation}
These conditions must be checked using the radius dependence of the running couplings and mass thresholds in the chosen model. They do not follow from the smallness of the residual CP angle.

Both stabilisation mechanisms can therefore be compatible with axion quality under the conditions above. The neutral masses determine the Casimir contribution supporting the minimum, while a larger charged mass can suppress the unwanted axion potential at that same radius. The appropriate $V''(\beta_0)$ is supplied by the preceding stabilisation analysis, in the scalar-winding construction, the parameters must remain sufficiently far from the boundary where the radion mass vanishes.

Preserving the desired vacuum energy is a separate requirement. For a tuned Minkowski or small positive vacuum energy, its value must be checked again after including the QCD and additional axion contributions, even when the radion displacement is negligible. The analysis here establishes local compatibility of the stationary axion and radion minima. Cosmological evolution towards that vacuum requires the coupled equations \eqref{eq:beta-axion-background-equations}.

\section{Cosmological and Phenomenological Implications}
\label{sec:Cosmological and Phenomenological Implications}
\subsection{Axion Quality vs Radion}
\label{sec:axion-quality-versus-radion}

In section~\ref{sec:beta-quality-compatibility}, we examined the stability of the coupled axion--radion system and derived the conditions for negligible radion displacement. We now apply these conditions to the minima constructed in section~\ref{sec:beta-stabilisation} and obtain explicit numerical bounds from axion quality and radion stability. The neutral and charged masses are independent inputs, allowing a heavier charged scalar to suppress the Wilson-line potential while the neutral fields support stabilisation. Magnetic gauging requires a further check as introducing $K_3$ changes the bulk spectrum and hence the Casimir potential supporting the minimum. We include this change before evaluating the quality and stability bounds.

\subsubsection{Radion masses}

The radion mass at each minimum is given by \eqref{eq:beta-ds-conditions},
\begin{equation}
    m_\beta^2=\frac{8V''(\beta_0)}{3b^2M_{\rm Pl}^2}.
    \label{eq:quality-radion-mass-scale}
\end{equation}
Keeping the dimensionless parameters of the potential fixed gives $V''(\beta_0)/b^2\propto R^{-4}$, and hence $m_\beta\propto R^{-2}$. For the examples in figures~\ref{fig:withoutwindingstabilisedbeta} and~\ref{fig:betawithwindingpotential}, we choose
\begin{equation}
    R^{-1}=2.55\times10^{12}\,\mathrm{GeV},
    \qquad
    M_{\rm Pl}=2.4\times10^{18}\,\mathrm{GeV}.
    \label{eq:quality-radion-physical-scale}
\end{equation}
The resulting masses are given in table~\ref{tab:quality-radion-masses}. In particular, the Minkowski minima at $\beta_0=0$ have $m_\beta\simeq169.6\,\mathrm{TeV}$ without scalar winding and $m_\beta\simeq120.9\,\mathrm{TeV}$ with scalar winding.

\begin{table}[htbp]
    \centering
    \begin{tabular}{lrrr}
        \hline
        Minimum & $b\beta_0$ & $V''(\beta_0)/(b^2V_*)$ & $m_\beta\,[\mathrm{TeV}]$ \\
        \hline
        Without winding: Minkowski & $0$ & $30.1318$ & $169.6$ \\
        Without winding: dS & $0.00923$ & $28.6260$ & $165.3$ \\
        With winding: Minkowski & $0$ & $15.3078$ & $120.9$ \\
        With winding: dS & $0$ & $15.2328$ & $120.6$ \\
        \hline
    \end{tabular}
    \caption{Radion masses at the minima of figures~\ref{fig:withoutwindingstabilisedbeta} and~\ref{fig:betawithwindingpotential}, with \(R^{-1}=2.55\times10^{12}\,\mathrm{GeV}\) and \(V_*=3/(64\pi^6R^4)\). The masses are calculated from the stabilising potential \(V(\beta)\).}
    \label{tab:quality-radion-masses}
\end{table}

\begin{figure}[tbp]
    \centering
    \includegraphics[width=0.8\linewidth]{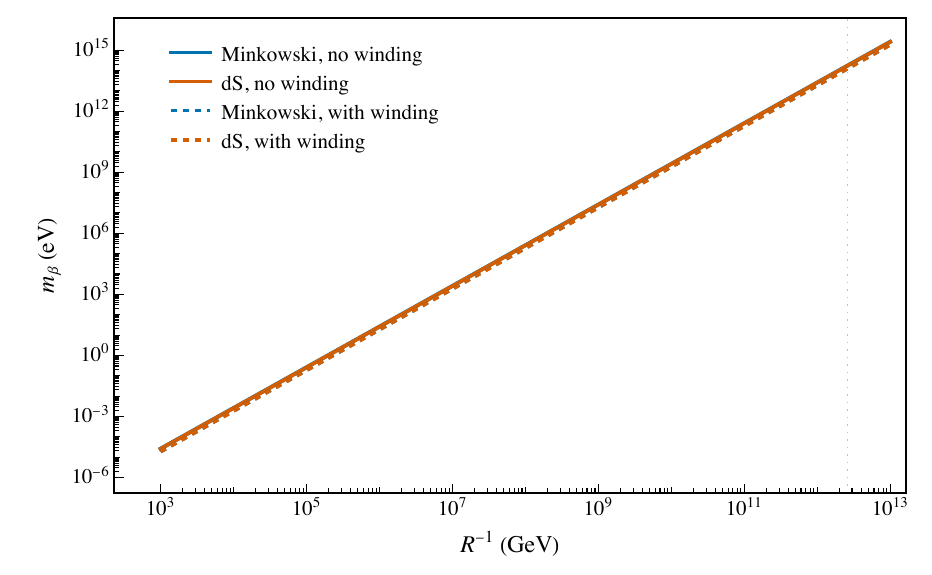}
    \caption{Radion masses at the Minkowski (blue) and dS (orange) minima, without scalar winding (solid) and with scalar winding (dashed). The dimensionless parameters of each potential remain fixed along the curves, giving \(m_\beta\propto R^{-2}\). The vertical line marks the chosen scale \(R^{-1}=2.55\times10^{12}\,\mathrm{GeV}\).}
    \label{fig:quality-radion-mass-radius}
\end{figure}

The dependence on $R^{-1}$ is shown in figure~\ref{fig:quality-radion-mass-radius}. Along each curve, the products $m_iR$ and the dimensionless coefficients of the potential remain fixed. Thus changing $R$ along these curves also changes the proper masses of the stabilising fields. This differs from varying $\beta$ within a given theory, where $R$, the proper masses and the five-dimensional couplings are held fixed.

For the scale in \eqref{eq:quality-radion-physical-scale}, the dS minima have $H\simeq8.04\,\mathrm{TeV}$ and $H\simeq3.45\,\mathrm{TeV}$, respectively. The ordinary zero-temperature QCD potential in flat spacetime is not a controlled approximation at these values of $H$.\footnote{The flat-spacetime approximation requires $H$ to be well below the relevant hadronic scales, conventionally expressed as $H\ll\Lambda_{\rm QCD}$ \cite{TakahashiYinGuth2018}. The de Sitter horizon has the Gibbons--Hawking temperature $T_{\rm dS}=H/(2\pi)$ \cite{GibbonsHawking1977}, giving approximately $1.28\,\mathrm{TeV}$ and $0.55\,\mathrm{TeV}$ for these examples, both far above the QCD scale.}. In the following, we therefore evaluate axion quality at the Minkowski minima. The same analysis applies to sufficiently small uplifts, provided the QCD description remains valid.

\subsubsection{Effect on $\beta$ stabilisation after magnetic gauging}

Gauging the magnetic symmetry introduces a dynamical $K_3$, so its effect on stabilisation is not limited to the branch potential $V_{K,n}$. The coupled $C$ and $K_3$ system has four physical polarisations, corresponding to a massive five-dimensional vector with proper mass $|M|g_5e_{K,5}/(2\pi)$. Here $M\in\mathbb Z$ is the coefficient in \eqref{eq:magnetic-gauging-five}. The original massless $C$ field has three polarisations. For the small coupling considered below, the vector mass is negligible compared with $R^{-1}$, and magnetic gauging therefore adds one effectively massless bosonic degree of freedom to the Casimir energy.

The bosonic count becomes $N_b=33$ without scalar winding and $N_b=34$ with scalar winding. The scalar $\sigma$ remains independent of the scalar dual to $K_3$. We retain $N_f=48$ in both constructions and impose the Minkowski conditions again with these modified spectra.

Without scalar winding, \eqref{eq:fluxless-radius-equation} gives $2\pi mR\simeq1.96401$, after which \eqref{eq:fluxless-minkowski-conditions} fixes $\Lambda_5$. With scalar winding, we keep $2\pi mR=2.4$ and adjust $V_\sigma(0)$ and $\Lambda_5$ so that $V(0)=V'(0)=0$. The radion masses are then
\begin{equation}
    m_\beta\simeq
    \begin{cases}
        165.9\,\mathrm{TeV}, & \text{without scalar winding},\\
        104.4\,\mathrm{TeV}, & \text{with scalar winding}.
    \end{cases}
    \label{eq:quality-magnetic-retuned-masses}
\end{equation}
Both second derivatives remain positive. Thus the additional bosonic contribution changes the parameters needed for stabilisation but does not remove the minima.

\begin{figure}[tbp]
    \centering
    \includegraphics[width=\linewidth]{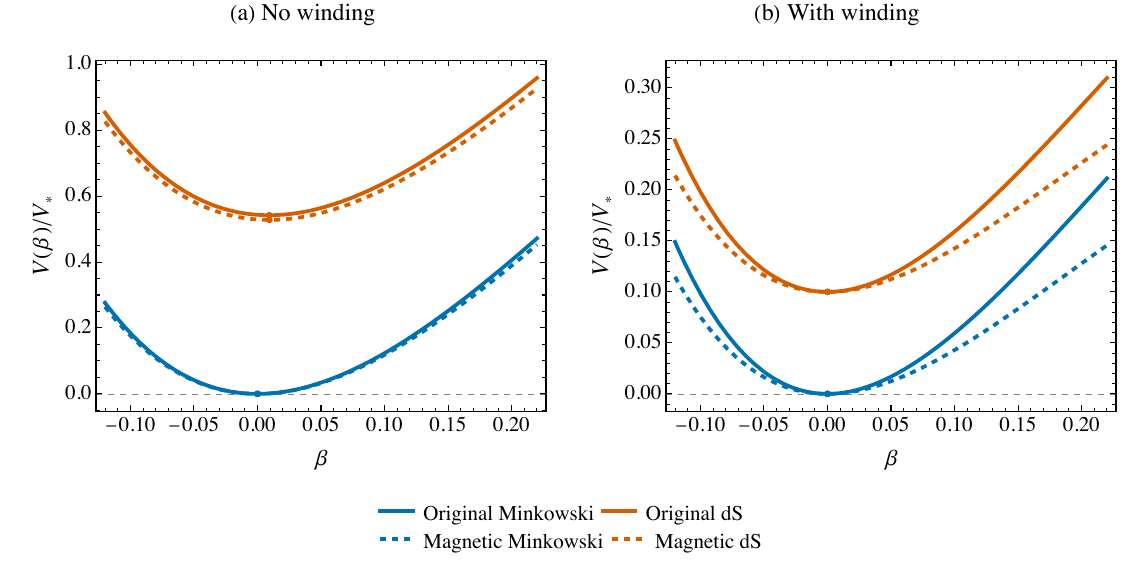}
    \caption{Radion potentials without scalar winding in panel (a) and with scalar winding in panel (b), for \(b=1\). Solid curves use the original spectra. Dashed curves include the additional bosonic degree of freedom from the \(C\)--\(K_3\) system, with the stabilising parameters adjusted as described in the text. Blue and orange curves have Minkowski and dS minima, respectively and dots mark the stationary points. The plotted quantity is \(V(\beta)/V_*\), before adding the magnetic branch potential and the QCD and charged-scalar contributions.}
    \label{fig:quality-magnetic-potentials}
\end{figure}

Figure~\ref{fig:quality-magnetic-potentials} shows how the additional bosonic degree of freedom changes the stabilising potentials. For each dashed curve, we impose the vacuum conditions again using the modified spectrum. Both constructions retain Minkowski minima and nearby dS minima. These potentials include the change in the Casimir energy with the additional contribution from the magnetic branch potential is considered below.

\subsubsection{Axion quality at the minimum}

\begin{figure}[tbp]
    \centering
    \includegraphics[width=\linewidth]{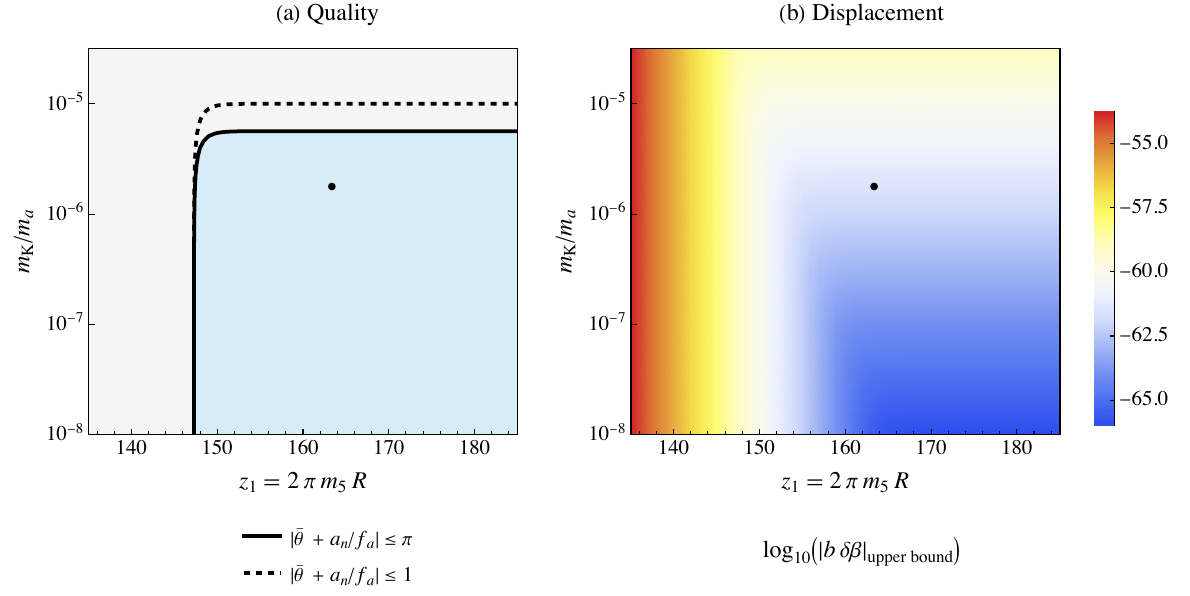}
    \caption{Axion quality and radion displacement for \(q=M=1\). In panel (a), the shaded region satisfies \(|\theta_{\rm eff}|\leq10^{-10}\) for any electric phase and branch misalignment \(|\bar\theta+a_n/f_a|\leq\pi\). The dashed boundary restricts the branch misalignment to one. Panel (b) shows the logarithm of the upper bound on \(|b\delta\beta|\) from the electric and magnetic potentials, taking the larger bound from the two modified Minkowski minima and allowing branch misalignment up to \(\pi\). The combined first and second radion derivatives of the independent QCD and brane vacuum terms are set to zero. Black dots mark the parameters in \eqref{eq:quality-electric-benchmark} and \eqref{eq:quality-magnetic-benchmark}.}
    \label{fig:quality-joint}
\end{figure}

We take the axion parameters at $\beta=0$ to be \cite{GrillidiCortona:2015jxo}
\begin{equation}
    f_{\rm eff}(0)=f_a=10^{14}\,\mathrm{GeV},
    \qquad
    \chi_{\rm QCD}(0)\simeq(75.5\,\mathrm{MeV})^4.
    \label{eq:quality-low-energy-inputs}
\end{equation}
The QCD axion mass is then $m_a\simeq5.70\times10^{-8}\,\mathrm{eV}$. To determine whether the additional potentials spoil the PQ minimum, we compare their slopes with the QCD contribution as in \eqref{eq:beta-general-quality}. Consider first the charged complex scalar, with magnetic gauging absent. At $\beta=0$, the first-winding amplitude gives
\begin{equation}
    \frac{\Lambda_{\rm el}^4}{\chi_{\rm QCD}}
    =\frac{3e^{-z_1}}{32\pi^6R^4\chi_{\rm QCD}}
    \left(1+z_1+\frac{z_1^2}{3}\right),
    \qquad
    z_1=2\pi m_5R.
    \label{eq:quality-electric-numerical-condition}
\end{equation}
For $q=1$ and a generic phase, the strong CP bound requires $z_1\gtrsim147.3$. Choosing
\begin{equation}
    q=1,
    \qquad
    m_5R=26,
    \qquad
    z_1=52\pi\simeq163.36
    \label{eq:quality-electric-benchmark}
\end{equation}
gives $m_5=6.63\times10^{13}\,\mathrm{GeV}$. This mass must lie below the five-dimensional cutoff. At the chosen radius,
\begin{equation}
    \frac{\Lambda_{\rm el}^4}{\chi_{\rm QCD}}
    \simeq1.30\times10^{-17},
    \qquad
    |\theta_{\rm eff}|\lesssim1.30\times10^{-17}.
    \label{eq:quality-electric-benchmark-results}
\end{equation}
Higher windings are suprressed. The charged scalar therefore leaves the axion sufficiently close to the QCD minimum for any value of its relative phase. When magnetic gauging is present, the magnetic branch can also displace the axion. In the regime where both corrections are small, a sufficient bound on their combined effect is
\begin{equation}
    \frac{|q|\Lambda_{\rm el}^4}{\chi_{\rm QCD}}
    +\frac{m_K^2}{m_a^2}
    \left|\bar\theta+\frac{a_n}{f_a}\right|
    \lesssim10^{-10}.
    \label{eq:quality-electric-magnetic-budget}
\end{equation}
Taking the absolute values separately avoids relying on a cancellation between the two contributions. Using \eqref{eq:beta-magnetic-mass},
\begin{equation}
    \frac{m_K^2}{m_a^2}
    =\frac{M^2e_{K,5}^2}{8\pi^3R\chi_{\rm QCD}(\beta)}
    e^{-3b\beta/2}.
    \label{eq:quality-magnetic-ratio}
\end{equation}
For $|\bar\theta+a_n/f_a|=1$ at $\beta=0$, and with the charged contribution already suppressed, this requires
\begin{equation}
    e_{K,5}^2\lesssim
    \frac{3.16\times10^{-25}}{M^2}\,\mathrm{GeV}^3.
    \label{eq:quality-magnetic-parameter-bound}
\end{equation}
Thus magnetic gauging is allowed if its coupling is sufficiently small \cite{CraigKongsore2024}. For the modified spectra of the preceding subsection, keeping the charged parameters in \eqref{eq:quality-electric-benchmark} and taking
\begin{equation}
    M=1,
    \qquad
    n=0,
    \qquad
    \bar\theta=1,
    \qquad
    e_{K,5}^2=10^{-26}\,\mathrm{GeV}^3.
    \label{eq:quality-magnetic-benchmark}
\end{equation}
We then obtain
\begin{equation}
    m_K\simeq1.01\times10^{-13}\,\mathrm{eV},
    \qquad
    \frac{m_K^2}{m_a^2}\simeq3.16\times10^{-12},
    \qquad
    |\theta_{\rm eff}|\simeq3.16\times10^{-12}.
    \label{eq:quality-magnetic-benchmark-results}
\end{equation}
For the chosen parameters, the magnetic contribution dominates the residual angle, which remains below the strong CP bound. Figure~\ref{fig:quality-joint}(a) shows the allowed region in the $(z_1,m_K/m_a)$ plane. The shaded region satisfies \eqref{eq:quality-electric-magnetic-budget} for all branch misalignments with $|\bar\theta+a_n/f_a|\leq\pi$, while the dashed boundary corresponds to the restricted range $|\bar\theta+a_n/f_a|\leq1$. As $z_1$ increases, the electric contribution becomes negligible and the solid boundary approaches $m_K/m_a=\sqrt{10^{-10}/\pi}\simeq5.64\times10^{-6}$. The benchmark lies below both boundaries and gives $|\theta_{\rm eff}|\lesssim9.94\times10^{-12}$ even when the branch misalignment is allowed to reach $\pi$.

\subsubsection{Radius dependence and stability}

The bounds are evaluated at $\beta=0$. Since the additional potentials also act on the radion, we must check how these bounds change if the radius moves. This requires the $\beta$ dependence of QCD as well as that of the electric and magnetic contributions. Equation~\eqref{eq:beta-colour-coupling} gives the tree-level colour coupling at the KK scale $m_{\rm KK,E}=R^{-1}e^{-3b\beta/4}$. To obtain the low-energy QCD scale, we run this coupling through the quark thresholds. We take the chiral quarks and the Higgs to be localised on a four-dimensional brane, while colour propagates in the bulk. The quarks couple to the gluon zero mode, whose axion coupling is supplied by the Chern--Simons term. Fixed proper quark masses then give Einstein-frame masses proportional to $e^{-b\beta/4}$.\footnote{Periodic bulk fermions on a smooth circle give a vector-like zero-mode spectrum. The localised matter sector supplies the chiral quarks used here and also contributes to the brane vacuum energy discussed below.}

\begin{figure}[tbp]
    \centering
    \includegraphics[width=\linewidth]{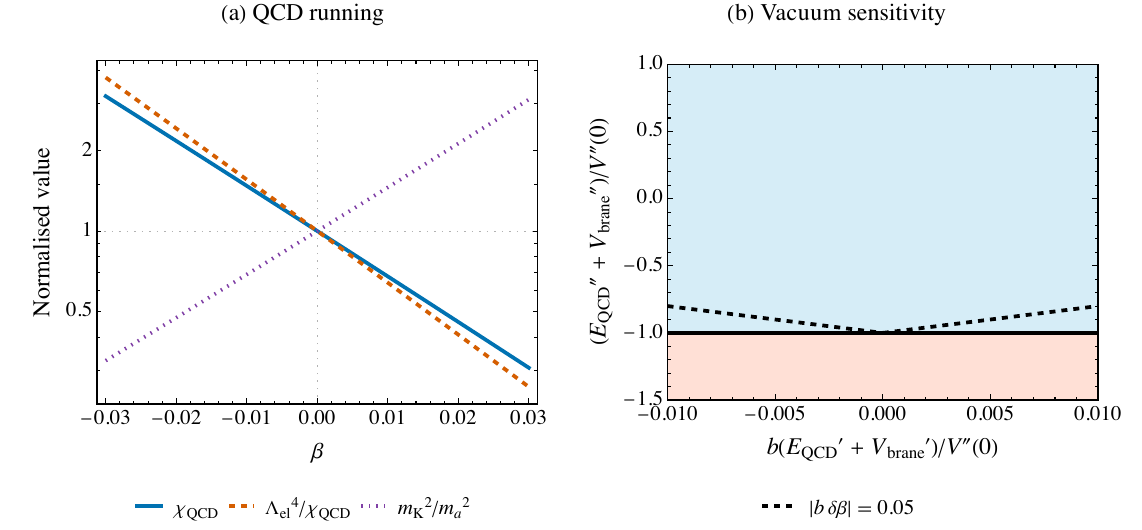}
    \caption{Radius dependence and sensitivity to the independent vacuum terms, for \(b=1\). Panel (a) shows \(\chi_{\rm QCD}\), \(\Lambda_{\rm el}^4/\chi_{\rm QCD}\) and \(m_K^2/m_a^2\), each divided by its value at \(\beta=0\), with the proper masses and five-dimensional couplings held fixed. In panel (b), \(V_{\rm brane}=T_{\rm br}e^{-b\beta}\), and primes denote derivatives at \(\beta=0\). Blue and red indicate positive and negative values of the radial stability expression in \eqref{eq:beta-coupled-stability}, evaluated in the local quadratic approximation. Dashed curves mark \(|b\delta\beta|=0.05\) on the positive side.}
    \label{fig:quality-qcd-vacuum}
\end{figure}

At a fixed radion background, the one-loop running of the strong coupling with $n_f$ active quark flavours gives
\begin{equation}
    \frac{8\pi^2}{g_{s,n_f}^2(\mu)}
    =B_{n_f}\log\left(\frac{\mu}{\Lambda_{n_f}}\right),
    \qquad
    B_{n_f}=11-\frac{2n_f}{3},
    \label{eq:quality-qcd-one-loop-running}
\end{equation}
where $\Lambda_{n_f}$ is the corresponding QCD scale. At a heavy-quark threshold $\mu=m_h$, we match the couplings above and below the threshold. Since $B_{n_f-1}-B_{n_f}=2/3$, continuity of the coupling at this order gives
\begin{equation}
    \Lambda_{n_f-1}^{B_{n_f-1}}
    =m_h^{2/3}\Lambda_{n_f}^{B_{n_f}}.
    \label{eq:quality-qcd-threshold-matching}
\end{equation}
Applying this relation successively at the top, bottom and charm thresholds, with $B_6=7$ and $B_3=9$, yields
\begin{equation}
    \Lambda_3^9
    =\Lambda_6^7(m_tm_bm_c)^{2/3}
    =m_{\rm KK,E}^{\,7}(m_tm_bm_c)^{2/3}
    \exp\left[-\frac{8\pi^2}{g_s^2(\beta)}\right].
    \label{eq:quality-qcd-three-flavour-scale}
\end{equation}
Here all masses and scales are expressed in the Einstein frame, and $g_s(\beta)$ is the six-flavour coupling evaluated at the matching scale $m_{\rm KK,E}(\beta)$. At leading order in the light-quark masses, the susceptibility is $\chi_{\rm QCD}\simeq\bar m\Sigma$, where $\bar m=(m_u^{-1}+m_d^{-1}+m_s^{-1})^{-1}$ and $\Sigma$ is the magnitude of the chiral condensate. Keeping this leading mass dependence and taking $\Sigma\propto\Lambda_3^3$ gives
\begin{equation}
    \chi_{\rm QCD}(\beta)
    \propto
    \bar m\,m_{\rm KK,E}^{\,7/3}(m_tm_bm_c)^{2/9}
    \exp\left[-\frac{8\pi^2}{3g_s^2(\beta)}\right].
    \label{eq:quality-qcd-susceptibility-scaling}
\end{equation}
The radius dependence now follows from $m_{\rm KK,E}=R^{-1}e^{-3b\beta/4}$ and $m_q(\beta)=m_q(0)e^{-b\beta/4}$. Since all light-quark masses acquire the same factor, their ratios remain fixed and $\bar m(\beta)=\bar m(0)e^{-b\beta/4}$. The KK matching scale therefore contributes $e^{-7b\beta/4}$, the heavy-quark thresholds contribute $e^{-b\beta/6}$, and the light-quark factor contributes $e^{-b\beta/4}$. Their product is $e^{-13b\beta/6}$. Using \eqref{eq:beta-colour-coupling} for the remaining coupling dependence and normalising at $\beta=0$, we obtain
\begin{equation}
    \frac{\chi_{\rm QCD}(\beta)}{\chi_{\rm QCD}(0)}
    \simeq
    \exp\left[
        -\frac{13b\beta}{6}
        -\frac{8\pi^2}{3g_s^2(0)}
        \left(e^{b\beta/2}-1\right)
    \right].
    \label{eq:quality-qcd-running}
\end{equation}
Here $g_s(0)$ is evaluated at the reference KK scale $R^{-1}$. We use $g_s^2=4\pi(0.0292)$ at $10^{12}\,\mathrm{GeV}$, which gives $g_s^2(0)\simeq0.356$ at the scale in \eqref{eq:quality-radion-physical-scale}. The normalisation $\chi_{\rm QCD}(0)$ is fixed by \eqref{eq:quality-low-energy-inputs}. Combining this result with \eqref{eq:beta-electric-log-derivative} and \eqref{eq:quality-magnetic-ratio} gives
\begin{align}
    \left.\frac1b\frac{\partial}{\partial\beta}
    \log\chi_{\rm QCD}\right|_{\beta=0}
    &\simeq-39.12,
    \nonumber\\
    \left.\frac1b\frac{\partial}{\partial\beta}
    \log\left(\frac{\Lambda_{\rm el}^4}{\chi_{\rm QCD}}\right)\right|_{\beta=0}
    &\simeq-44.57,
    \nonumber\\
    \left.\frac1b\frac{\partial}{\partial\beta}
    \log\left(\frac{m_K^2}{m_a^2}\right)\right|_{\beta=0}
    &\simeq37.62.
    \label{eq:quality-magnetic-radius-condition}
\end{align}
Thus increasing the proper radius suppresses the electric contribution relative to QCD, but increases the relative importance of magnetic breaking. The magnetic potential itself decreases but QCD decreases faster. These dependences are shown in figure~\ref{fig:quality-qcd-vacuum}(a), where each quantity is normalised to its value at $\beta=0$. The same potentials also exert a force on the radion. Using the derivatives of the full first-winding amplitude, including its polynomial prefactor, we find
\begin{equation}
    |b\delta\beta_{\rm el}|\lesssim1.5\times10^{-66},
    \qquad
    \frac{|\partial_\beta^2\Delta V_{\rm el}|}{V''(0)}
    \lesssim1.3\times10^{-64}.
    \label{eq:quality-electric-radion-corrections}
\end{equation}
These bounds use the largest correction among the four Minkowski minima, with and without magnetic gauging. For the magnetic parameters in \eqref{eq:quality-magnetic-benchmark}, \eqref{eq:beta-backreaction-magnetic-bounds} gives
\begin{equation}
    |b\delta\beta_K|\lesssim3.3\times10^{-63},
    \qquad
    \frac{|\partial_\beta^2V_{K,n}|}{V''(0)}
    \lesssim5.0\times10^{-63}.
    \label{eq:quality-magnetic-radion-corrections}
\end{equation}
Here we take the larger correction of the two modified minima. The mixing term in \eqref{eq:beta-coupled-stability}, divided by $V''(0)$, is below $10^{-70}$ for these parameters. The electric and magnetic potentials therefore give negligible corrections to both the position of the radion minimum and its second derivative.

Figure~\ref{fig:quality-joint}(b) shows the displacement bound over the same range of parameters as the quality plot. Allowing branch misalignment up to $\pi$ gives $|b\delta\beta|\lesssim3.3\times10^{-62}$ at the marked point. The corresponding fractional change in the proper radius is $b\delta\beta/2$. With such a small displacement, the logarithmic derivatives in \eqref{eq:quality-magnetic-radius-condition} also imply a negligible change in the quality ratios.

These shifts come from the electric and magnetic axion potentials. As discussed in section~\ref{sec:beta-quality-compatibility}, the QCD vacuum energy $E_{\rm QCD}(\beta,0)$ can exert an additional force even when the axion sits at its CP-preserving minimum. An important point to note is that standard model must be supported on branes in this construction and a localised matter sector also contributes a renormalised brane term $T_{\rm br}e^{-b\beta}$.\footnote{The quarks and Higgs are localised at a fixed position on the circle, so their vacuum energy is also localised on this four-dimensional hypersurface. Its induced metric follows from \eqref{eq:beta-einstein-metric} as $\gamma_{\mu\nu}=e^{-b\beta/2}g^E_{\mu\nu}$, giving $\sqrt{-\gamma}=e^{-b\beta}\sqrt{-g_E}$ which is also discussed in section \ref{sec:localised matter sector}. The local vacuum term $-T_{\rm br}\int\mathrm{d}^4x\sqrt{-\gamma}$ therefore contributes $T_{\rm br}e^{-b\beta}$ to the Einstein-frame potential. Quantum corrections from the localised matter renormalise $T_{\rm br}$, whose finite value is an independent parameter \cite{PontonPoppitz2001}.} Its radius dependence differs from that of the bulk cosmological constant, so it cannot be absorbed into $\Lambda_5$.

The numerical comparison above sets the combined first and second radion derivatives of these independent vacuum contributions to zero. For non-zero values, the small-correction conditions become
\begin{align}
    \left|
        \frac{b\left[\partial_\beta E_{\rm QCD}(0,0)-bT_{\rm br}\right]}
        {V''(0)}
    \right|&\ll1,
    \nonumber\\
    \left|
        \frac{\partial_\beta^2E_{\rm QCD}(0,0)+b^2T_{\rm br}}
        {V''(0)}
    \right|&\ll1.
    \label{eq:quality-independent-vacuum-conditions}
\end{align}
The derivatives act on the first argument of $E_{\rm QCD}$. Once these contributions are included, the displacement must remain within the expansion used in \eqref{eq:beta-backreaction-displacement}, and the stability and quality conditions must be evaluated at the shifted minimum.

Figure~\ref{fig:quality-qcd-vacuum}(b) illustrates the effect of the independent vacuum terms. The horizontal axis measures their additional slope and the vertical axis their additional second derivative, normalised to the original stabilising potential. Blue and red indicate positive and negative values of the radial stability expression in the local quadratic approximation. At a vertical value of $-1$, the quadratic term vanishes and this approximation no longer determines the displacement. The dashed curves mark $|b\delta\beta|=0.05$ within the blue region, with smaller displacements above the curves. This value is an illustrative measure of the displacement, rather than an observational bound. A nearby minimum requires both a positive stability expression and a displacement small enough for the expansion to remain valid.

The neutral fields can therefore fix the radius while the heavier charged scalar leaves the QCD minimum almost unchanged. For magnetic gauging, the corresponding suppression comes from the small coupling in \eqref{eq:quality-magnetic-benchmark}, after including the additional bosonic contribution to the Casimir energy. Provided the remaining vacuum terms satisfy \eqref{eq:quality-independent-vacuum-conditions} and the expansion remains valid, both constructions retain a local minimum with $|\theta_{\rm eff}|<10^{-10}$ and a negligible displacement of the radius.

\subsection{The graviphoton and its couplings}
\label{sec:graviphoton}

The metric \eqref{eq:beta-einstein-metric} retains the four-dimensional graviton and the radion, but sets the off-diagonal components $g_{\mu y}$ to zero. These components contain a four-dimensional vector, the Kaluza--Klein graviphoton. Although its background value can vanish in the stabilisation analysis, its fluctuations must be included when determining the radion decay channels. We restore this field by writing
\begin{equation}
    \mathrm{d}s_5^2
    =e^{-b\beta/2}g^E_{\mu\nu}\mathrm{d}x^\mu\mathrm{d}x^\nu
    -e^{b\beta}\left(\mathrm{d}y+\frac{\sqrt2}{M_{\rm Pl}}\gamma_\mu\mathrm{d}x^\mu\right)^2,
    \qquad y\sim y+2\pi R.
    \label{eq:graviphoton-metric}
\end{equation}
The normalisation is chosen so that $\gamma_\mu$ has a canonical kinetic term at $\beta=0$. A coordinate transformation along the circle acts as
\begin{equation}
    y\to y-\frac{\sqrt2}{M_{\rm Pl}}\lambda(x),
    \qquad
    \gamma_\mu\to\gamma_\mu+\partial_\mu\lambda.
    \label{eq:graviphoton-gauge-transformation}
\end{equation}
The combination in parentheses in \eqref{eq:graviphoton-metric} is invariant. Thus the gauge symmetry of the graviphoton descends from five-dimensional diffeomorphism invariance. Reducing the Einstein--Hilbert term gives
\begin{equation}
    \mathcal L_\gamma
    =-\frac14e^{3b\beta/2}F^{(\gamma)}_{\mu\nu}F^{(\gamma)\mu\nu},
    \qquad
    F^{(\gamma)}_{\mu\nu}=\partial_\mu\gamma_\nu-\partial_\nu\gamma_\mu.
    \label{eq:graviphoton-kinetic}
\end{equation}
All four-dimensional contractions below use $g^E_{\mu\nu}$. Before a mass is generated, the five polarisations of the five-dimensional graviton become the two graviton polarisations, two graviphoton polarisations and the radion. The graviphoton is thus already included in the gravitational contribution to the Casimir spectrum.

\subsubsection{The axion contribution}

We consider the vector sector without magnetic gauging, as used in the reheating calculation below. Keeping the graviphoton, the zero modes of $C$ can be written as
\begin{equation}
    C=c_\mu\mathrm{d}x^\mu
    +\frac{a}{2\pi Rf_a}
    \left(\mathrm{d}y+\frac{\sqrt2}{M_{\rm Pl}}\gamma_\mu\mathrm{d}x^\mu\right).
    \label{eq:graviphoton-c-decomposition}
\end{equation}
Here $c_\mu$ is the four-dimensional vector descending from $C$, while the integral around the circle remains $\oint C=a/f_a$. Both $c_\mu$ and $a$ are invariant under \eqref{eq:graviphoton-gauge-transformation}. Substituting \eqref{eq:graviphoton-c-decomposition} into the five-dimensional gauge kinetic term gives
\begin{equation}
    \mathcal L_C
    =-\frac{2\pi R}{4g_5^2}e^{b\beta/2}
    \left(F^{(c)}_{\mu\nu}
    +\frac{\sqrt2a}{2\pi Rf_aM_{\rm Pl}}F^{(\gamma)}_{\mu\nu}\right)^2
    +\frac12e^{-b\beta}\partial_\mu a\,\partial^\mu a,
    \label{eq:graviphoton-c-kinetic}
\end{equation}
where $F^{(c)}_{\mu\nu}=\partial_\mu c_\nu-\partial_\nu c_\mu$. At a constant axion background, the vector kinetic mixing is removed by shifting $c_\mu$ by $\sqrt2\langle a\rangle\gamma_\mu/(2\pi Rf_aM_{\rm Pl})$. The axion kinetic term still contains an ordinary derivative, rather than a covariant derivative involving $\gamma_\mu$. Hence the graviphoton does not eat the Wilson-line axion, and this mixing does not by itself generate an axion potential. We now look into how the graviphoton can get mass due to the presence of other scalars via the St\"uckelberg mechanism.  An important point to note is that scalars can provide a St\"uckelberg mode only if it shifts under the vector's gauge transformation unlike our QCD axion as we see below. 

\subsubsection{The localised matter sector}
\label{sec:localised matter sector}
The brane supporting the localised matter introduces a position field $Y(x)$ along the circle. We take its leading action to be the Nambu--Goto action with positive renormalised tension $T_{\rm br}$ \cite{DobadoMaroto2001}. Pulling back \eqref{eq:graviphoton-metric} to the brane gives the induced metric
\begin{equation}
    g^{\rm ind}_{\mu\nu}
    =e^{-b\beta/2}g^E_{\mu\nu}
    -e^{b\beta}
    \left(\partial_\mu Y+\frac{\sqrt2}{M_{\rm Pl}}\gamma_\mu\right)
    \left(\partial_\nu Y+\frac{\sqrt2}{M_{\rm Pl}}\gamma_\nu\right).
    \label{eq:graviphoton-induced-metric}
\end{equation}
Under \eqref{eq:graviphoton-gauge-transformation}, the position transforms as $Y\to Y-\sqrt2\lambda/M_{\rm Pl}$, so the combination $\partial_\mu Y+\sqrt2\gamma_\mu/M_{\rm Pl}$ is gauge invariant. Expanding the brane action to quadratic order in this combination gives
\begin{align}
    S_{\rm br}
    &=-T_{\rm br}\int\mathrm{d}^4x\sqrt{-g_{\rm ind}}
    \nonumber\\
    &=\int\mathrm{d}^4x\sqrt{-g_E}
    \left[
        -T_{\rm br}e^{-b\beta}
        +\frac{T_{\rm br}}2e^{b\beta/2}
        \left(\partial_\mu Y+\frac{\sqrt2}{M_{\rm Pl}}\gamma_\mu\right)^2
        +\cdots
    \right].
    \label{eq:graviphoton-brane-action}
\end{align}
The first term is precisely the brane vacuum contribution discussed in the preceding subsection. The same renormalised tension also fixes the kinetic term of the position field. We work at leading order in the effect of the brane on the compactification. The localised matter sector is taken to live on a brane with positive renormalised tension $T_{\rm br}/V_*=10^{-22}$.\footnote{At the reference radius, this corresponds to $T_{\rm br}^{1/4}\simeq674\,\mathrm{TeV}$. The tension is an independent input specifying the localised sector. Its value is not determined by radion stabilisation i.e the neutral parameters are subsequently adjusted to retain the Minkowski minimum.}

Choosing the gauge $Y=0$ transfers the position fluctuation into the longitudinal polarisation of the graviphoton. This is the St\"uckelberg mechanism. Dividing the resulting mass term by the kinetic coefficient in \eqref{eq:graviphoton-kinetic}, the physical mass at a fixed radion background is
\begin{equation}
    m_\gamma^2(\beta_0)
    =\frac{2T_{\rm br}}{M_{\rm Pl}^2}e^{-b\beta_0}.
    \label{eq:graviphoton-brane-mass}
\end{equation}
Thus a positive brane tension gives a positive graviphoton mass. The scalar which is eaten is $Y$, while the Wilson-line axion remains in the spectrum.

The two factors $e^{3b\beta/2}$ and $e^{b\beta/2}$ in \eqref{eq:graviphoton-kinetic} and \eqref{eq:graviphoton-brane-action} determine the radion couplings to the transverse and longitudinal polarisations. At $\beta=0$, the canonically normalised brane-position fluctuation is $\sqrt{T_{\rm br}}Y$. Its derivative coupling to the radion therefore remains finite when the graviphoton mass is small compared with $m_\beta$. Equivalently, at energies much larger than $m_\gamma$, production of longitudinal graviphotons is described by production of the eaten position mode. This explains why both polarisations contribute to the decay width in \eqref{eq:reheating-hidden-widths}, even for a very light graviphoton.

\subsubsection{The effect of scalar winding}

Scalar winding provides another contribution to the graviphoton mass. Restoring the four-dimensional fluctuation of the scalar in \eqref{eq:beta-winding-background}, we write $\sigma=n_\sigma y/R+s(x)$. Reduction of \eqref{eq:beta-winding-action} then gives
\begin{equation}
    \mathcal L_\sigma
    =\frac{2\pi Rf_{\sigma,5}^3}{2}
    \left(\partial_\mu s-\frac{\sqrt2n_\sigma}{RM_{\rm Pl}}\gamma_\mu\right)^2
    -V_\sigma(\beta).
    \label{eq:graviphoton-winding-action}
\end{equation}
The fluctuation transforms as $s\to s+\sqrt2n_\sigma\lambda/(RM_{\rm Pl})$, making the derivative combination gauge invariant. Its four-dimensional kinetic coefficient is independent of $\beta$, whereas the vector kinetic coefficient is not. After normalising the vector, the winding contribution to its mass is $4V_\sigma(\beta_0)/M_{\rm Pl}^2$. If both the brane and scalar winding are present, the total mass is therefore
\begin{equation}
    m_\gamma^2(\beta_0)
    =\frac{4V_\sigma(\beta_0)+2T_{\rm br}e^{-b\beta_0}}{M_{\rm Pl}^2}.
    \label{eq:graviphoton-total-mass}
\end{equation}

With both sources present, the graviphoton eats only one combination of $s$ and $Y$. The gauge-invariant combination $s+n_\sigma Y/R$ remains a physical scalar and is massless in the actions written above unless further interactions generate a potential for it. Scalar winding consequently changes the vector mass, the scalar spectrum and the radion couplings to these fields. The decay channels must then be determined using this spectrum. The reheating calculation below instead uses the construction without scalar winding, for which the longitudinal mode comes entirely from the brane position.

\subsection{Radion reheating and axion dark matter}
\label{sec:radion-reheating-axion-dark-matter}

A stable radion minimum also determines the mass and couplings of the radion excitation. If the radion is initially displaced from this minimum, it oscillates as the universe expands. Once these oscillations lie in the quadratic part of the potential and $H\ll m_\beta$, their averaged energy density redshifts as pressureless matter. We consider a more phenomenologically interesting history in which this component dominates before decaying. Its decay then produces visible radiation, together with relativistic hidden particles, and changes the axion abundance through the accompanying entropy production.

We use the Minkowski construction without scalar winding or magnetic gauging, at the radius in \eqref{eq:quality-radion-physical-scale}. The localised matter sector is taken to live on a brane with positive renormalised tension $T_{\rm br}/V_*=10^{-22}$. Including its contribution $T_{\rm br}e^{-b\beta}$ and readjusting the neutral fermion mass and $\Lambda_5$ preserves the minimum at $\beta=0$, with $m_\beta\simeq169.6\,\mathrm{TeV}$. The correction to this mass is negligible for the chosen tension. The independent QCD vacuum contribution remains subject to \eqref{eq:quality-independent-vacuum-conditions}.

\subsubsection{Radion decay channels}

The localisation brane also affects the light spectrum. Its position mode is eaten by the graviphoton, giving it a mass $\sqrt{2T_{\rm br}}/M_{\rm Pl}\simeq268\,\mathrm{eV}$. Both its transverse and longitudinal polarisations are therefore accessible in radion decay. The neutral stabilising fermions, the charged scalar and the KK excitations are much heavier than the radion and cannot be produced. We assume that the colour Wilson-line scalars are also above threshold and that there are no additional light brane excitations. These choices specify the spectrum used in the following calculation.

To calculate the decay channels, we write the canonically normalised radion as $\phi=\sqrt{3/8}\,bM_{\rm Pl}\beta$ and expand the reduced kinetic terms around $\beta=0$. After removing the vector mixing at the constant axion background and canonically normalising the vector fields, the relevant interactions are
\begin{equation}
    \mathcal L_{\rm int}\supset
    \frac{\phi}{\sqrt6M_{\rm Pl}}
    \left[
        -2(\partial a)^2
        -\frac12(F^{(C)})^2
        -\frac32(F^{(\gamma)})^2
        +(D Y_c)^2
    \right],
    \label{eq:reheating-hidden-interactions}
\end{equation}
where $F^{(C)}$ is the field strength of the canonically normalised four-dimensional vector descending from $C$. We have also defined $Y_c=\sqrt{T_{\rm br}}Y$ and $D_\mu Y_c=\partial_\mu Y_c+m_\gamma\gamma_\mu$, with $m_\gamma$ evaluated at $\beta=0$. The last term follows from the brane-position kinetic term and determines the coupling to the longitudinal graviphoton.

Neglecting the final-state masses relative to $m_\beta$, the two-body decay width into identical particles is $\Gamma=\sum_{\rm pol}|\mathcal M|^2/(32\pi m_\beta)$, where the sum runs over final-state polarisations. Applying this to the interactions above gives
\begin{equation}
    \begin{aligned}
        \Gamma_0&=\frac{m_\beta^3}{96\pi M_{\rm Pl}^2},
        \qquad
        \Gamma_{aa}=2\Gamma_0,
        \qquad
        \Gamma_{CC}=\Gamma_0,\\
        \Gamma_{\rm graviphoton}
        &=\left(9+\frac12\right)\Gamma_0,
        \qquad
        \Gamma_{\rm dark}=\frac{25}{2}\Gamma_0.
    \end{aligned}
    \label{eq:reheating-hidden-widths}
\end{equation}
The transverse graviphoton coupling is three times the $C$-vector coupling, giving $9\Gamma_0$ after summing over its two transverse polarisations. For $m_\beta\gg m_\gamma$, the longitudinal amplitude can be calculated using the eaten position mode $Y_c$. Its derivative coupling is half the axion coupling in magnitude, so the longitudinal width is $\Gamma_{aa}/4=\Gamma_0/2$. This explains why the longitudinal contribution remains finite for a light graviphoton. The total hidden width $\Gamma_{\rm dark}$ is the sum of the axion, $C$-vector and graviphoton channels.

For the visible sector, we retain the decays into gluons and the Higgs doublet. We allow a non-minimal coupling of the brane Higgs to the induced Ricci scalar, with coefficient $\xi$ and conformal value $\xi=1/6$. Neglecting electroweak masses relative to $m_\beta$, the interaction with the canonically normalised radion can be written on shell as
\begin{equation}
    \mathcal L_{\phi H}
    \simeq-\frac{1-6\xi}{\sqrt6M_{\rm Pl}}
    \phi\sum_{i=1}^{4}(\partial h_i)^2,
    \label{eq:reheating-higgs-interaction}
\end{equation}
where $h_i$ are the four canonically normalised real components of the Higgs doublet. Each component contributes $(1-6\xi)^2\Gamma_0/2$, giving a total Higgs width $2(1-6\xi)^2\Gamma_0$. This includes the physical Higgs and the three Goldstone modes describing the longitudinal $W^\pm$ and $Z$ polarisations at high energy. At $\xi=1/6$, this leading Higgs contribution vanishes.

For gluons, the radion coupling also depends on the running between the KK and decay scales. With six active quark flavours, the one-loop QCD coefficient is $11-2n_f/3=7$. Taking the bulk coupling at the KK scale as the matching condition gives
\begin{equation}
    \frac{1}{g_s^2(\mu,\beta)}
    =\frac{e^{b\beta/2}}{g_s^2(R^{-1})}
    +\frac{7}{8\pi^2}
    \log\left(\frac{\mu}{m_{\rm KK,E}(\beta)}\right),
    \qquad
    m_{\rm KK,E}(\beta)=R^{-1}e^{-3b\beta/4},
    \label{eq:reheating-colour-running}
\end{equation}
where $g_s(R^{-1})$ is the reference coupling at $\beta=0$. Differentiating at fixed external Einstein-frame scale $\mu$ gives
\begin{equation}
    \left.
    \frac{\partial}{\partial\beta}
    \frac{1}{g_s^2(\mu,\beta)}
    \right|_{\beta=0}
    =\frac{b}{2g_s^2(R^{-1})}
    +\frac{21b}{32\pi^2}.
    \label{eq:reheating-colour-derivative}
\end{equation}
The first term comes from the bulk gauge kinetic coefficient, while the second follows from the radion dependence of the KK matching scale. After canonically normalising the gluons and setting $\mu=m_\beta$, the decay amplitude for each colour is multiplied by the factor $\mathcal R_g$ below relative to the single-vector channel. Summing over the eight gluon colours therefore gives
\begin{equation}
    \begin{aligned}
        \Gamma_{\rm vis}
        &=\left[2(1-6\xi)^2+8\mathcal R_g^2\right]\Gamma_0,
        \qquad
        \Gamma_\beta=\Gamma_{\rm vis}+\Gamma_{\rm dark},\\
        \mathcal R_g
        &=\frac{\alpha_s(m_\beta)}{\alpha_s(R^{-1})}
        +\frac{21\alpha_s(m_\beta)}{4\pi}.
    \end{aligned}
    \label{eq:reheating-baseline-widths}
\end{equation}
Using the same colour input as in the preceding subsection gives $\mathcal R_g\simeq2.19$.\footnote{Both scales lie above the top threshold. We use $\alpha_s^{-1}(\mu)=0.0292^{-1}+[7/(2\pi)]\log[\mu/(10^{12}\,\mathrm{GeV})]$, with no additional coloured thresholds or sizeable brane gauge kinetic terms.} Increasing $|1-6\xi|$ enhances the Higgs contribution to the visible width and reduces the fraction $\Gamma_{\rm dark}/\Gamma_\beta$ of radion decays into hidden radiation. \footnote{Enhanced modulus decays into the Higgs sector can also suppress axionic dark radiation in LARGE-volume compactifications \cite{CicoliEtAl2022Hydra}.}

\subsubsection{Reheating and dark radiation}

We follow the decay continuously, writing $A(t)$ for the cosmological scale factor and $H=\dot A/A$. The averaged radion density and the visible and hidden radiation densities obey
\begin{align}
    \dot\rho_\beta+3H\rho_\beta
    &=-\Gamma_\beta\rho_\beta,
    \nonumber\\
    \dot\rho_{\rm vis}+3H(\rho_{\rm vis}+p_{\rm vis})
    &=\Gamma_{\rm vis}\rho_\beta,
    \nonumber\\
    \dot\rho_{\rm dark}+4H\rho_{\rm dark}
    &=\Gamma_{\rm dark}\rho_\beta,
    \qquad
    H^2=\frac{\rho_\beta+\rho_{\rm vis}+\rho_{\rm dark}}{3M_{\rm Pl}^2}.
    \label{eq:reheating-continuous-equations}
\end{align}
The initial state is radion dominated, with a small radiation component supplied by its decay. We use the temperature-dependent Standard Model equation of state from \cite{Saikawa:2018rcs}, including the change in relativistic degrees of freedom across the QCD crossover. The visible decay products are assumed to thermalise, while the hidden products remain decoupled. The coherent axion contribution is negligible in the Friedmann equation during reheating for the solutions considered below.

We define the visible reheating temperature at the time when the total radiation density equals the remaining radion density,
\begin{equation}
    T_{\rm RH,vis}
    =\left.T_{\rm vis}\right|_{\rho_{\rm vis}+\rho_{\rm dark}=\rho_\beta}.
    \label{eq:reheating-temperature-definition}
\end{equation}
This definition is used throughout the plots. Since the decay takes place over a finite interval, it need not coincide with the visible temperature at $H=\Gamma_\beta$.

After the radion has decayed, the hidden radiation contributes to the effective number of relativistic species. Evaluating the radiation densities before neutrino decoupling and accounting for the subsequent visible entropy release gives \footnote{The suprression of $\Delta N_{\rm eff}$ via stringy mechanisms after brane-antibrane inflation was explored in \cite{Chakraborty:2026oac} which gives an interesting alternative mechanism.}
\begin{equation}
    \Delta N_{\rm eff}
    =\frac47g_{*\rho}(T)
    \frac{\rho_{\rm dark}}{\rho_{\rm vis}}
    \left(\frac{10.75}{g_{*s}(T)}\right)^{4/3},
    \label{eq:reheating-dark-radiation}
\end{equation}
where $\rho_{\rm vis}=\pi^2g_{*\rho}(T)T^4/30$ and $s_{\rm vis}=2\pi^2g_{*s}(T)T^3/45$. All hidden decay products in \eqref{eq:reheating-hidden-widths} are included. In particular, the produced graviphotons remain relativistic at recombination for the displayed parameters.

\begin{figure}[tbp]
    \centering
    \includegraphics[width=\linewidth]{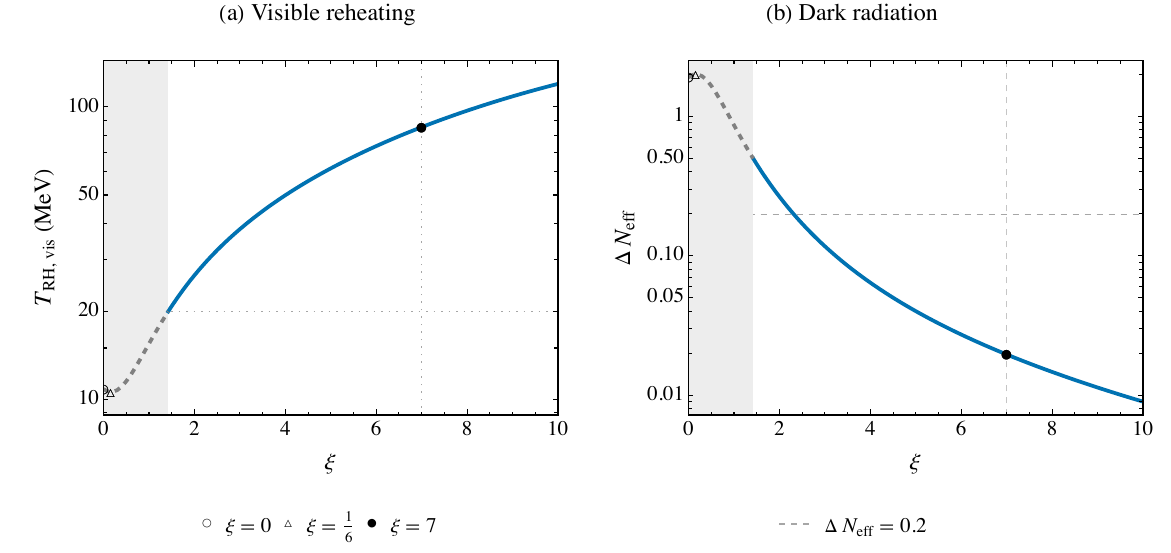}
    \caption{Visible reheating temperature in panel (a) and dark radiation in panel (b), as functions of the Higgs coupling \(\xi\), for the minimum without scalar winding or magnetic gauging at \(R^{-1}=2.55\times10^{12}\,\mathrm{GeV}\). The markers identify \(\xi=0,1/6,7\). The horizontal dashed line in panel (b) gives \(\Delta N_{\rm eff}=0.2\). Grey regions and dashed continuations have \(T_{\rm RH,vis}<20\,\mathrm{MeV}\), where a treatment of neutrino thermalisation is required.}
    \label{fig:radion-reheating-dark-radiation}
\end{figure}

For $\xi=7$, we obtain
\begin{equation}
    T_{\rm RH,vis}\simeq85.3\,\mathrm{MeV},
    \qquad
    \Delta N_{\rm eff}\simeq0.0196.
    \label{eq:reheating-benchmark}
\end{equation}
Figure~\ref{fig:radion-reheating-dark-radiation} shows how these quantities change with $\xi$. We used $\Delta N_{\rm eff}\leq0.2$ (which is within current experimental bounds \cite{Planck2018Inflation, AtacamaCosmologyTelescope:2025blo, AtacamaCosmologyTelescope:2025nti}) as an illustrative criterion for the plots. We also mark the region with $T_{\rm RH,vis}<20\,\mathrm{MeV}$, using this threshold as a conservative lower limit for our equilibrium treatment. As reheating approaches the neutrino decoupling epoch, weak interactions may be too slow to establish thermal neutrino distributions, while hadrons produced in radion decay can modify neutron--proton conversion rates. These effects require a computation beyond the equilibrium equation of state used here \cite{Hasegawa2019Reheating}.

\subsubsection{Axion dark matter}

The relativistic axions produced directly in radion decay contribute to $\rho_{\rm dark}$. A separate cold component can arise from coherent oscillations of the homogeneous axion field. We denote its initial QCD angle by $\vartheta_i$, retaining $\vartheta=\bar\theta+a/f_a$. This initial displacement is distinct from the residual vacuum angle $\theta_{\rm eff}$ constrained by axion quality.

During the late oscillatory regime considered here, the radion is averaged about $\beta=0$, so $f_{\rm eff}=f_a$. The temperature dependence of the QCD susceptibility then determines when the axion starts to move. We approximate the angular potential by $\chi_{\rm QCD}(T)[1-\cos\vartheta]$ and solve
\begin{equation}
    \ddot\vartheta+3H\dot\vartheta+m_a^2(T)\sin\vartheta=0,
    \qquad
    m_a^2(T)=\frac{\chi_{\rm QCD}(T)}{f_a^2},
    \label{eq:reheating-coherent-axion}
\end{equation}
with initial conditions $\vartheta=\vartheta_i$ and $\dot\vartheta=0$. We use the temperature dependence of the QCD susceptibility in \cite{Borsanyi2016Cosmology}, normalised to $\chi_{\rm QCD}(0)=(75.5\,\mathrm{MeV})^4$.\footnote{We interpolate the central susceptibility values logarithmically, match to a low-temperature plateau and use a $T^{-8.16}$ continuation above $3\,\mathrm{GeV}$. The cosine gives an approximation to the finite-angle potential; the susceptibility fixes its quadratic term. The calculation assumes homogeneous misalignment and does not include axion production from strings or domain walls.}

\begin{figure}[tbp]
    \centering
    \includegraphics[width=\linewidth]{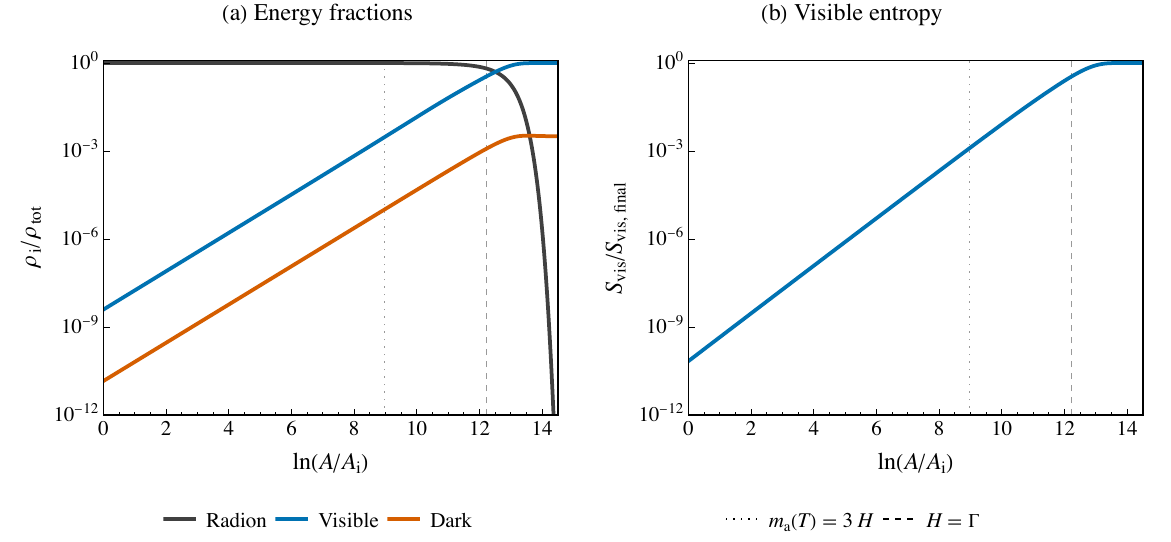}
    \caption{Reheating history for \(\xi=7\) at the reference radius. Panel (a) shows the radion, visible-radiation and hidden-radiation energy fractions. Panel (b) shows the comoving visible entropy \(S_{\rm vis}=A^3s_{\rm vis}\), divided by its final value. The dotted line marks \(m_a(T)=3H\) for \(f_a=10^{14}\,\mathrm{GeV}\), while the dashed line marks \(H=\Gamma_\beta\). Entropy production continues after the axion begins to oscillate.}
    \label{fig:radion-reheating-history}
\end{figure}

At high temperature, $m_a(T)\ll H$ and the axion remains approximately fixed. As the universe cools, its mass increases and it begins to oscillate. For $\xi=7$ and $f_a=10^{14}\,\mathrm{GeV}$, the condition $m_a(T)=3H$ occurs at $T\simeq0.277\,\mathrm{GeV}$, before reheating is complete. Figure~\ref{fig:radion-reheating-history} shows that visible entropy continues to grow after this time. The axion abundance must therefore be evolved through the remaining radion decay.

We obtain the abundance by integrating \eqref{eq:reheating-coherent-axion} through the onset of oscillations. Once the oscillations are adiabatic, meaning that the scale factor and axion mass change little during each cycle, their comoving action determines the approximately conserved comoving axion number $N_a=A^3n_a$. In the quadratic regime, $n_a\simeq\rho_a/m_a(T)$, where $\rho_a$ is the oscillation-averaged energy density of the coherent field above its minimum. Meanwhile, radion decay continues to increase the comoving visible entropy $S_{\rm vis}=A^3s_{\rm vis}$, so the ratio $n_a/s_{\rm vis}=N_a/S_{\rm vis}$ decreases even though $N_a$ remains approximately constant. This is the dilution mechanism associated with low-temperature reheating \cite{VisinelliGondolo2010}. After entropy production ends, this ratio is conserved and fixes the present axion number density through $n_{a,0}=s_0(n_a/s_{\rm vis})_{\rm after\ reheating}$. Multiplying by the present axion mass $m_a(0)$ and normalising to the critical density therefore gives
\begin{equation}
    \Omega_a h^2
    =\frac{m_a(0)s_0}{\rho_{c,0}/h^2}
    \left.\frac{n_a}{s_{\rm vis}}\right|_{\rm after\ reheating},
    \label{eq:reheating-axion-abundance}
\end{equation}
where $s_0$ is the present entropy density and $\rho_{c,0}$ is the present critical density. The condition $m_a(T)=3H$ is used only to indicate the onset in the history plot; the final abundance follows from the numerical evolution and the subsequent entropy dilution.

\begin{figure}[tbp]
    \centering
    \includegraphics[width=\linewidth]{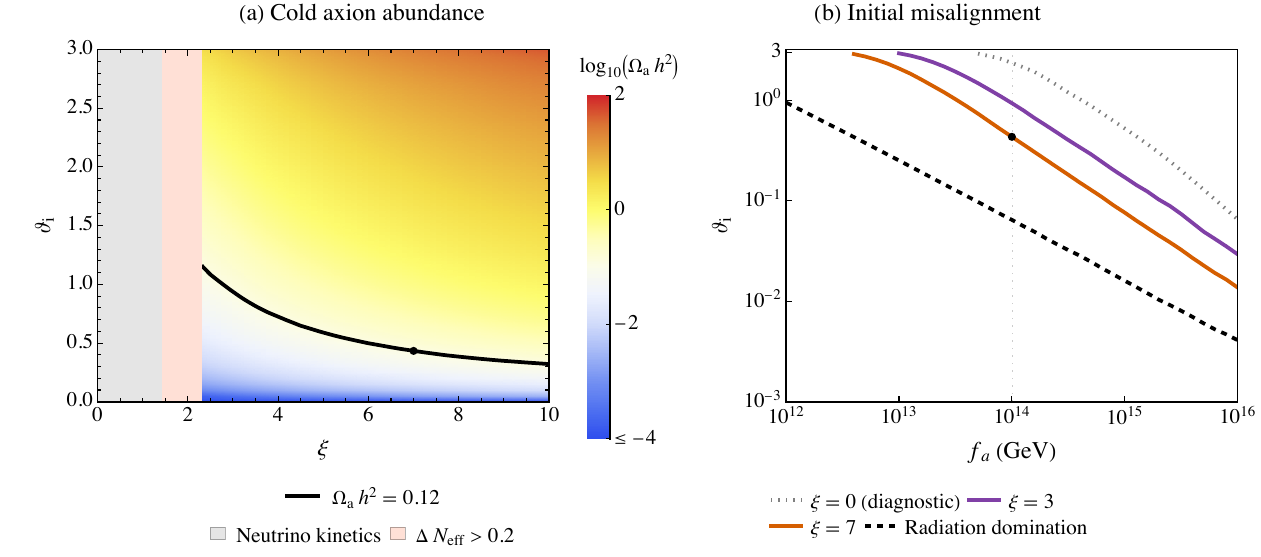}
    \caption{Cold-axion abundance from homogeneous misalignment during radion reheating. Panel (a) shows \(\log_{10}(\Omega_a h^2)\) at \(f_a=10^{14}\,\mathrm{GeV}\) and the black contour gives \(\Omega_a h^2=0.12\). Panel (b) shows the initial angle giving this abundance as a function of \(f_a\), for the indicated values of \(\xi\) and for radiation domination with the same QCD inputs. Grey and orange shading mark the low-temperature region and \(\Delta N_{\rm eff}>0.2\), respectively. The grey \(\xi=0\) curve is an equilibrium continuation outside the adopted temperature range. Black dots mark \eqref{eq:reheating-axion-benchmark}.}
    \label{fig:axion-reheating-abundance}
\end{figure}

For the parameters in \eqref{eq:reheating-benchmark}, the observed dark-matter abundance is obtained with
\begin{equation}
    f_a=10^{14}\,\mathrm{GeV},
    \qquad
    \vartheta_i\simeq0.433,
    \qquad
    \Omega_a h^2\simeq0.12.
    \label{eq:reheating-axion-benchmark}
\end{equation}
With the same QCD inputs but ordinary radiation domination, the required angle is $\vartheta_i\simeq0.0646$. Radion reheating therefore permits a larger initial displacement at the same decay constant. Figure~\ref{fig:axion-reheating-abundance} shows the dependence on $\xi$, $\vartheta_i$ and $f_a$. The calculation is restricted to $\vartheta_i\leq3$, without extrapolation towards the maximum of the angular potential. \footnote{A radion-dependent enhancement of the axion decay constant during inflation has been studied in warped five-dimensional models \cite{ChoiGherghetta2026Isocurvature}. The corresponding inflationary evolution in our circle construction remains to be determined.}

The homogeneous initial condition used above assumes that the circle compactification is established before inflation, that any initial defects are diluted and that no defect network is produced afterwards. If the axion is light during inflation and the radion varies slowly, its typical angular fluctuation is $H_I/[2\pi f_{\rm eff}(\beta_I)]$, where $H_I$ and $\beta_I$ denote the inflationary Hubble scale and radion value. We assume that this angle perturbation is preserved until axion oscillations begin and that radion reheating produces no additional isocurvature.

For axions constituting all dark matter, the approximately quadratic dependence of the abundance on $\vartheta_i$ then gives
\begin{equation}
    H_I\lesssim1.2\times10^9\,\mathrm{GeV}
    \left(\frac{f_{\rm eff}(\beta_I)}{10^{14}\,\mathrm{GeV}}\right)
    \left(\frac{|\vartheta_i|}{0.433}\right),
    \label{eq:reheating-axion-isocurvature}
\end{equation}
using the Planck bound on nearly scale-invariant, uncorrelated cold-dark-matter isocurvature \cite{Planck2018Inflation}.\footnote{We use the upper limit $0.038$ on the isocurvature fraction and the scalar curvature power $2.1\times10^{-9}$. The quadratic approximation is adequate near our benchmark; at larger angles, the bound must instead use the dependence of the numerical abundance on $\vartheta_i$.} The relevant decay constant is $f_{\rm eff}(\beta_I)=f_a e^{-b\beta_I/2}$, which need not equal its late-time value. Note that, the inflationary radion background and its fluctuations are not determined by the reheating calculation. In particular, the late-time mass $m_\beta\simeq169.6\,\mathrm{TeV}$ does not ensure that the radion remains fixed during inflation near the upper limit in \eqref{eq:reheating-axion-isocurvature}.

\subsubsection{Dependence on the stabilised radius}

\begin{figure}[tbp]
    \centering
    \includegraphics[width=0.9\linewidth]{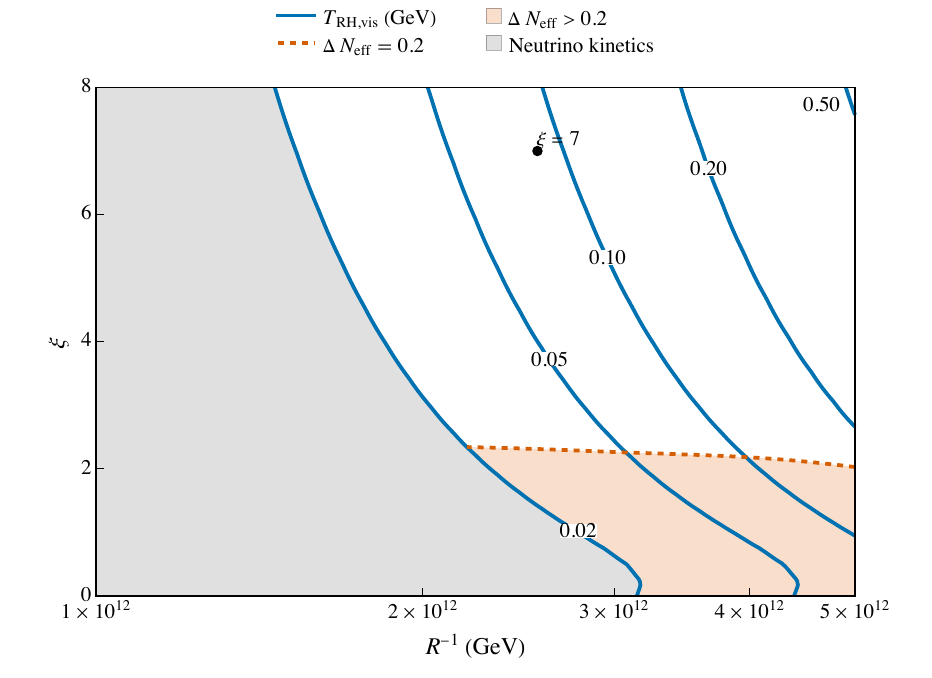}
    \caption{Visible reheating temperature across the family of Minkowski minima without scalar winding or magnetic gauging. Blue contours give \(T_{\rm RH,vis}\) in GeV. Orange shading has \(\Delta N_{\rm eff}>0.2\), and grey shading has \(T_{\rm RH,vis}<20\,\mathrm{MeV}\). The black dot marks \(R^{-1}=2.55\times10^{12}\,\mathrm{GeV}\) and \(\xi=7\). The dimensionless stabilising spectrum and brane tension remain fixed as specified in the text, while the proper masses and matching parameters change with \(R\).}
    \label{fig:radion-reheating-radius}
\end{figure}

Finally, figure~\ref{fig:radion-reheating-radius} shows how reheating changes across the family of stabilised radii. We keep the dimensionless neutral spectrum, $m_5R=26$ and $T_{\rm br}/V_*=10^{-22}$ fixed, together with $M_{\rm Pl}$ and $f_a=10^{14}\,\mathrm{GeV}$. The proper particle masses and vacuum parameters are consequently adjusted at each radius, and the colour coupling is matched to the same low-energy input. As in figure~\ref{fig:quality-radion-mass-radius}, this compares different theories rather than following $\beta(t)$ within one theory.

The scaling $m_\beta\propto R^{-2}$ makes the radion heavier at smaller radius. Its decay rate then grows approximately as $R^{-6}$, with a logarithmic correction from the gluon contribution, and reheating occurs at a higher temperature. The contours in figure~\ref{fig:radion-reheating-radius} include this running and the temperature-dependent equation of state. The marked example gives a reheating temperature above the adopted low-temperature region and a small dark-radiation contribution, while the entropy produced during decay allows the axion abundance in \eqref{eq:reheating-axion-benchmark}.

\section{Outlook}
\label{sec:outlookl}

The PQ solution to the strong CP problem requires an axion whose potential is sufficiently insensitive to additional explicit symmetry breaking contributions. Higher-form symmetries and extra dimensions provide a way to achieve this protection. When the axion descends from a higher-dimensional gauge field, gauge invariance and locality restrict the contributions that can generate its potential. Charged particles can generate a potential by propagating around the compact direction, but their contribution is exponentially suppressed when their mass is large compared with the inverse proper radius. The possible sources of PQ breaking are therefore more restricted than for a generic four-dimensional scalar carrying a global symmetry \cite{CraigKongsore2024,Reece:2023czb,Reece2024Expectations}.

In this work, we have extended this description to a dynamical compactification radius. Building on the higher-form analysis of \cite{CraigKongsore2024,Reece:2023czb,Reece2024Expectations}, we have determined how the axion kinetic term, the charged-particle contribution and the magnetic branch potential depend on $\beta$. This dependence makes radion stabilisation part of the axion-quality question. The radius determines the suppression of unwanted breaking, while the resulting potentials also contribute to the radion equation. A satisfactory construction must therefore fix the radius and preserve the QCD minimum within the same effective theory.

We have constructed two classes of stabilised examples using neutral Casimir contributions, with and without scalar winding. Without winding, the massive spectrum determines a unique finite Minkowski radius when $\sum_iN_i>N_b$, and the radion second derivative is positive. Scalar winding gives a second construction with different vacuum-energy requirements. In both cases, the neutral fields supporting the minimum can be lighter than the charged field responsible for the unwanted Wilson-line potential. The latter can consequently be suppressed without removing the contribution needed for stabilisation. We have checked the axion displacement, the radion displacement and the coupled stability conditions at explicit minima. Under the stated conditions on the independent vacuum terms, good axion quality and a stable radius coexist.

Related aspects of radion stabilisation and axion physics have been explored in earlier work. In particular, \cite{AbeEtAl2014} studied radion stabilisation in the presence of a Wilson-line phase, while \cite{KitanoSuzukiYin2023} investigated the role of QCD in radion stabilisation during inflation and its consequences for axion dark matter. Our analysis connects these questions to the higher-form symmetry protection of the QCD axion. We identify stabilised examples in which the electric and magnetic contributions satisfy the strong CP bound and preserve the coupled axion--radion minimum, subject to the stated conditions on the independent vacuum terms. For the construction without scalar winding or magnetic gauging, we further determine the radion decay channels and their consequences for reheating and axion dark matter.

Magnetic gauging requires a separate assessment. Its additional degree of freedom changes the Casimir spectrum, but the minima persist after readjusting the stabilising parameters. The magnetic coupling must nevertheless remain sufficiently small. Moreover, the radius dependence of QCD matters when comparing the different contributions: for the matching prescription considered here, increasing the proper radius improves the suppression of electric breaking relative to QCD, while increasing the relative importance of magnetic breaking. Radion evolution therefore does not provide a universal suppression of every unwanted axion potential.

The stabilised theory also determines the radion mass and its decay channels. For the construction without scalar winding or magnetic gauging, we have included the graviphoton and followed radion decay continuously through reheating. The example with $\xi=7$ gives $T_{\rm RH,vis}\simeq85.3\,\mathrm{MeV}$ and $\Delta N_{\rm eff}\simeq0.0196$. The accompanying entropy production allows $f_a=10^{14}\,\mathrm{GeV}$ and $\vartheta_i\simeq0.433$ to reproduce the full observed cold-dark-matter abundance. Axion dilution by late matter decay and dark radiation from moduli decay are established mechanisms \cite{VisinelliGondolo2010,CicoliConlonQuevedo2013}. Here they follow from the spectrum and couplings of the same effective theory used to study axion quality and radion stabilisation.

\subsection{Ingredients in UV Physics/String Theory}

String compactifications provide several possible origins for the ingredients considered here. Axions arise from periods of higher-form gauge fields over internal cycles, while gauge fields on branes can also supply Wilson-line scalars when suitable one-cycles are present. Reducing bulk and D-brane Chern--Simons couplings can generate their anomalous coupling to QCD \cite{SvrcekWitten2006}. Wrapped branes, charged states and additional gauge sectors then determine the allowed sources of explicit breaking. A string embedding would have to identify the surviving QCD axion and derive these contributions together with the potential fixing the internal geometry.

KKLT and LVS illustrate that geometric stabilisation can also generate potentials for the associated axions \cite{KKLT2003,LVS2005,Conlon2006Axion, Cicoli:2012sz}. Perturbative moduli stabilisation constructions can fix geometric directions while leaving their axionic partners flat at leading order \cite{CicoliEtAl2024Brane,HaiEtAl2025Perturbative,Chakraborty:2025wqn}. Subleading non-perturbative effects then stabilise these axions. As in our construction, a viable QCD axion must survive any St\"uckelberg couplings, acquire the required QCD coupling and satisfy the quality bound on all additional contributions to its potential.

Warping can modify the axion decay constant and suppress couplings to distant PQ-breaking sources \cite{Choi2004GaugeAxion,CoxGherghettaNguyen2020}.Axion quality has also been studied in warped five-dimensional orbifolds \cite{ChoiLeeShin2026} and on a flat interval \cite{FernandezNavarro2026}. A warped orbifold GUT construction constrains the axion decay constant through perturbative gauge coupling unification \cite{ChoiGherghetta2025}. On a related note, our stabilisation mechanism also offers a possible route to stabilising the evolving internal scale $\beta$ in time-dependent type IIB backgrounds and doubly warped type II solutions \cite{KamalMahanta2026,ArkoEtAl2026}, providing another interesting application of this work. 

Mixing a higher-form axion with a complex scalar phase can preserve axion quality while allowing post-inflationary axion strings, with possible heterotic and type II realisations \cite{LoladzePlatschorreReig2025}. These constructions suggest an extension of our defect sector. A complete ultraviolet description should determine the axion decay constant together with the string spectrum and the sources of PQ breaking \cite{Reece2024Expectations}.

\subsection{Limitations and Open Directions}

Our examples rely on lighter neutral fields to stabilise the radius and a heavier charged field to suppress the unwanted axion potential. Deriving this spectrum and its matching to the localised sector is therefore a natural next step towards an ultraviolet realisation. In particular, the state with $m_5R=26$ must remain below the cutoff with perturbative interactions. For the magnetic construction, such a realisation would also need to explain the small coupling used here and identify any additional sources of PQ breaking.

The persistence of these minima also depends on the independent vacuum terms. The susceptibility does not determine $E_{\rm QCD}(\beta,0)$ or its radion derivatives, and the brane tension remains an independent input. Equation~\eqref{eq:quality-independent-vacuum-conditions} specifies when these contributions leave our minima approximately unchanged. Matching them in a concrete model would allow us to impose the vacuum-energy, stationarity and stability conditions together, and then assess the lifetime of any metastable vacuum.

The entropy dilution giving our axion dark-matter abundance assumes that the radion reaches the quadratic part of its potential and dominates before decaying. Following the coupled fields and their perturbations from inflation would establish whether this history is realised, determine $f_{\rm eff}$ during inflation and test whether subsequent radion evolution changes the isocurvature constraint. Low-scale inflation and stochastic axion evolution could also account for the initial misalignment \cite{Reig2021Stochastic}.

The same cosmological analysis can be extended to our winding and magnetically gauged vacua, using their own fluctuation spectra and decay couplings. Winding changes the graviphoton mass and can leave an additional scalar when the brane is present, so the resulting entropy dilution and dark radiation may differ from the example studied here. Extending the thermal treatment to include neutrino thermalisation and hadronic injection at low temperatures, together with improved QCD inputs, would give more precise constraints on these histories.

\section*{Acknowledgements}
We thank Nabil Iqbal for discussions on higher-form symmetries. ARK thanks Ondrej Hulik for helpful discussions on generalised symmetries and Steven Hsia for painstakingly explaining various aspects of the subject. We thank the ICTP Physics Without Frontiers Bangladesh: Summer Internship Program 2025, during which this work began. ARK is supported by the Czech Science Foundation (GA\v{C}R) through the grant ``One loop in ten and eleven dimensions'' (GA26-22343S). YJ and ARK also acknowledge support from BRAC University under grant BRACURSGI25014. We acknowledge the use of ChatGPT 6 Astra and Claude Opus 5.5 for various checks, suggestions on plot presentation, grammatical corrections and reference searches. All calculations in this manuscript were performed by the authors, and we take full responsibility for the material presented here.

\appendix

\section{Lightning Introduction to Higher Form Symmetries}
\label{app:higher-form-symmetries}

We introduce the higher-form symmetry language used throughout the paper, starting from ordinary global symmetries and then applying the same ideas to the five-dimensional gauge field $C$. The main distinction is between an operator carrying a charge and a topological operator implementing the corresponding symmetry. After compactification, this distinction explains which symmetry protects the axion and how the sources of breaking considered in the main text generate its potential. We restrict to invertible symmetries, following \cite{Gaiotto2015Generalized}. Introductory accounts can be found in \cite{Iqbal2024Lectures,Gomes2023HigherForm,Bhardwaj:2023kri}.

\subsection{Symmetry as a topological operator}

Consider first an ordinary global $U(1)$ symmetry, under which a local operator of charge $q$ transforms as $\mathcal O_q(x)\to e^{i\alpha q}\mathcal O_q(x)$. In $D$ spacetime dimensions, its Noether current can be written as a one-form $j_1$, with conservation equation $\mathrm{d}(*_D j_1)=0$. Integrating over a spatial hypersurface gives our Noether charge, and exponentiating it implements a finite transformation,
\begin{equation}
    Q(\Sigma_{D-1})=\int_{\Sigma_{D-1}}*_D j_1,
    \qquad
    U_\alpha(\Sigma_{D-1})=e^{i\alpha Q(\Sigma_{D-1})}.
    \label{eq:app-ordinary-charge}
\end{equation}
Current conservation means that this hypersurface can be deformed without changing the charge, provided no charged insertion is crossed and there is no flux through any boundary. In a correlation function, $U_\alpha$ can therefore be regarded as an operator supported on a hypersurface whose precise shape does not matter. Such an operator is called a topological operator or a symmetry defect operator. 

A $p$-form symmetry extends this description to charged operators supported on $p$-dimensional submanifolds. Ordinary symmetries have $p=0$ because their charged operators are local with one-form symmetries act on line operators, while two-form symmetries act on surface operators. For a continuous $U(1)^{(p)}$ symmetry, the current is a $(p+1)$-form $j_{p+1}$. We instead use its Hodge dual as the closed current, as in the main text,
\begin{equation}
    J_{D-p-1}=*_D j_{p+1},
    \qquad
    \mathrm{d}J_{D-p-1}=0,
    \qquad
    U_\alpha(\Sigma)=
    \exp\left(i\alpha\int_\Sigma J_{D-p-1}\right).
    \label{eq:app-symmetry-defect}
\end{equation}
The integration surface $\Sigma$ is closed and oriented, with dimension $D-p-1$. Thus the symmetry defect has codimension $p+1$, whereas the operator on which it acts has dimension $p$. We take charges to be integer-normalised, so $\alpha\sim\alpha+2\pi$.

The topological property can be seen directly from Stokes' theorem. If a deformation from $\Sigma$ to $\Sigma'$ sweeps out a region $V$, so that $\Sigma'-\Sigma=\partial V$, then
\begin{equation}
    \int_{\Sigma'}J-\int_\Sigma J
    =\int_V\mathrm{d}J=0,
    \label{eq:app-topological-deformation}
\end{equation}
provided $V$ does not intersect a charged insertion. The defect can consequently be moved or distorted without changing the correlation function. Crossing a charged operator is different as it implements the symmetry action on that operator.

For an ordinary symmetry, we measure the charge of a local operator on a sphere surrounding its insertion. For a one-form symmetry, the same idea applies to a charged line, which we surround in the transverse directions. More generally, a small sphere $S^{D-p-1}$ links the support of a charge-$q$ operator $W_q(\gamma_p)$ once. Placing the symmetry defect on this sphere gives the phase $e^{i\alpha q}$. If the defect instead has oriented linking number $\ell$ with $\gamma_p$, then
\begin{equation}
    \left\langle U_\alpha(\Sigma)W_q(\gamma_p)\cdots\right\rangle
    =e^{i\alpha q\ell}
    \left\langle W_q(\gamma_p)\cdots\right\rangle.
    \label{eq:app-linking-action}
\end{equation}
Here we take $\Sigma$ to lie in a contractible region and to link none of the omitted insertions. We can deform the symmetry defect without changing the correlator as long as it crosses no charged insertions. This property belongs to the defect and need not hold for the charged operator itself i.e a Wilson loop, for example, can still depend on its shape.

\begin{figure}[tbp]
    \centering
    \includegraphics[width=0.9\linewidth]{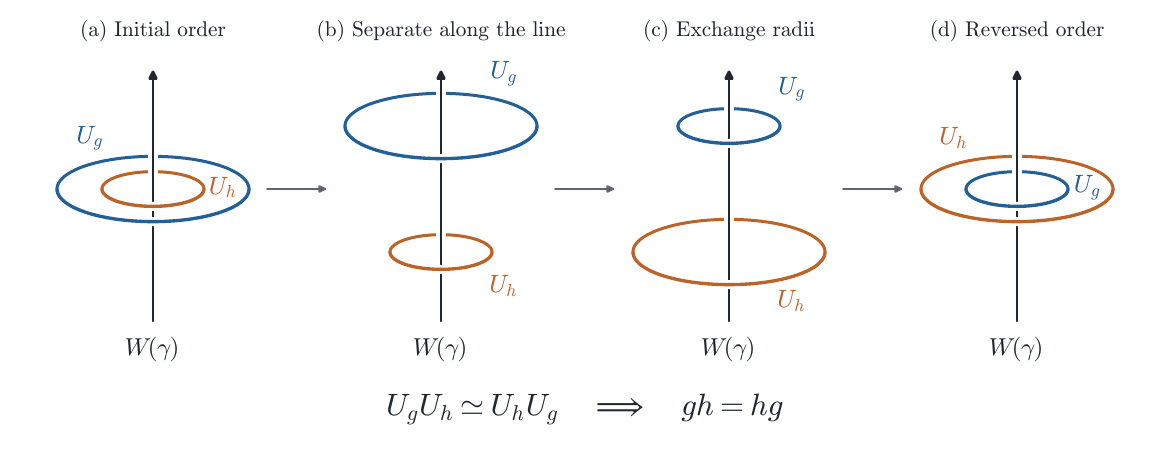}
    \caption{This figure demonstrates how one-form symmetries are commutative/abelian only which ofcourse can be extended to higher forms. The black line supports the charged operator $W(\gamma)$, while the blue and orange loops support the topological symmetry defects $U_g$ and $U_h$. In (a), the loops have a definite radial order. They are separated along the line in (b), their radii are exchanged at different heights in (c), and they are returned to the same height in (d). Throughout this deformation, the loops remain disjoint from each other and from $W(\gamma)$, and each continues to link the charged line once. The correlation function is unchanged, so fusing the defects in either order gives the same symmetry transformation: $U_{gh}=U_{hg}$.}
    \label{fig:one_form_3d_color}
\end{figure}

\begin{figure}[tbp]
    \centering
    \includegraphics[width=0.9\linewidth]{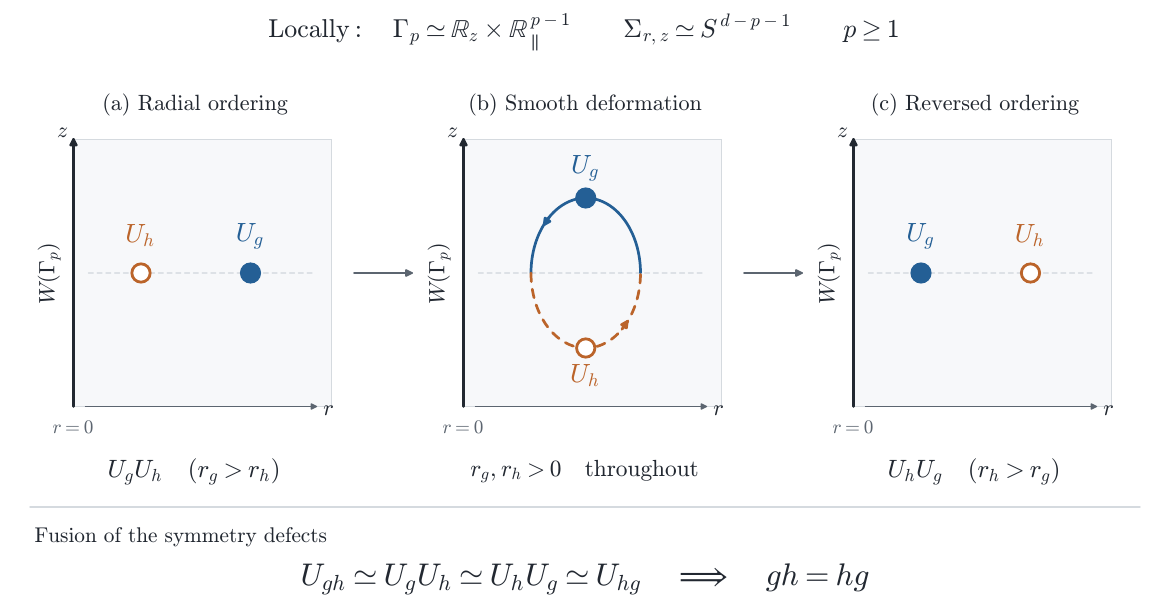}
    \caption{The same exchange for a $p$-form symmetry with $p\geq1$. The figure uses $d=D$ for the spacetime dimension. The axis $r=0$ represents the charged support $\Gamma_p$, with one tangent coordinate $z$ displayed and the remaining $p-1$ tangent directions suppressed. Each coloured point represents an entire linking sphere $S^{D-p-1}$ of radius $r$ in the normal space. In (a), the two spheres have the same $z$ coordinate and a definite radial order. The paths in (b) separate them along $z$ and exchange their radii while keeping $r>0$. When their radii coincide, their different $z$ coordinates keep the spheres disjoint. Returning to the same $z$ coordinate gives the reversed order in (c), so the two fusion orders define the same group element.}
    \label{fig:general_p_charged_operator}
\end{figure}

More generally, let $U_g$ denote the defects of an invertible symmetry group $G$. Bringing two defects together implements composition of the transformations. This operation is called fusion: for group elements $g$ and $h$, it gives $U_gU_h=U_{gh}$. Invertibility means that $U_g$ has an inverse $U_{g^{-1}}$. For $U(1)$, these statements reduce to $U_\alpha U_{\alpha'}=U_{\alpha+\alpha'}$ and $U_\alpha U_{-\alpha}=1$. For $p\geq1$, the geometry also explains why the symmetry group is Abelian. There is at least one direction along the charged operator in which two linking defects can be separated. They can then exchange their radial order without crossing each other or the charged operator, before being brought back together. The two fusion orders are therefore equivalent, giving $U_{gh}=U_{hg}$. Figure~\ref{fig:one_form_3d_color} shows the argument for a line in three dimensions, and figure~\ref{fig:general_p_charged_operator} illustrates the same deformation for a general extended operator. A point insertion has no tangent direction along which to perform this separation, so the argument does not force ordinary zero-form symmetries to be Abelian \cite{Gaiotto2015Generalized,Bhardwaj:2023kri}.

This description also applies to discrete symmetries. They have no infinitesimal Noether current, but their transformations are still implemented by topological defects with the appropriate fusion rules. We will use this fact when a continuous symmetry is reduced to a finite subgroup.

\subsection{The five-dimensional gauge theory}

We now apply these ideas to the compact $U(1)$ connection $C$ of \eqref{eq:circle-axion}, with field strength $F=\mathrm{d}C$. Initially, we omit dynamical electric and magnetic sources, the QCD coupling and higher-form gauging. The Maxwell equation and the Bianchi identity then give
\begin{equation}
    J_e=\frac{1}{g_5^2}*_5F,
    \qquad \mathrm{d}J_e=0,
    \qquad
    J_m=\frac{F}{2\pi},
    \qquad \mathrm{d}J_m=0.
    \label{eq:app-maxwell-currents}
\end{equation}
The electric conservation law follows from the equation of motion, while the magnetic one follows locally from $\mathrm{d}F=\mathrm{d}^2C=0$. Since $D=5$, the three-form $J_e$ has degree $D-p-1$ with $p=1$, and therefore generates an electric one-form symmetry. The two-form $J_m$ instead corresponds to $p=2$ and generates a magnetic two-form symmetry. Their symmetry defects are
\begin{equation}
    U^e_\alpha(\Sigma_3)=
    \exp\left(\frac{i\alpha}{g_5^2}\int_{\Sigma_3}*_5F\right),
    \qquad
    U^m_{\alpha_m}(\Sigma_2)=
    \exp\left(\frac{i\alpha_m}{2\pi}\int_{\Sigma_2}F\right).
    \label{eq:app-electric-magnetic-defects}
\end{equation}
Here $\alpha_m$ is the magnetic symmetry parameter. These global transformations act on gauge-invariant extended operators. The electrically charged operators are Wilson lines,
\begin{equation}
    W_q(\gamma_1)=\exp\left(iq\oint_{\gamma_1}C\right),
    \qquad q\in\mathbb Z.
    \label{eq:app-wilson-line}
\end{equation}
A small $S^3$ surrounding the line in its four transverse directions measures the electric flux and gives the phase $e^{i\alpha q}$ for unit linking. The magnetic charged operators are 't Hooft surface defects $T_m(\gamma_2)$. Such an insertion is defined by requiring
\begin{equation}
    \frac{1}{2\pi}\int_{S^2}F=m,
    \qquad m\in\mathbb Z,
    \label{eq:app-magnetic-probe-charge}
\end{equation}
on a small sphere in the three directions normal to $\gamma_2$. In five dimensions, a magnetic source is a string, and $\gamma_2$ is its worldsheet. The magnetic defect measures this flux and gives the phase $e^{i\alpha_m m}$ for unit linking. These probe insertions do not explicitly break the symmetry. Current conservation holds away from their support, and the contact terms at the insertions express their charges through the Ward identities. Introducing dynamical charged matter into the theory is a separate operation and it does break the symmetries. 

\subsection{Reduction to the axion symmetries}

On $M_4\times S^1$, an operator wrapped around the internal circle can become local from the four-dimensional viewpoint. In particular,
\begin{equation}
    W_q(S^1_x)=\exp\left(iq\frac{a(x)}{f_a}\right),
    \qquad
    \frac{a}{f_a}=\oint_{S^1}C.
    \label{eq:app-wrapped-operator}
\end{equation}
The loop extends only along the internal circle and is labelled by a single point $x$ in four dimensions. An electric defect supported on a three-dimensional hypersurface in $M_4$, at a fixed point on the circle, acts on this operator by the phase $e^{i\alpha q}$. It therefore implements an ordinary zero-form symmetry of the four-dimensional axion. The same conclusion follows directly from the gauge field. A shift $C\to C+\lambda_1$ by a closed one-form leaves $F$ unchanged, but if $\oint_{S^1}\lambda_1=\alpha$, it changes the holonomy by
\begin{equation}
    a\to a+\alpha f_a,
    \qquad
    e^{iqa/f_a}\to e^{i\alpha q}e^{iqa/f_a}.
    \label{eq:app-axion-shift}
\end{equation}
This is the PQ shift inherited from the electric one-form symmetry. Large gauge transformations identify $a$ with $a+2\pi f_a$, while a general $\alpha$ acts non-trivially on the wrapped Wilson line. An exact continuous shift forbids any non-constant potential for $a$ \cite{CraigKongsore2024}.

In the axion sector with vanishing four-dimensional vector backgrounds, the kinetic term is canonical at the reference radius $\beta=0$, giving $J_{\rm PQ}=f_a*_4\mathrm{d}a$ with $f_a^2=1/(2\pi Rg_5^2)$. For the dynamical radius considered in the main text, the kinetic coefficient instead gives
\begin{equation}
    J_{\rm PQ}=f_a e^{-b\beta}*_4\mathrm{d}a,
    \qquad
    \mathrm{d}J_{\rm PQ}=0,
    \label{eq:app-pq-current}
\end{equation}
when the sources of PQ breaking are absent, with $*_4$ defined using the Einstein-frame metric. The conservation equation is the axion equation of motion. Thus the evolving radius changes the current normalisation but does not by itself break the shift symmetry. The magnetic current gives a different symmetry. The axion zero mode contributes $F_{\mu y}=\partial_\mu a/(2\pi Rf_a)$, so integrating $J_m$ over the circle gives
\begin{equation}
    J_{\rm w}=\int_{S^1}J_m
    =\frac{1}{2\pi}\mathrm{d}\left(\frac{a}{f_a}\right),
    \qquad
    U^{\rm w}_{\alpha_m}(\gamma_1)=
    \exp\left(\frac{i\alpha_m}{2\pi}
    \oint_{\gamma_1}\mathrm{d}\left(\frac{a}{f_a}\right)\right).
    \label{eq:app-winding-defect}
\end{equation}
This generator descends from $U^m_{\alpha_m}(\gamma_1\times S^1)$. Although $J_{\rm w}$ is a one-form, it generates a two-form symmetry in four dimensions, since $D-p-1=1$ gives $p=2$. Its charged operators are axion-string worldsheets. If the angle $a/f_a$ changes by $2\pi n$ around a contour linking such a string, then $\oint_{\gamma_1}J_{\rm w}=n$, and the defect gives the phase $e^{i\alpha_m n}$.

Away from string insertions, $\mathrm{d}J_{\rm w}=0$ follows from $\mathrm{d}^2=0$ and does not use the axion equation of motion. This local identity therefore survives the addition of a periodic axion potential. Winding conservation measures the string charge; the electric symmetry descending to the PQ shift is what forbids an axion potential.

\subsection{Breaking, gauging and axion quality}

We can now interpret the modifications studied in the main text. Consider first adding dynamical matter of electric charge $q$. A Wilson line can end on a charged matter insertion, so its charge can be screened. The electric current equation becomes $\mathrm{d}J_e=j_{\rm matter}$, where $j_{\rm matter}$ is the four-form electric source. Deforming a generic electric defect can now encounter dynamical charge, and the continuous one-form symmetry is explicitly broken.

If charged matter is the only source of breaking, the transformations satisfying $e^{i\alpha q}=1$ survive as $\mathbb Z_{|q|}^{(1)}$. With several charged species, this condition must hold for every charge, so the surviving group is fixed by their greatest common divisor. In particular, a unit-charged particle leaves no non-trivial electric subgroup. The breaking is explicit: charged matter changes which transformations are symmetries of the theory, whereas spontaneous breaking leaves the symmetry exact but selects a vacuum that does not respect it. The effect of this explicit breaking on the axion potential can nevertheless be small. After compactification, virtual charged particles generate the harmonics in \eqref{eq:charged-scalar-potential} by winding around $S^1$. A worldline with winding number $w$ has length $2\pi R|w|$, giving the suppression $e^{-2\pi Rm_5|w|}$ at fixed radius. Thus a sufficiently heavy particle breaks the continuous symmetry while generating only an exponentially small axion potential.

The QCD coupling supplies the breaking needed for the PQ mechanism. Equation~\eqref{eq:five-dimensional-abj} gives $\mathrm{d}J_e=\mathcal I_4$, where $\mathcal I_4$ is the QCD instanton-density four-form. An electric defect can consequently acquire an instanton-dependent contribution when it is deformed. Locally writing $\mathcal I_4=\mathrm{d}\omega_3$ and replacing $J_e$ by $J_e-\omega_3$ restores closure, but the colour Chern--Simons form $\omega_3$ is gauge dependent. A generic continuous defect is therefore not restored by this redefinition. This ABJ breaking makes the vacuum energy depend on $\bar\theta+a/f_a$, allowing QCD to select the CP-preserving minimum. The quality problem arises because other sources of breaking can favour a different axion value \cite{CraigKongsore2024}.

Gauging changes the theory in a different way. To probe a $p$-form symmetry, we first couple its current to a background $(p+1)$-form gauge field through
\begin{equation}
    S_{\rm background}\supset
    \int A_{p+1}\wedge J_{D-p-1}.
    \label{eq:app-background-coupling}
\end{equation}
The degrees add to $D$, as required for a spacetime integral. Gauging means promoting this background to a dynamical field and summing over its configurations, with a consistent gauge symmetry. Electric one-form gauging therefore introduces a two-form $B_2$, whereas magnetic two-form gauging introduces a three-form $K_3$.

The electric and magnetic symmetries have a mixed 't Hooft anomaly: their background gauge invariances cannot both be maintained without additional ingredients. This does not remove either global symmetry in the original free Maxwell theory. It obstructs gauging them simultaneously, and gauging one can reduce the other \cite{Gaiotto2015Generalized,Iqbal2024Lectures}.

Magnetic gauging provides an explicit example. With the other breaking terms switched off, \eqref{eq:magnetic-gauging-five} gives $\mathrm{d}J_e=(M/2\pi)\mathrm{d}K_3$. The combination $J_e-(M/2\pi)K_3$ is closed, suggesting the improved electric defect
\begin{equation}
    \widehat U^e_\alpha(\Sigma_3)=
    \exp\left[i\alpha\int_{\Sigma_3}
    \left(J_e-\frac{M}{2\pi}K_3\right)\right].
    \label{eq:app-improved-defect}
\end{equation}
However, a physical symmetry operator must also be gauge invariant. Under a large $K_3$ gauge transformation with $\int_{\Sigma_3}\delta K_3=2\pi n$, the candidate defect transforms as
\begin{equation}
    \widehat U^e_\alpha\to
    e^{-i\alpha Mn}\widehat U^e_\alpha.
    \label{eq:app-defect-large-gauge-transformation}
\end{equation}
Invariance for every integer $n$ requires $\alpha M\in2\pi\mathbb Z$. For $M\ne0$, only $\mathbb Z_{|M|}^{(1)}$ survives. Thus a formally conserved current does not guarantee a continuous global symmetry if its defect is not gauge invariant. The remaining discrete symmetry permits the branch potential \eqref{eq:magnetic-branch-potential}; unlike the charged-particle contribution, its coefficient has no winding exponential.

Electric gauging of the free theory instead replaces $F$ by $F-kB_2$, invariant under $C\to C+k\Lambda_1$ and $B_2\to B_2+\mathrm{d}\Lambda_1$. The reduced vector is $\mathcal A_\mu=\int_0^{2\pi R}\mathrm{d}y\,B_{\mu y}$, so the axion derivative becomes $\mathrm{d}a-kf_a\mathcal A_1$. Indeed, writing $\lambda(x)=\oint_{S^1}\Lambda_1$ gives $\delta a=kf_a\lambda$ and $\delta\mathcal A_1=\mathrm{d}\lambda$, leaving this combination invariant. For $k\ne0$, the vector eats the axion through the St\"uckelberg mechanism, so no independent axion scalar remains. Restoring the QCD coupling also requires the anomaly-cancelling ingredients discussed in the main text.

The connection to axion quality is therefore direct. The electric one-form symmetry supplies the protective PQ shift, while charged matter, the QCD anomaly and magnetic gauging modify that protection in different ways. Heavy electric charges give an exponentially small correction, whereas magnetic gauging allows a potential without this suppression. The size and relative phase of every additional contribution must consequently satisfy \eqref{eq:general-extra-dimensional-quality} \cite{CraigKongsore2024}.
\section{One-loop derivation of the Casimir potential}
\label{app:radion-one-loop}

We derive the finite Casimir contribution of a neutral real scalar with constant proper mass $m$ and periodic boundary conditions around the circle \cite{ArkaniHamed2003Extranatural,PontonPoppitz2001,CraigKongsore2024}. To extract the potential, we hold $\beta$ constant and evaluate the determinant in a flat four-dimensional background. Terms involving spacetime derivatives of $\beta$ or spacetime curvature are therefore not included. The calculation separates the local contribution, which renormalises the bulk cosmological constant, from the finite contribution associated with propagation around the circle.

The KK masses follow from \eqref{eq:beta-charged-kk-masses} by setting the charge to zero:
\begin{equation}
    m_{k,E}^2=m^2e^{-b\beta/2}
    +\frac{k^2}{R^2}e^{-3b\beta/2},
    \qquad k\in\mathbb Z.
    \label{eq:beta-kk-masses}
\end{equation}
The first term comes from the proper five-dimensional mass, while the second comes from momentum along the compact direction. Since $\sqrt{|g_5|}g_5^{\mu\nu}=\sqrt{-g_E}g_E^{\mu\nu}$, the four-dimensional scalar kinetic term has no additional $\beta$ dependence.

Integrating out a real scalar gives the one-loop effective action $\Gamma_1=\frac12\operatorname{Tr}\log[(-\partial_E^2+m_{k,E}^2)/\mu^2]$, where the trace includes the KK sum and $\mu$ is a reference mass. Dividing by the four-dimensional volume and extracting the finite Casimir contribution gives
\begin{equation}
    V_B(\beta;m)
    =\left.\frac12\sum_{k\in\mathbb Z}
    \int\frac{\mathrm{d}^4p_E}{(2\pi)^4}
    \log\left(\frac{p_E^2+m_{k,E}^2}{\mu^2}\right)
    \right|_{\rm finite}.
    \label{eq:beta-boson-determinant}
\end{equation}
The factor $1/2$ follows from the Gaussian integral over a real field. We now specify the finite prescription by retaining the local contribution until its dependence on $\beta$ has been identified.

We use the proper-time representation of the determinant, with the short-proper-time region regulated in five-dimensional proper units. The normalisation of the determinant is included in the local counterterms. With this prescription,
\begin{equation}
    \Gamma_1=-\frac12\int_0^\infty\frac{\mathrm{d}u}{u}\,
    \operatorname{Tr}e^{-u(-\partial_E^2+m_{k,E}^2)},
    \qquad
    \int\frac{\mathrm{d}^4p_E}{(2\pi)^4}e^{-up_E^2}
    =\frac{1}{16\pi^2u^2}.
    \label{eq:beta-proper-time-identities}
\end{equation}
All proper-time expressions before the local subtraction are understood to be regulated. Performing the momentum integral, the contribution per unit four-dimensional volume becomes
\begin{equation}
    -\frac{1}{32\pi^2}\sum_{k\in\mathbb Z}
    \int_0^\infty\mathrm{d}u\,u^{-3}
    \exp\left[-um^2e^{-b\beta/2}
    -\frac{uk^2}{R^2}e^{-3b\beta/2}\right].
    \label{eq:beta-proper-time-sum}
\end{equation}
The power $u^{-3}$ combines the $u^{-1}$ from the proper-time representation with the $u^{-2}$ from the four-dimensional momentum integral.

The KK sum can be rewritten using Poisson resummation:
\begin{equation}
    \sum_{k\in\mathbb Z}
    e^{-uk^2e^{-3b\beta/2}/R^2}
    =\frac{\sqrt{\pi}R e^{3b\beta/4}}{\sqrt u}
    \sum_{w\in\mathbb Z}
    \exp\left[-\frac{\pi^2R^2w^2e^{3b\beta/2}}{u}\right].
    \label{eq:beta-poisson-resummation}
\end{equation}
This replaces the sum over KK momenta with a sum over trajectories winding around the circle. Substituting into \eqref{eq:beta-proper-time-sum} gives
\begin{equation}
    -\frac{R e^{3b\beta/4}}{32\pi^{3/2}}
    \sum_{w\in\mathbb Z}
    \int_0^\infty\mathrm{d}u\,u^{-7/2}
    \exp\left[-um^2e^{-b\beta/2}
    -\frac{\pi^2R^2w^2e^{3b\beta/2}}{u}\right].
    \label{eq:beta-resummed-determinant}
\end{equation}
Here $w$ counts the winding of a virtual particle trajectory. It is independent of the classical scalar winding $n_\sigma$ introduced in \eqref{eq:beta-winding-background}.

The separation between local and finite contributions is now explicit. For $w=0$, there is no suppression as $u\to0$, and this term contains the bulk ultraviolet divergence. Setting $u=e^{b\beta/2}\tau$, where $\tau$ is the five-dimensional proper-time parameter, gives
\begin{equation}
    -\frac{R e^{-b\beta/2}}{32\pi^{3/2}}
    \int_0^\infty\mathrm{d}\tau\,\tau^{-7/2}e^{-m^2\tau}
    =2\pi R\,\delta\Lambda_5\,e^{-b\beta/2}.
    \label{eq:beta-zero-winding-renormalisation}
\end{equation}
The regulator is fixed in $\tau$, so the remaining integral is independent of $\beta$. This contribution therefore has exactly the radius dependence of the bulk cosmological constant in \eqref{eq:beta-bulk-potential} \cite{PontonPoppitz2001}. We absorb it, together with its counterterm, into the renormalised parameter $\Lambda_5$ used in the main text. Its contribution to the radion potential is retained through this parameter.

For $w\ne0$, the exponential containing $1/u$ suppresses the short-proper-time region, and each winding contribution is ultraviolet finite. We can therefore remove the regulator after separating the $w=0$ term. Using the winding action $z_w=2\pi wmR e^{b\beta/2}$, the required integral for $m>0$ and $w\geq1$ is
\begin{equation}
    \begin{aligned}
        &\int_0^\infty\mathrm{d}u\,u^{-7/2}
        \exp\left[-um^2e^{-b\beta/2}
        -\frac{\pi^2R^2w^2e^{3b\beta/2}}{u}\right]\\
        &\qquad=2\left(\frac{m}{\pi Rw}\right)^{5/2}
        e^{-5b\beta/2}K_{5/2}(z_w).
    \end{aligned}
    \label{eq:beta-bessel-integral}
\end{equation}
Here $K_{5/2}$ is a modified Bessel function. For this half-integer order, it has the explicit form
\begin{equation}
    K_{5/2}(z_w)=\sqrt{\frac{\pi}{2z_w}}e^{-z_w}
    \left(1+\frac3{z_w}+\frac3{z_w^2}\right).
    \label{eq:beta-half-integer-bessel}
\end{equation}
The contributions from $w$ and $-w$ are equal for a neutral field. Combining them and substituting \eqref{eq:beta-half-integer-bessel} into \eqref{eq:beta-resummed-determinant}, with the zero-winding term removed, gives
\begin{align}
    V_B(\beta;m)
    &=-\frac{m^2e^{-2b\beta}}{16\pi^4R^2}
    \sum_{w=1}^\infty\frac{e^{-z_w}}{w^3}
    \left(1+\frac3{z_w}+\frac3{z_w^2}\right)\nonumber\\
    &=-\frac{3e^{-3b\beta}}{64\pi^6R^4}
    \sum_{w=1}^\infty\frac{e^{-z_w}}{w^5}
    \left(1+z_w+\frac{z_w^2}{3}\right).
    \label{eq:beta-casimir-evaluation}
\end{align}
The second line follows from $z_w^2=4\pi^2w^2m^2R^2e^{b\beta}$ and reproduces \eqref{eq:beta-finite-casimir}. The exponential is determined by the proper length of the winding trajectory multiplied by the particle mass. The massless limit is regular in the second line: the winding sum becomes $\sum_{w=1}^\infty w^{-5}=\zeta(5)$, giving \eqref{eq:beta-massless-casimir}.

For a complex scalar, the result is multiplied by two. If the scalar carries charge $q$ under $C$, the shifted KK spectrum introduces the phase $\exp(iwqa/f_a)$ after Poisson resummation. The two orientations then combine as $\exp(iwqa/f_a)+\exp(-iwqa/f_a)=2\cos(wqa/f_a)$. Including this cosine and the two real degrees of freedom gives \eqref{eq:beta-charged-potential}, which reduces to \eqref{eq:charged-scalar-potential} at $\beta=0$.

The fermionic contribution follows from the Grassmann determinant. For the vacuum-energy term, a five-dimensional Dirac field gives $-\frac12\operatorname{Tr}_{\rm spin}\log(-\partial_E^2+m_{k,E}^2)$. The spin trace has four components, so its contribution is $-4V_B(\beta;m)$. This relation assumes the same mass and periodic boundary conditions as the scalar calculation. Different boundary conditions change the phases in the winding sum.

Finally, the derivatives needed for radion stabilisation follow directly from the finite expression. Holding $m$ and $R$ fixed, we have
\begin{equation}
    z_w'=\frac b2z_w,\qquad
    \frac{\mathrm{d}}{\mathrm{d}\beta}
    \left[e^{-z_w}\left(1+z_w+\frac{z_w^2}{3}\right)\right]
    =-\frac b6e^{-z_w}z_w^2(1+z_w).
    \label{eq:beta-winding-derivative}
\end{equation}
Differentiating both the winding sum and the prefactor $e^{-3b\beta}$ therefore gives
\begin{align}
    V_B'(\beta;m)
    &=\frac{3b\,e^{-3b\beta}}{64\pi^6R^4}
    \sum_{w=1}^\infty\frac{e^{-z_w}}{w^5}
    \left(3+3z_w+\frac76z_w^2+\frac16z_w^3\right),\nonumber\\
    V_B''(\beta;m)
    &=-\frac{3b^2e^{-3b\beta}}{64\pi^6R^4}
    \sum_{w=1}^\infty\frac{e^{-z_w}}{w^5}
    \left(9+9z_w+\frac{23}{6}z_w^2+\frac56z_w^3+\frac1{12}z_w^4\right).
    \label{eq:beta-casimir-derivatives}
\end{align}
The derivatives of $V_{\rm Cas}$ are obtained by summing these expressions with the same masses, multiplicities and signs as in \eqref{eq:fluxless-potential}. They determine the stationary-point and radion-mass conditions for both stabilisation mechanisms discussed in section~\ref{sec:beta-stabilisation}.

\bibliographystyle{JHEP}
\bibliography{biblio}

\providecommand{\href}[2]{#2}\begingroup\raggedright\begin{thebibliography}{10}

\bibitem{GellMann1961Eightfold}
M.~Gell-Mann, \emph{{The Eightfold Way: A Theory of Strong Interaction Symmetry}},  Tech. Rep. CTSL-20, California Institute of Technology (1961), \href{https://doi.org/10.2172/4008239}{DOI}.

\bibitem{Neeman1961}
Y.~Ne'eman, \emph{{Derivation of Strong Interactions from a Gauge Invariance}}, \href{https://doi.org/10.1016/0029-5582(61)90134-1}{\emph{Nucl. Phys.} {\bfseries 26} (1961) 222}.

\bibitem{Barnes:1964pd}
V.E.~Barnes et~al., \emph{{Observation of a Hyperon with Strangeness Minus Three}}, \href{https://doi.org/10.1103/PhysRevLett.12.204}{\emph{Phys. Rev. Lett.} {\bfseries 12} (1964) 204}.

\bibitem{GellMann1964Quarks}
M.~Gell-Mann, \emph{{A Schematic Model of Baryons and Mesons}}, \href{https://doi.org/10.1016/S0031-9163(64)92001-3}{\emph{Phys. Lett.} {\bfseries 8} (1964) 214}.

\bibitem{Zweig1964}
G.~Zweig, \emph{{An SU(3) Model for Strong Interaction Symmetry and Its Breaking. Version 1}},  Tech. Rep. CERN-TH-401, CERN (1964), \href{https://doi.org/10.17181/CERN-TH-401}{DOI}.

\bibitem{Greenberg:1964pe}
O.W.~Greenberg, \emph{{Spin and Unitary-Spin Independence in a Paraquark Model of Baryons and Mesons}}, \href{https://doi.org/10.1103/PhysRevLett.13.598}{\emph{Phys. Rev. Lett.} {\bfseries 13} (1964) 598}.

\bibitem{Han:1965pf}
M.Y.~Han and Y.~Nambu, \emph{{Three-Triplet Model with Double SU(3) Symmetry}}, \href{https://doi.org/10.1103/PhysRev.139.B1006}{\emph{Phys. Rev.} {\bfseries 139} (1965) B1006}.

\bibitem{Breidenbach:1969kd}
M.~Breidenbach et~al., \emph{{Observed Behavior of Highly Inelastic Electron-Proton Scattering}}, \href{https://doi.org/10.1103/PhysRevLett.23.935}{\emph{Phys. Rev. Lett.} {\bfseries 23} (1969) 935}.

\bibitem{Fritzsch:1973pi}
H.~Fritzsch, M.~Gell-Mann and H.~Leutwyler, \emph{{Advantages of the Color Octet Gluon Picture}}, \href{https://doi.org/10.1016/0370-2693(73)90625-4}{\emph{Phys. Lett. B} {\bfseries 47} (1973) 365}.

\bibitem{Gross:1973id}
D.J.~Gross and F.~Wilczek, \emph{{Ultraviolet Behavior of Non-Abelian Gauge Theories}}, \href{https://doi.org/10.1103/PhysRevLett.30.1343}{\emph{Phys. Rev. Lett.} {\bfseries 30} (1973) 1343}.

\bibitem{Politzer:1973fx}
H.D.~Politzer, \emph{{Reliable Perturbative Results for Strong Interactions?}}, \href{https://doi.org/10.1103/PhysRevLett.30.1346}{\emph{Phys. Rev. Lett.} {\bfseries 30} (1973) 1346}.

\bibitem{Jackiw:1976pf}
R.~Jackiw and C.~Rebbi, \emph{{Vacuum Periodicity in a Yang--Mills Quantum Theory}}, \href{https://doi.org/10.1103/PhysRevLett.37.172}{\emph{Phys. Rev. Lett.} {\bfseries 37} (1976) 172}.

\bibitem{Abel:2020gbr}
C.~Abel et~al., \emph{{Measurement of the Permanent Electric Dipole Moment of the Neutron}}, \href{https://doi.org/10.1103/PhysRevLett.124.081803}{\emph{Phys. Rev. Lett.} {\bfseries 124} (2020) 081803} [\href{https://arxiv.org/abs/2001.11966}{{\ttfamily 2001.11966}}].

\bibitem{GrillidiCortona:2015jxo}
G.~Grilli~di Cortona, E.~Hardy, J.~Pardo~Vega and G.~Villadoro, \emph{{The QCD axion, precisely}}, \href{https://doi.org/10.1007/JHEP01(2016)034}{\emph{JHEP} {\bfseries 01} (2016) 034} [\href{https://arxiv.org/abs/1511.02867}{{\ttfamily 1511.02867}}].

\bibitem{Peccei:1977hh}
R.D.~Peccei and H.R.~Quinn, \emph{{CP Conservation in the Presence of Pseudoparticles}}, \href{https://doi.org/10.1103/PhysRevLett.38.1440}{\emph{Phys. Rev. Lett.} {\bfseries 38} (1977) 1440}.

\bibitem{Peccei:1977ur}
R.D.~Peccei and H.R.~Quinn, \emph{{Constraints Imposed by CP Conservation in the Presence of Pseudoparticles}}, \href{https://doi.org/10.1103/PhysRevD.16.1791}{\emph{Phys. Rev. D} {\bfseries 16} (1977) 1791}.

\bibitem{Weinberg:1977ma}
S.~Weinberg, \emph{{A New Light Boson?}}, \href{https://doi.org/10.1103/PhysRevLett.40.223}{\emph{Phys. Rev. Lett.} {\bfseries 40} (1978) 223}.

\bibitem{Wilczek:1977pj}
F.~Wilczek, \emph{{Problem of Strong P and T Invariance in the Presence of Instantons}}, \href{https://doi.org/10.1103/PhysRevLett.40.279}{\emph{Phys. Rev. Lett.} {\bfseries 40} (1978) 279}.

\bibitem{MohapatraSenjanovic1978}
R.N.~Mohapatra and G.~Senjanovi{\'c}, \emph{{Natural Suppression of Strong P and T Non-invariance}}, \href{https://doi.org/10.1016/0370-2693(78)90243-5}{\emph{Phys. Lett. B} {\bfseries 79} (1978) 283}.

\bibitem{Kim:1979if}
J.E.~Kim, \emph{{Weak-Interaction Singlet and Strong CP Invariance}}, \href{https://doi.org/10.1103/PhysRevLett.43.103}{\emph{Phys. Rev. Lett.} {\bfseries 43} (1979) 103}.

\bibitem{Shifman:1979if}
M.A.~Shifman, A.I.~Vainshtein and V.I.~Zakharov, \emph{{Can Confinement Ensure Natural CP Invariance of Strong Interactions?}}, \href{https://doi.org/10.1016/0550-3213(80)90209-6}{\emph{Nucl. Phys. B} {\bfseries 166} (1980) 493}.

\bibitem{Banks:2010zn}
T.~Banks and N.~Seiberg, \emph{{Symmetries and Strings in Field Theory and Gravity}}, \href{https://doi.org/10.1103/PhysRevD.83.084019}{\emph{Phys. Rev. D} {\bfseries 83} (2011) 084019} [\href{https://arxiv.org/abs/1011.5120}{{\ttfamily 1011.5120}}].

\bibitem{Kamionkowski:1992mf}
M.~Kamionkowski and J.~March-Russell, \emph{{Planck-Scale Physics and the Peccei--Quinn Mechanism}}, \href{https://doi.org/10.1016/0370-2693(92)90492-M}{\emph{Phys. Lett. B} {\bfseries 282} (1992) 137} [\href{https://arxiv.org/abs/hep-th/9202003}{{\ttfamily hep-th/9202003}}].

\bibitem{Holman:1992us}
R.~Holman, S.D.H.~Hsu, T.W.~Kephart, E.W.~Kolb, R.~Watkins and L.M.~Widrow, \emph{{Solutions to the strong CP problem in a world with gravity}}, \href{https://doi.org/10.1016/0370-2693(92)90491-L}{\emph{Phys. Lett. B} {\bfseries 282} (1992) 132} [\href{https://arxiv.org/abs/hep-ph/9203206}{{\ttfamily hep-ph/9203206}}].

\bibitem{Choi2004GaugeAxion}
K.~Choi, \emph{{QCD Axion from a Higher Dimensional Gauge Field Theory}}, \href{https://doi.org/10.1103/PhysRevLett.92.101602}{\emph{Phys. Rev. Lett.} {\bfseries 92} (2004) 101602} [\href{https://arxiv.org/abs/hep-ph/0308024}{{\ttfamily hep-ph/0308024}}].

\bibitem{Reece:2023czb}
M.~Reece, \emph{{TASI Lectures: (No) Global Symmetries to Axion Physics}}, \href{https://doi.org/10.22323/1.439.0008}{\emph{PoS} {\bfseries TASI2022} (2024) 008} [\href{https://arxiv.org/abs/2304.08512}{{\ttfamily 2304.08512}}].

\bibitem{Reece2024Expectations}
M.~Reece, \emph{{Extra-Dimensional Axion Expectations}}, \href{https://doi.org/10.1007/JHEP07(2025)130}{\emph{JHEP} {\bfseries 07} (2025) 130} [\href{https://arxiv.org/abs/2406.08543}{{\ttfamily 2406.08543}}].

\bibitem{CraigKongsore2024}
N.~Craig and M.~Kongsore, \emph{{High-Quality Axions from Higher-Form Symmetries in Extra Dimensions}}, \href{https://doi.org/10.1103/PhysRevD.111.015047}{\emph{Phys. Rev. D} {\bfseries 111} (2025) 015047} [\href{https://arxiv.org/abs/2408.10295}{{\ttfamily 2408.10295}}].

\bibitem{Tong2017LineOperators}
D.~Tong, \emph{{Line Operators in the Standard Model}}, \href{https://doi.org/10.1007/JHEP07(2017)104}{\emph{JHEP} {\bfseries 07} (2017) 104} [\href{https://arxiv.org/abs/1705.01853}{{\ttfamily 1705.01853}}].

\bibitem{ChoiLamShao2022}
Y.~Choi, H.T.~Lam and S.-H.~Shao, \emph{{Non-invertible Global Symmetries in the Standard Model}}, \href{https://doi.org/10.1103/PhysRevLett.129.161601}{\emph{Phys. Rev. Lett.} {\bfseries 129} (2022) 161601} [\href{https://arxiv.org/abs/2205.05086}{{\ttfamily 2205.05086}}].

\bibitem{Gaiotto2015Generalized}
D.~Gaiotto, A.~Kapustin, N.~Seiberg and B.~Willett, \emph{{Generalized Global Symmetries}}, \href{https://doi.org/10.1007/JHEP02(2015)172}{\emph{JHEP} {\bfseries 02} (2015) 172} [\href{https://arxiv.org/abs/1412.5148}{{\ttfamily 1412.5148}}].

\bibitem{KamalMahanta2026}
A.R.~Kamal and R.~Mahanta, \emph{Time-dependent flux backgrounds in type {IIB} supergravity}, \href{https://doi.org/10.1007/JHEP06(2026)188}{\emph{JHEP} {\bfseries 06} (2026) 188} [\href{https://arxiv.org/abs/2512.19793}{{\ttfamily 2512.19793}}].

\bibitem{ArkoEtAl2026}
K.F.R.~Arko, W.~Hasan, M.F.~Hoque and A.R.~Kamal, \emph{Big crunching your big bang: Doubly warped time-dependent type {II} flux backgrounds and scale separation},  \href{https://arxiv.org/abs/2610.30248}{{\ttfamily 2610.30248}}.

\bibitem{Fujikawa:1979ay}
K.~Fujikawa, \emph{{Path-Integral Measure for Gauge-Invariant Fermion Theories}}, \href{https://doi.org/10.1103/PhysRevLett.42.1195}{\emph{Phys. Rev. Lett.} {\bfseries 42} (1979) 1195}.

\bibitem{vanBeest:2021lhn}
M.~van Beest, J.~Calder{\'o}n-Infante, D.~Mirfendereski and I.~Valenzuela, \emph{{Lectures on the Swampland Program in String Compactifications}}, \href{https://doi.org/10.1016/j.physrep.2022.09.002}{\emph{Phys. Rept.} {\bfseries 989} (2022) 1} [\href{https://arxiv.org/abs/2102.01111}{{\ttfamily 2102.01111}}].

\bibitem{Witten1984Superstrings}
E.~Witten, \emph{{Some Properties of O(32) Superstrings}}, \href{https://doi.org/10.1016/0370-2693(84)90422-2}{\emph{Phys. Lett. B} {\bfseries 149} (1984) 351}.

\bibitem{ChoiKim1985Axions}
K.~Choi and J.E.~Kim, \emph{{Compactification and Axions in E(8) x E(8)-prime Superstring Models}}, \href{https://doi.org/10.1016/0370-2693(85)90693-8}{\emph{Phys. Lett. B} {\bfseries 165} (1985) 71}.

\bibitem{SvrcekWitten2006}
P.~Svrcek and E.~Witten, \emph{{Axions In String Theory}}, \href{https://doi.org/10.1088/1126-6708/2006/06/051}{\emph{JHEP} {\bfseries 06} (2006) 051} [\href{https://arxiv.org/abs/hep-th/0605206}{{\ttfamily hep-th/0605206}}].

\bibitem{Chakraborty:2025wqn}
D.~Chakraborty, M.~Hai, S.T.~Jahan, A.R.~Kamal and M.S.J.~Shuvo, \emph{{Effect of moduli redefinitions on fibre inflation}}, \href{https://doi.org/10.1088/1475-7516/2026/06/025}{\emph{JCAP} {\bfseries 06} (2026) 025} [\href{https://arxiv.org/abs/2511.19610}{{\ttfamily 2511.19610}}].

\bibitem{Conlon2006Axion}
J.P.~Conlon, \emph{{The QCD Axion and Moduli Stabilisation}}, \href{https://doi.org/10.1088/1126-6708/2006/05/078}{\emph{JHEP} {\bfseries 05} (2006) 078} [\href{https://arxiv.org/abs/hep-th/0602233}{{\ttfamily hep-th/0602233}}].

\bibitem{BurgessChoiQuevedo2024}
C.P.~Burgess, G.~Choi and F.~Quevedo, \emph{{UV and IR Effects in Axion Quality Control}}, \href{https://doi.org/10.1007/JHEP03(2024)051}{\emph{JHEP} {\bfseries 03} (2024) 051} [\href{https://arxiv.org/abs/2301.00549}{{\ttfamily 2301.00549}}].

\bibitem{BurgessQuevedo2025}
C.P.~Burgess and F.~Quevedo, \emph{{Dual is Different: EFTs, Axions and Nonpropagating Form Fields in Cosmology}}, \href{https://doi.org/10.1007/JHEP10(2026)069}{\emph{JHEP} {\bfseries 10} (2026) 069} [\href{https://arxiv.org/abs/2509.11340}{{\ttfamily 2509.11340}}].

\bibitem{ArkaniHamed2003Extranatural}
N.~Arkani-Hamed, H.-C.~Cheng, P.~Creminelli and L.~Randall, \emph{{Extranatural Inflation}}, \href{https://doi.org/10.1103/PhysRevLett.90.221302}{\emph{Phys. Rev. Lett.} {\bfseries 90} (2003) 221302} [\href{https://arxiv.org/abs/hep-th/0301218}{{\ttfamily hep-th/0301218}}].

\bibitem{KaloperSorbo2009}
N.~Kaloper and L.~Sorbo, \emph{{A Natural Framework for Chaotic Inflation}}, \href{https://doi.org/10.1103/PhysRevLett.102.121301}{\emph{Phys. Rev. Lett.} {\bfseries 102} (2009) 121301} [\href{https://arxiv.org/abs/0811.1989}{{\ttfamily 0811.1989}}].

\bibitem{RubinRoth1983}
M.A.~Rubin and B.D.~Roth, \emph{{Fermions and stability in five-dimensional Kaluza-Klein theory}}, \href{https://doi.org/10.1016/0370-2693(83)91629-5}{\emph{Phys. Lett. B} {\bfseries 127} (1983) 55}.

\bibitem{PontonPoppitz2001}
E.~Ponton and E.~Poppitz, \emph{{Casimir Energy and Radius Stabilization in Five and Six Dimensional Orbifolds}}, \href{https://doi.org/10.1088/1126-6708/2001/06/019}{\emph{JHEP} {\bfseries 06} (2001) 019} [\href{https://arxiv.org/abs/hep-ph/0105021}{{\ttfamily hep-ph/0105021}}].

\bibitem{GreeneLevin2007}
B.R.~Greene and J.~Levin, \emph{{Dark Energy and Stabilization of Extra Dimensions}}, \href{https://doi.org/10.1088/1126-6708/2007/11/096}{\emph{JHEP} {\bfseries 11} (2007) 096} [\href{https://arxiv.org/abs/0707.1062}{{\ttfamily 0707.1062}}].

\bibitem{BlancoPilladoShlaer2010}
J.J.~Blanco-Pillado and B.~Shlaer, \emph{{Bubbles of Nothing in Flux Compactifications}}, \href{https://doi.org/10.1103/PhysRevD.82.086015}{\emph{Phys. Rev. D} {\bfseries 82} (2010) 086015} [\href{https://arxiv.org/abs/1002.4408}{{\ttfamily 1002.4408}}].

\bibitem{TakahashiYinGuth2018}
F.~Takahashi, W.~Yin and A.H.~Guth, \emph{{QCD} axion window and low-scale inflation}, \href{https://doi.org/10.1103/PhysRevD.98.015042}{\emph{Phys. Rev. D} {\bfseries 98} (2018) 015042} [\href{https://arxiv.org/abs/1805.08763}{{\ttfamily 1805.08763}}].

\bibitem{GibbonsHawking1977}
G.W.~Gibbons and S.W.~Hawking, \emph{Cosmological event horizons, thermodynamics, and particle creation}, \href{https://doi.org/10.1103/PhysRevD.15.2738}{\emph{Phys. Rev. D} {\bfseries 15} (1977) 2738}.

\bibitem{DobadoMaroto2001}
A.~Dobado and A.L.~Maroto, \emph{{The dynamics of the Goldstone bosons on the brane}}, \href{https://doi.org/10.1016/S0550-3213(00)00574-5}{\emph{Nucl. Phys. B} {\bfseries 592} (2001) 203} [\href{https://arxiv.org/abs/hep-ph/0007100}{{\ttfamily hep-ph/0007100}}].

\bibitem{CicoliEtAl2022Hydra}
M.~Cicoli, A.~Hebecker, J.~Jaeckel and M.~Wittner, \emph{{Axions in string theory --- slaying the Hydra of dark radiation}}, \href{https://doi.org/10.1007/JHEP09(2022)198}{\emph{JHEP} {\bfseries 09} (2022) 198} [\href{https://arxiv.org/abs/2203.08833}{{\ttfamily 2203.08833}}].

\bibitem{Saikawa:2018rcs}
K.~Saikawa and S.~Shirai, \emph{Primordial gravitational waves, precisely: The role of thermodynamics in the standard model}, \href{https://doi.org/10.1088/1475-7516/2018/05/035}{\emph{JCAP} {\bfseries 05} (2018) 035} [\href{https://arxiv.org/abs/1803.01038}{{\ttfamily 1803.01038}}].

\bibitem{Chakraborty:2026oac}
D.~Chakraborty and A.R.~Kamal, \emph{{Fire at the tip of the throat: Hagedorn phase after brane-antibrane inflation?}}, \href{https://doi.org/10.1140/epjc/s10052-026-16470-w}{\emph{Eur. Phys. J. C} {\bfseries 86} (2026) 1153} [\href{https://arxiv.org/abs/2606.08767}{{\ttfamily 2606.08767}}].

\bibitem{Planck2018Inflation}
{\scshape Planck} collaboration, \emph{{Planck 2018 results. X. Constraints on inflation}}, \href{https://doi.org/10.1051/0004-6361/201833887}{\emph{Astron. Astrophys.} {\bfseries 641} (2020) A10} [\href{https://arxiv.org/abs/1807.06211}{{\ttfamily 1807.06211}}].

\bibitem{AtacamaCosmologyTelescope:2025blo}
{\scshape Atacama Cosmology Telescope} collaboration, \emph{{The Atacama Cosmology Telescope: DR6 power spectra, likelihoods and {\ensuremath{\Lambda}}CDM parameters}}, \href{https://doi.org/10.1088/1475-7516/2025/11/062}{\emph{JCAP} {\bfseries 11} (2025) 062} [\href{https://arxiv.org/abs/2503.14452}{{\ttfamily 2503.14452}}].

\bibitem{AtacamaCosmologyTelescope:2025nti}
{\scshape Atacama Cosmology Telescope} collaboration, \emph{{The Atacama Cosmology Telescope: DR6 constraints on extended cosmological models}}, \href{https://doi.org/10.1088/1475-7516/2025/11/063}{\emph{JCAP} {\bfseries 11} (2025) 063} [\href{https://arxiv.org/abs/2503.14454}{{\ttfamily 2503.14454}}].

\bibitem{Hasegawa2019Reheating}
T.~Hasegawa, N.~Hiroshima, K.~Kohri, R.S.L.~Hansen, T.~Tram and S.~Hannestad, \emph{{MeV-scale reheating temperature and thermalization of oscillating neutrinos by radiative and hadronic decays of massive particles}}, \href{https://doi.org/10.1088/1475-7516/2019/12/012}{\emph{JCAP} {\bfseries 12} (2019) 012} [\href{https://arxiv.org/abs/1908.10189}{{\ttfamily 1908.10189}}].

\bibitem{Borsanyi2016Cosmology}
S.~Borsanyi et~al., \emph{{Calculation of the axion mass based on high-temperature lattice quantum chromodynamics}}, \href{https://doi.org/10.1038/nature20115}{\emph{Nature} {\bfseries 539} (2016) 69} [\href{https://arxiv.org/abs/1606.07494}{{\ttfamily 1606.07494}}].

\bibitem{VisinelliGondolo2010}
L.~Visinelli and P.~Gondolo, \emph{{Axion cold dark matter in non-standard cosmologies}}, \href{https://doi.org/10.1103/PhysRevD.81.063508}{\emph{Phys. Rev. D} {\bfseries 81} (2010) 063508} [\href{https://arxiv.org/abs/0912.0015}{{\ttfamily 0912.0015}}].

\bibitem{ChoiGherghetta2026Isocurvature}
G.~Choi and T.~Gherghetta, \emph{{Suppressing extra-dimensional axion isocurvature dynamically}}, \href{https://doi.org/10.1103/pbs4-gg4r}{\emph{Phys. Rev. D} (2026) } [\href{https://arxiv.org/abs/2607.09969}{{\ttfamily 2607.09969}}].

\bibitem{AbeEtAl2014}
Y.~Abe, T.~Inami, Y.~Kawamura and Y.~Koyama, \emph{{Radion stabilization in the presence of a Wilson line phase}}, \href{https://doi.org/10.1093/ptep/ptu097}{\emph{PTEP} {\bfseries 2014} (2014) 073B04} [\href{https://arxiv.org/abs/1404.5125}{{\ttfamily 1404.5125}}].

\bibitem{KitanoSuzukiYin2023}
R.~Kitano, M.~Suzuki and W.~Yin, \emph{Role of {QCD} in moduli stabilization during inflation and axion dark matter}, \href{https://doi.org/10.1007/JHEP11(2023)103}{\emph{JHEP} {\bfseries 11} (2023) 103} [\href{https://arxiv.org/abs/2307.15059}{{\ttfamily 2307.15059}}].

\bibitem{CicoliConlonQuevedo2013}
M.~Cicoli, J.P.~Conlon and F.~Quevedo, \emph{{Dark Radiation in LARGE Volume Models}}, \href{https://doi.org/10.1103/PhysRevD.87.043520}{\emph{Phys. Rev. D} {\bfseries 87} (2013) 043520} [\href{https://arxiv.org/abs/1208.3562}{{\ttfamily 1208.3562}}].

\bibitem{KKLT2003}
S.~Kachru, R.~Kallosh, A.~Linde and S.P.~Trivedi, \emph{{de Sitter Vacua in String Theory}}, {\emph{Phys. Rev. D} {\bfseries 68} (2003) 046005} [\href{https://arxiv.org/abs/hep-th/0301240}{{\ttfamily hep-th/0301240}}].

\bibitem{LVS2005}
V.~Balasubramanian, P.~Berglund, J.P.~Conlon and F.~Quevedo, \emph{{Systematics of Moduli Stabilisation in Calabi-Yau Flux Compactifications}}, {\emph{JHEP} {\bfseries 03} (2005) 007} [\href{https://arxiv.org/abs/hep-th/0502058}{{\ttfamily hep-th/0502058}}].

\bibitem{Cicoli:2012sz}
M.~Cicoli, M.~Goodsell and A.~Ringwald, \emph{{The type IIB string axiverse and its low-energy phenomenology}}, \href{https://doi.org/10.1007/JHEP10(2012)146}{\emph{JHEP} {\bfseries 10} (2012) 146} [\href{https://arxiv.org/abs/1206.0819}{{\ttfamily 1206.0819}}].

\bibitem{CicoliEtAl2024Brane}
M.~Cicoli, C.~Hughes, A.R.~Kamal, F.~Marino, F.~Quevedo, M.~Ramos-Hamud et~al., \emph{{Back to the origins of brane-antibrane inflation}}, {\emph{Eur. Phys. J. C} {\bfseries 85} (2025) 315} [\href{https://arxiv.org/abs/2410.00097}{{\ttfamily 2410.00097}}].

\bibitem{HaiEtAl2025Perturbative}
M.~Hai, A.R.~Kamal, N.F.~Shamma and M.S.J.~Shuvo, \emph{{Perturbative K\"ahler Moduli Inflation}}, \href{https://doi.org/10.1142/S0218271826500719}{\emph{Int. J. Mod. Phys. D} {\bfseries 35} (2026) 2650071} [\href{https://arxiv.org/abs/2506.08083}{{\ttfamily 2506.08083}}].

\bibitem{CoxGherghettaNguyen2020}
P.~Cox, T.~Gherghetta and M.D.~Nguyen, \emph{{A Holographic Perspective on the Axion Quality Problem}}, {\emph{JHEP} {\bfseries 01} (2020) 188} [\href{https://arxiv.org/abs/1911.09385}{{\ttfamily 1911.09385}}].

\bibitem{ChoiLeeShin2026}
K.~Choi, C.H.~Lee and C.S.~Shin, \emph{Axion quality in warped extra-dimension},  \href{https://arxiv.org/abs/2604.08700}{{\ttfamily 2604.08700}}.

\bibitem{FernandezNavarro2026}
M.~Fern{\'a}ndez~Navarro, \emph{The minimal high-quality {QCD} axion},  \href{https://arxiv.org/abs/2607.08814}{{\ttfamily 2607.08814}}.

\bibitem{ChoiGherghetta2025}
G.~Choi and T.~Gherghetta, \emph{An extra-dimensional axion in a {5D} warped orbifold {GUT}}, \href{https://doi.org/10.1007/JHEP03(2026)169}{\emph{JHEP} {\bfseries 03} (2026) 169} [\href{https://arxiv.org/abs/2512.15666}{{\ttfamily 2512.15666}}].

\bibitem{LoladzePlatschorreReig2025}
V.~Loladze, A.~Platschorre and M.~Reig, \emph{{Higher Axion Strings}}, \href{https://doi.org/10.1007/JHEP08(2025)182}{\emph{JHEP} {\bfseries 08} (2025) 182} [\href{https://arxiv.org/abs/2503.18707}{{\ttfamily 2503.18707}}].

\bibitem{Reig2021Stochastic}
M.~Reig, \emph{{The Stochastic Axiverse}}, {\emph{JHEP} {\bfseries 09} (2021) 207} [\href{https://arxiv.org/abs/2104.09923}{{\ttfamily 2104.09923}}].

\bibitem{Iqbal2024Lectures}
N.~Iqbal, \emph{{Jena lectures on generalized global symmetries: principles and applications}},  \href{https://arxiv.org/abs/2407.20815}{{\ttfamily 2407.20815}}.

\bibitem{Gomes2023HigherForm}
P.R.S.~Gomes, \emph{{An Introduction to Higher-Form Symmetries}}, \href{https://doi.org/10.21468/SciPostPhysLectNotes.74}{\emph{SciPost Phys. Lect. Notes} {\bfseries 74} (2023) 1} [\href{https://arxiv.org/abs/2303.01817}{{\ttfamily 2303.01817}}].

\bibitem{Bhardwaj:2023kri}
L.~Bhardwaj, L.E.~Bottini, L.~Fraser-Taliente, L.~Gladden, D.S.W.~Gould, A.~Platschorre et~al., \emph{{Lectures on generalized symmetries}}, \href{https://doi.org/10.1016/j.physrep.2023.11.002}{\emph{Phys. Rept.} {\bfseries 1051} (2024) 1} [\href{https://arxiv.org/abs/2307.07547}{{\ttfamily 2307.07547}}].

\end{thebibliography}\endgroup
\end{document}